\documentclass[12pt]{article}

\usepackage{ amssymb }
\usepackage{latexsym}
\usepackage{graphicx}
\usepackage[reqno]{amsmath}
\usepackage{natbib}
\usepackage{multirow}
\usepackage{url}
\usepackage{color}
\usepackage{breqn}
\usepackage{hyperref}
\usepackage{leftidx}
\usepackage{setspace}

\usepackage{subcaption}

\begin{document}

\setlength{\abovedisplayskip}{4pt}
\setlength{\belowdisplayskip}{4pt}

\title{\vspace{-2cm} \Large \bf Infinite Mixture Models for Improved Modeling of Across-Site Evolutionary Variation}

\vspace{0.2cm}

\author{Mandev S.~Gill$^{1,2}$, Guy Baele$^{3}$, Marc A.~Suchard$^{4,5,6}$ and Philippe Lemey$^{3}$}

\date{}

\maketitle

\begin{center}
\small
$^1$Department of Statistics, University of Georgia, Athens, GA, USA \\
$^2$Institute of Bioinformatics, University of Georgia, Athens, GA, USA\\
$^3$Department of Microbiology, Immunology and Transplantation, Rega Institute, KU Leuven, Leuven, Belgium\\
$^4$Department of Human Genetics, David Geffen School of Medicine, University of California, Los Angeles, CA, USA\\
$^5$Department of Biomathematics, David Geffen School of Medicine, University of California, Los Angeles, CA, USA\\
$^6$Department of Biostatistics, Jonathan and Karin Fielding School of Public Health, University of California, Los Angeles, CA, USA\\
 \end{center}
\medskip
%\begin{center}
 % {\bf Running head}: Infinite Mixture Models for Improved Modeling of Across-Site Evolutionary Variation \\ \ \\
  %{\bf Keywords}: Phylogenetics, Bayesian Statistics, Bayesian Nonparametrics, Substitution Models,
  %Across-Site Variation, Model Selection, Virus Evolution \\ \ \\
  %{\bf Corresponding Author}: E-mail: philippe.lemey@kuleuven.be \\
 
%\end{center}

%\clearpage

%\doublespacing

\begin{abstract}
Scientific studies in many areas of biology routinely employ evolutionary analyses based on the probabilistic inference of phylogenetic trees from molecular sequence data. Evolutionary processes that act at the molecular level are highly variable, and properly accounting for heterogeneity in evolutionary processes is crucial for more accurate phylogenetic inference. Nucleotide substitution rates and patterns are known to vary among sites in multiple sequence alignments, and such variation can be modeled by partitioning alignments into categories corresponding to different substitution models. Determining \textit{a priori} appropriate partitions can be difficult, however, and better model fit can be achieved through flexible Bayesian infinite mixture models that simultaneously infer the number of partitions, the partition that each site belongs to, and the evolutionary parameters corresponding to each partition. Here, we consider several different types of infinite mixture models, including classic Dirichlet process mixtures, as well as novel approaches for modeling across-site evolutionary variation: hierarchical models for data with a natural group structure, and infinite hidden Markov models that account for spatial patterns in alignments. In analyses of several viral data sets, we find that different types of infinite mixture models emerge as the best choices in different scenarios. To enable these models to scale efficiently to large data sets, we adapt efficient Markov chain Monte Carlo algorithms and exploit opportunities for parallel computing. We implement this infinite mixture modeling framework in BEAST X, a widely-used software package for Bayesian phylogenetic inference.
\end{abstract}

\clearpage

\renewcommand{\baselinestretch}{1}

\section{Introduction}

Probabilistic phylogenetic inference requires statistical models for molecular sequence evolution. Evolutionary processes are typically described using Markov models for the substitution of discrete molecular characters, such as DNA bases \citep{Felsenstein2004}. Each observed molecular sequence can be thought of as corresponding to the tip of an unobserved phylogenetic tree and produced by a Markov model that starts at the root of the tree and proceeds down its branches. Calculation of the observed data likelihood under such a model forms the basis for both maximum likelihood and Bayesian phylogenetic inference \citep{Felsenstein81}.

Evolutionary processes are known to be highly variable \citep{Yang2006}, and evolutionary modeling has been gradually refined from early, restrictive approaches to better account for such variability and in turn enable more accurate phylogenetic inferences. The simplest nucleotide substitution model assumes equal DNA base equilibrium frequencies as well as equal relative exchange rates between bases, with an overall substitution rate being the only model parameter \citep{Jukes1969}. Numerous extensions of this model have been developed to enable variation in relative exchange rates \citep{Kimura1980}, equilibrium frequencies \citep{Felsenstein81}, and both \citep{Hasegawa1985, Lanave1984, Tamura1992, Tamura1993}. While early evolutionary modeling approaches assumed that such substitution models remained constant across the branches of the phylogenetic tree and the sites of the multiple sequence alignment, these restrictions have since also been relaxed in various ways \citep{Yang2006}. 

Different alignment sites in molecular sequences have different functional and structural roles and are subject to different selective pressures and thus do not necessarily evolve in the same ways \citep{Yang2006}. Modeling advances for across-site variation initially focused on allowing the overall substitution rates of substitution models to vary along alignment sites. However, alignment sites may also evolve in qualitatively different ways. That is, the substitution pattern, which is characterized by the relative exchange rates of molecular characters in a substitution model, can also vary. A widely used method for modeling variation of substitution rates and/or relative character exchange rates is to model the parameters as random variables that are drawn from a common distribution for all sites \citep{Nei1976, Golding1983, Yang1993, Yang1994, Waddell1997, Huelsenbeck1999}. A common alternative approach is to partition alignment sites into a fixed number of different categories and estimate substitution rates and/or relative character exchange rates independently for each category. While it is possible to treat each site as belonging to its own category \citep{Bruno1996, Swofford1996, Nielsen1997}, this can lead to overfitting. More commonly used partitioning schemes are biologically informed. For example, sites can be partitioned by gene, by stems and loops in ribosomal data, or by codon position for protein coding sequences \citep{PagelMeade2004}. However, it may be not be clear how to best partition a given data set \textit{a priori}, and moreover, any fixed partition would fail to account for partition uncertainty. There may be some sites that appear to clearly belong to one of the fixed partitions while other sites may be more appropriately modeled as belonging to different partitions with different probabilities. In the face of such challenges, finite mixture models have emerged as popular and effective approaches \citep{PagelMeade2004, Yang1994, Huelsenbeck1999, Venditti2008}. 

A major limitation of finite mixture models is the necessity to fix an \textit{a priori} number of categories. Researchers have overcome this obstacle through the development of Bayesian infinite mixture models that treat the number of categories as an unknown and unbounded parameter \citep{Lartillot2004, Huelsenbeck2007b, Wu2013}. Notably, by inferring the number of categories for an evolutionary mixture model along with all other model parameters, these approaches enable the data to determine the number of alignment partitions that best captures the heterogeneity of the evolutionary process that generated the data. Infinite mixture models for evolutionary heterogeneity have thus far relied on Dirichlet processes \citep{Ferguson1973} to specify prior probabilities on a space of discrete distributions that can generate site-to-category assignments and evolutionary parameters corresponding to each category. This prior specification affords the flexibility to adequately model evolutionary variation while also providing a built-in mechanism that guards against overfitting \citep{Ghahramani2013}, and Dirichlet process mixture models have been shown to routinely achieve better model fit than standard models for across-site evolutionary variation \citep{Huelsenbeck2007b, Wu2013}. However, Dirichlet processes have limitations that may hamper their effectiveness at modeling across-site variation in certain scenarios. Specifically, Dirichlet process mixture models are unable to account for any spatial correlation in evolutionary parameters, which may be expected for evolutionary rates at adjacent sites. To model spatially correlated patterns of evolutionary rates along multiple sequence alignments, \citet{Yang1995} and \citet{Felsenstein1996} use hidden Markov models to specify finite mixture models.

Here, we build on the success of Dirichlet process mixtures in modeling across-site variation by exploiting methods from Bayesian nonparametrics that can overcome some of the Dirichlet process' limitations. To model spatial patterns along alignments, we employ an infinite hidden Markov model \citep{Beal2002} that expands the hidden Markov model to a countably infinite state space. We also use a hierarchical Dirichlet process framework \citep{Teh2006} that posits different Dirichlet processes for different groups of sites while linking the separate Dirichlet processes to pool information and share statistical strength. We compare the performance of the resulting infinite mixture models as well as standard approaches for modeling across-site evolutionary variation through analyses of respiratory syncytial virus subgroup A data, hepatitis C subtype 4 data, and rabies virus data. We find that infinite mixture models based on infinite hidden Markov models and hierarchical Dirichlet processes can yield substantially better model fit than Dirichlet process mixtures. While infinite mixture models emerge as a clearly preferable alternative to standard approaches, the best performing infinite mixture model varies. In particular, for each of the three types of infinite mixtures, we observe at least one scenario in which it outperforms the other two.

\section{Methods}

Consider a multiple sequence alignment of molecular sequence data $\textbf X$. Let $n$ denote the number of sequences and  $s$ the number of sites, and let $\textbf X_i$ denote the observed molecular sequence characters at alignment site $i$.  We assume that the data $\textbf X$ are generated by continuous-time Markov chains that act independently at the different alignment sites. Each Markov chain begins at the root node of an unobserved phylogenetic tree $\tau$ and acts independently along its lineages to ultimately produce the observed molecular sequence data at its external nodes. The Markov chain for site $i$ is characterized by evolutionary parameters $\boldsymbol \theta_i = (\rho_i,\textbf Q_i)$. Here, $\textbf Q_i$ is a matrix specifying the relative exchange rates of molecular characters. The matrix $\textbf Q_i$ is normalized so that its expected substitution rate is 1, the parameter $\rho_i$ is the overall substitution rate, and the product $\rho_i \textbf Q_i$ is the Markov chain's infinitesimal rate matrix. 

The free parameters that characterize the rate matrix $\textbf Q_i$ depend on the substitution model. For example, the popular HKY model for nucleotide substitution \citep{Hasegawa1985} posits a bias between the rates of \textit{transitions} (substitutions between two purines, bases $A$ and $G$, or between two pyrimidines, bases $T$ and $C$) and \textit{transversions} (all other types of substitutions). In addition to the transition/transversion ratio $\kappa$, the free parameters also include nucleotide base equilibrium frequencies $\pi_A,\pi_C,\pi_G,\pi_T$. The HKY model is one of several popular time-reversible nucleotide substitution models. Although there is no biological reason to believe that the substitution process should be time-reversible, the computational convenience of time-reversible models has led to their widespread use \citep{Yang2006}. The most flexible model for nucleotide substitution that ensures time-reversibility of the Markov chain is the general time-reversible (GTR) model \citep{Lanave1984, Tavare1986}. Its $\textbf Q_i$ matrix can be specified with four nucleotide base equilibrium frequencies and five additional relative rate parameters. 

Under these modeling assumptions, the full data likelihood can be expressed as
\begin{equation*}
P(\textbf X | \tau, \boldsymbol \theta) = \prod_{i=1}^s P(\textbf X_i | \tau, \boldsymbol \theta_i),
\end{equation*}
where $\boldsymbol \theta = (\boldsymbol \theta_1,\dots,\boldsymbol \theta_s)$. Allowing the evolutionary parameters $\boldsymbol \theta_1,\dots,\boldsymbol \theta_s$ to all assume distinct values would lead to overfitting. Instead, we want multiple $\boldsymbol \theta_i$ to be able to take on the same value. We think of each distinct value assumed by at least one $\boldsymbol \theta_i$ as determining an evolutionary category, where the number of categories as well as the assignment of alignment sites to categories depend on the model for across-site evolutionary variation. Suppose there are $K$ distinct evolutionary categories with corresponding parameters $\boldsymbol \phi_1,\dots,\boldsymbol \phi_K$, and let $z_i \in \{1,\dots,K \}$ denote the evolutionary category corresponding to alignment site $i$ for $i = 1,\dots,s$. Thus $\boldsymbol \theta_i = \boldsymbol \phi_{z_i}$. To account for uncertainty in the number of evolutionary categories and in the partitioning of alignment sites into evolutionary categories, we model $K$ and $z_1,\dots,z_s$ as random variables. This leads to the following Bayesian evolutionary model:
\begin{equation*}
P(\tau,\boldsymbol \phi, \textbf z, K | \textbf X) \propto \left[  \prod_{i=1}^s P(\textbf X_i | \tau, \phi_{z_i} )\right] P(\boldsymbol \phi, \textbf z, K) P(\tau),
\end{equation*}
where $\boldsymbol \phi = (\boldsymbol \phi_1,\dots,\boldsymbol \phi_K)$ and $\textbf z = (z_1,\dots,z_s)$. 
To specify a prior distribution for $\boldsymbol \phi, \textbf z$ and $K$, it suffices to specify a prior distribution for the $\boldsymbol \theta_i$. We desire a prior distribution $G$ for the $\boldsymbol \theta_i$ that is discrete and has a countably infinite support. The discreteness will allow the number of distinct values $K$ to be strictly less than $s$. While $K$ cannot be greater than the number of alignment sites $s$, the countably infinite support of $G$ will allow $K$ to assume any value between 1 and $s$ without having to make any adjustments for data sets with different numbers of sites. Because the number of evolutionary categories $K$ in this framework is not fixed, it is theoretically unbounded and is said to yield an \textit{infinite} mixture model. Rather than fix the distribution $G$, we model its uncertainty by specifying a prior distribution on $G$ itself. The subfield of Bayesian inference that focuses on such models without a fixed number of model parameters has come to be known as \textit{Bayesian nonparametrics}. We consider three different kinds of Bayesian nonparametric prior distributions for $G$: Dirichlet processes, hierarchical Dirichlet processes, and infinite hidden Markov models.

\subsection{Dirichlet Processes}

Dirichlet processes \citep{Ferguson1973} are widely used to specify infinite mixture models \citep{Antoniak1974}. A Dirichlet process defines a distribution for a random probability measure. Consider a positive scalar $\alpha$ known as the \textit{concentration parameter} and a probability distribution $G_0$ known as the \textit{base distribution}. We can specify a Dirichlet process in terms of the concentration parameter and base distribution, denoted $\text{DP}(\alpha,G_0)$, as follows. Let $\boldsymbol \phi_1, \boldsymbol \phi_2, \dots$ be a sequence of independent random variables distributed according to $G_0$, and let $\beta_1, \beta_2, \dots$ be a sequence of independent random variables that follow a $\text{Beta}(1,\alpha)$ distribution. Define the random variables $\pi_1,\pi_2,\dots$ by 
\begin{equation*}
\pi_k = \beta_k \prod_{l=1}^{k-1} (1-\beta_l)
\end{equation*}
for $k=1,2,\dots$, and let $\delta_{\boldsymbol \phi_k}$ denote the Dirac measure, where $\delta_{\boldsymbol \phi_k}(A) = 1$ if the set $A$ contains $\boldsymbol \phi_k$ and $\delta_{\phi_k}(A) = 0$ otherwise. Then the random probability measure
\begin{equation*}
G = \sum_{k=1}^{\infty} \pi_k \delta_{\boldsymbol \phi_k}
\end{equation*}
is distributed according to $\text{DP}(\alpha,G_0)$ \citep{Sethuraman1994}. Note that the measure is random because the $\pi_k$ and $\boldsymbol \phi_k$ are random variables rather than fixed values. From this construction, it is clear that a draw from a Dirichlet process will be a discrete distribution with a countably infinite support consisting of the \textit{atoms} $\boldsymbol \phi_1, \boldsymbol \phi_2, \dots$. Each atom $\boldsymbol \phi_k$ is associated with a \textit{weight} $\pi_k$, and $\sum_{k=1}^{\infty} \pi_k = 1$ with probability 1. The concentration parameter $\alpha$ determines the level of discretization: as $\alpha$ becomes smaller, draws from a Dirichlet process will have mass increasingly concentrated among a smaller number of atoms, while as $\alpha$ tends to $\infty$, the draws will be closer to continuous distributions. The value of $\alpha$ can be fixed beforehand, or it can be treated as a random variable and inferred from the data along with all other model parameters. 

An alternative perspective on Dirichlet processes can be obtained by considering a conditional distribution characterizing a sequence of independent and indentically distributed draws $\boldsymbol \theta_1, \dots, \boldsymbol \theta_n | G  \sim G$, where $G \sim \text{DP}(\alpha, G_0)$.  \citet{Blackwell1973} show that $G$ can be integrated out to obtain
\begin{eqnarray*}
\boldsymbol \theta_n | \boldsymbol \theta_1,\dots,\boldsymbol \theta_{n-1}, \alpha, G_0 \sim \sum_{k=1}^K \frac{m_k}{n-1 + \alpha} \delta_{\boldsymbol \phi_{k}} +  \frac{\alpha}{n-1 + \alpha} G_0.
\end{eqnarray*}
Here, $m_k$ denotes the number of $\boldsymbol \theta_i$, where $1 \leq i \leq n-1$, such that $\boldsymbol \theta_i = \boldsymbol \phi_k$. This conditional distribution can be understood through the metaphor of a Chinese restaurant, which emerged from the underlying distribution on partitions that became known as the \textit{Chinese restaurant process} \citep{Aldous1985}. In the metaphor, customers sequentially enter a Chinese restaurant with an infinite number of tables, each serving a unique dish.
Each customer $i$ corresponds to a $\boldsymbol \theta_i$ and each dish (and table) $k$ corresponds to a $\boldsymbol \phi_k$. The $n^{\text{th}}$ customer sits at an occupied table $k$ with probability proportional to the number of customers $m_k$ already sitting at the table, and sits at an unoccupied table with probability proportional to $\alpha$. The ``dishes'' $\boldsymbol \phi_k$ can be thought of as being independently sampled from the ``menu'' distribution $G_0$.

\subsection{Hierarchical Dirichlet Processes}

Many standard approaches for modeling variability in evolutionary processes partition multiple sequence alignment sites into different groups on the basis that the sites in each group are believed to exhibit similar evolutionary dynamics. From the perspective of infinite mixture models, it is then natural to wonder if, rather than assuming that evolutionary parameter values for all sites are distributed according to draws from a single Dirichlet process, different groups of sites are better represented by different Dirichlet processes. While it is possible to model different groups with independent mixture models, it is appealing to consider a hierarchical Bayesian framework that allows for variation of the model between groups while still sharing information across groups. 

\citet{Teh2006} propose a hierarchical Dirichlet process with a random probability measure $G_j$ for each group $j$, where each $G_j$ is distributed according to a Dirichlet process defined by concentration parameter $\alpha$ and base distribution $G_0$. The base distribution $G_0$ is itself distributed according to a Dirichlet process, characterized by concentration parameter $\gamma$ and base distribution $H$. Because the $G_j$ all inherit their sets of atoms from the same discrete base distribution $G_0$, they all share the same atoms. The $G_j$ differ from one another in the weights that they associate with the atoms. To summarize, if we divide the alignment sites into $J$ groups, where group $j$ has $n_j$ sites and $\boldsymbol \theta_{j1}, \dots, \boldsymbol \theta_{jn_j}$ are the evolutionary parameters associated with them, we have
\begin{eqnarray*}
G_0 | \gamma, H & \sim & \text{DP}(\gamma, H) \\
G_j | \alpha, G_0 & \sim & \text{DP}(\alpha, G_0) \text{ for } j = 1,\dots, J \\
\boldsymbol \theta_{ji} | G_j & \sim & G_j \text{ for } i = 1,\dots, n_j.
\end{eqnarray*}

\citet{Teh2006} extend the Chinese restaurant metaphor for Dirichlet processes to a \textit{Chinese restaurant franchise} for hierarchical Dirichlet processes. The restaurants in the franchise share a common menu of dishes, each table in each restaurant serves one dish, and multiple tables at multiple restaurants can feature the same dish. Each restaurant corresponds to a group $j$, and customer $i$ in restaurant $j$ corresponds to $\boldsymbol \theta_{ji}$. Each unique dish $k$ corresponds to a $\boldsymbol \phi_k$ and is drawn from the franchise-wide menu distribution $H$. We represent the dish served at table $t$ of restaurant $j$ by a new variable, $\boldsymbol \psi_{jt}$. Thus, each $\boldsymbol \theta_{ji}$ is associated with one $\boldsymbol \psi_{jt}$, and each $\boldsymbol \psi_{jt}$ is associated with one $\boldsymbol \phi_k$. The number of customers in restaurant $j$ at table $t$ who are being served dish $k$ is denoted $n_{jtk}$, and the number of tables in restaurant $j$ serving dish $k$ is denoted $m_{jk}$. A dot in a subscript represents summation over the corresponding index. For instance, $m_{j.}$ is the number of occupied tables in restaurant $j$.

The dynamics of the Chinese restaurant franchise can be illustrated by integrating out the random measures $G_j$ to obtain
\begin{eqnarray*}
\boldsymbol \theta_{j,n_{j..}+1} | \boldsymbol \theta_{j1},\dots,\boldsymbol \theta_{j,n_{j..}}, \alpha, G_0 \sim \sum_{t=1}^{m_{j.}} \frac{n_{jt.}}{n_{j..} + \alpha} \delta_{\boldsymbol \psi_{jt}} +  \frac{\alpha}{n_{j..} + \alpha} G_0,
\end{eqnarray*}
and integrating out $G_0$ to arrive at
\begin{eqnarray*}
\boldsymbol \psi_{j,m_{j.}+1} | \boldsymbol \psi_{11},\dots,\boldsymbol \psi_{1,m_{1.}},
\dots, \boldsymbol \psi_{J1},\dots,\boldsymbol \psi_{J,m_{J.}}, \gamma, H \sim \sum_{k=1}^K \frac{m_{.k}}{m_{..} + \gamma} \delta_{\boldsymbol \phi_{k}} +  \frac{\gamma}{m_{..} + \gamma} H.
\end{eqnarray*}
Thus a new customer enters restaurant $j$ and sits at occupied table $t$ serving dish $\boldsymbol \psi_{jt}$ with probability proportional to $n_{jt.}$, and the customer sits at an unoccupied table with probability proportional to $\alpha$, in which case a dish for the table is needed. The dish for the newly occupied table is equal to a dish $\boldsymbol \phi_k$ that is already being served at at least one table in the franchise with probability proportional to $n_{jt.}$, and it is equal to a new dish, drawn from menu $H$, with probability proportional to $\gamma$.

\subsection{Infinite Hidden Markov Models}

Hidden Markov models \citep{Baum1966} offer an alternative partitioning strategy that can account for spatial patterns along alignments. Starting at one end of the alignment, site-specific evolutionary parameters can be thought of as being generated sequentially according to a Markov chain with a finite state space \citep{Felsenstein1996}. Thus the evolutionary category for a specific site depends on the evolutionary category assumed by the preceding site.  In particular, for any site $i$, we have
\begin{equation*}
P(\boldsymbol \theta_i | \boldsymbol \theta_1,\dots,\boldsymbol \theta_{i-1}) = P(\boldsymbol \theta_i | \boldsymbol \theta_{i-1}),
\end{equation*}
and
\begin{equation*}
P(\textbf X_i | \tau, \boldsymbol \theta_1,\dots,\boldsymbol \theta_{i}) = P(\textbf X_i | \tau, \boldsymbol \theta_{i}).
\end{equation*}
To overcome the restriction of having to specify the dimension of the state space beforehand, \citet{Beal2002} introduce an infinite hidden Markov model with a countably infinite state space. \citet{Teh2006} show that an infinite hidden Markov model can in fact be achieved through an extension of the hierarchical Dirichlet process framework. The key difference is that rather than a fixed number of ``groups'' with the division of sites into groups determined beforehand, there is an unbounded number of groups, and the group membership of a given site is determined by the evolutionary category of the preceding site.

As in the hierarchical Dirichlet process, we have a collection of random probability measures that are distributed according to the same Dirichlet process, whose underlying base distribution is itself distributed according to a Dirichlet process. In particular, to each evolutionary category $k$, we associate a random probability measure $G_k$ where 
\begin{eqnarray*}
G_k | \alpha, G_0 \sim \text{DP}(\alpha, G_0) \text{ for } k = 1,2, \dots 
\end{eqnarray*}
and
\begin{eqnarray*}
G_0 | \gamma, H  \sim  \text{DP}(\gamma, H).
\end{eqnarray*}
Then the Markovian nature of the process is captured by
 \begin{eqnarray*}
\boldsymbol \theta_{i} | \boldsymbol \theta_{i-1}, G_1, G_2,\dots & \sim & G_{z_{i-1}} \text{ for } i = 1,\dots, s,
\end{eqnarray*}
where, as before, $z_i$ denotes the evolutionary category of site $i$. As in the hierarchical Dirichlet process, the common discrete base distribution $G_0$ ensures that the random measures $G_j$ share the same atoms, which ensures that any state (i.e., evolutionary category) can be reached from any other state.

\subsection{Posterior Inference}

We implement our infinite mixture model framework in the BEAST X (v10.5.0) software package for Bayesian evolutionary inference \citep{BEAST2018}. Our framework is currently implemented in the development branch, available at \url{https://github.com/beast-dev/beast-mcmc/}, and will be included in the next official release of BEAST X. Importantly, this implementation enables our framework for across-site variation to be employed in the wide range of phylogenetic and phylodynamic inference models that can be specified and run by BEAST X. We generate samples from the posterior distribution through Markov chain Monte Carlo (MCMC) simulation \citep{Metropolis1953, Hastings1970}. Standard MCMC methods for infinite mixture models can be hampered by slow mixing and high computational burden. To enable our model to scale efficiently to large genomic data sets that have become commonplace with advances in sequencing technology, we adapt a cost-effective ``data squashing'' MCMC sampling strategy put forth by \citet{Guha2010} to update the site-to-category assignments. This approach can be applied to infinite mixture models based on a wide class of Bayesian nonparametric prior distributions, including all priors that we consider in this study. We outline the main ideas of the sampling scheme here and refer to \citet{Guha2010} for further details.

The strategy of the algorithm is to simultaneously propose Metropolis-Hastings updates for the evolutionary category assignment variables $z_j$ for a group of alignment sites $j$ that have similar full conditional distributions for the $z_j$ at the current iteration. By working with sites that have category assignment variables with approximately identically distributed full conditionals, candidate category assignments can be jointly generated in an efficient manner by simply working with one representative member of the group of sites. For the current iteration $t$, suppose there are $K$ evolutionary categories with distinct evolutionary parameter values  $\boldsymbol \phi_{1}^{(t)}, \boldsymbol \phi_{2}^{(t)}, \dots, \boldsymbol \phi_{K}^{(t)}$. Rather than compute the full conditional distribution for each $z_j$, we adopt simpler mass functions $\pi_j(z_j)$ that approximate the full conditionals and are less computationally expensive. In the case of a Dirichlet process mixture, for example, we define $\pi_j(z_j) \propto n_{z_j}^{(t)} * P(\textbf X_j | \tau^{(t)}, \boldsymbol \phi_{z_j}^{(t)})$ for $z_j = 1, \dots, K$, and $\pi_j(z_j) = 0$ otherwise, where $n_k^{(t)}$ denotes the number of sites with parameter values equal to $\boldsymbol \phi_{k}^{(t)}$. Let $i$ be a randomly chosen site. To compare the mass functions for sites $i$ and $j$, $i \neq j$, we can compute a difference measure $\Delta_{ij}$ such as the squared Hellinger distance \citep{YangLeCam2000}, in which case  $\Delta_{ij} = 2(1 - \sum_{k=1}^K \sqrt{\pi_i (k) \pi_j (k)})$. Next, we specify a set of sites $D$ for which we will jointly propose category assignment updates. We form the set by including site $i$ along with sites with mass functions most similar to $\pi_i$ until $D$ has the desired size. In our analyses, we let $D$ assume different sizes in different iterations, ranging between 1 site and approximately 10$\%$ of the total number of alignment sites. To accommodate potential new evolutionary categories, we augment the state space with auxiliary parameters, as in the Gibbs sampling procedure introduced by \citet{Neal2000} for non-conjugate Dirichlet process mixture models. Next, we construct a proposal distribution for $z_i$ by approximating the conditional distribution of $z_i$ given the data and model parameters that correspond to sites that are \textit{not} in $D$. Finally, we generate candidate category assignments for all sites in $D$ as independent and identically distributed realizations from the aforementioned proposal distribution for $z_i$. This set of candidate category assignments is then jointly accepted or rejected according to the corresponding Metropolis-Hastings ratio.

Each iteration of this algorithm requires computationally expensive evaluation of the observed sequence data likelihoods $P(\textbf X_i | \tau, \boldsymbol \phi_k)$ for alignment sites $i =1, \dots, s$ and occupied and unoccupied evolutionary categories $k =1, \dots, c$. Here, $c$ can vary from one iteration to another. Fortunately, these $s \times c$ computations are independent and can be performed in parallel. To this end, we construct an interface between the BEAST X implementation of our model and BEAGLE \citep{Ayres2019}, a high-performance library for parallel phylogenetic likelihood evaluation.

We propose updates for evolutionary model parameters as well as Dirichlet process concentration parameters and base distribution hyperparameters using standard Metropolis-Hastings transition kernels. We implement Dirichlet process mixtures using the Chinese restaurant process representation that integrates out the random measure $G$ and thus do not need to generate parameters that characterize the weights of $G$. For hierarchical Dirichlet processes (and infinite hidden Markov models), however, \citet{Teh2006} note that starting with the Chinese restaurant franchise representation but explicitly instantiating the shared base distribution $G_0$ eases the implementation by enabling the posterior conditioned on $G_0$ to factor across groups. We follow the strategy of \citet{Teh2006} by implementing a Gibbs sampler to generate weights for $G_0$. This in turn necessitates the generation of ``table count'' variables $m_{jk}$, for which we also implement a Gibbs sampler. Finally, to propose updates to the phylogenetic tree and hyperparameters for the phylogenetic tree prior distribution, we employ transition kernels already available in BEAST X.

\section{Empirical Examples}

We evaluate different methods for modeling across-site variation in analyses of three data sets. Because BEAST X is often used for phylodynamic analyses of measurably evolving pathogens, incorporating dated-tip molecular clock models, we here focus on viral data sets, but the methods are broadly applicable in evolutionary biology. The methods include infinite mixture models as well as several commonly used ``standard'' approaches that employ fixed partitions and/or finite mixture models. For each method of modeling across-site variation, we conduct analyses using two different commonly used DNA substitution models: the HKY model \citep{Hasegawa1985}, and the GTR model \citep{Lanave1984, Tavare1986}.  

We consider several standard approaches for modeling across-site evolutionary variation. First, we employ a restrictive model with one substitution rate and one set of relative character exchange parameters for all alignment sites. We refer to this model for across-site variability as the ``No Variation'' model, and we adopt the convention to refer to the overall evolutionary model in terms of the substitution model and the across-site variability model (for example, ``HKY + No Variation'' or ``GTR + No Variation''). We relax the No Variation approach by allowing for substitution rate variation according to the popular finite mixture model proposed by \citet{Yang1994} while maintaining one set of relative character exchange rate parameters for all alignment sites. The \citet{Yang1994} model posits a fixed number of equally probable  substitution rate categories, with each rate drawn from a discretized gamma distribution. We use five different rate categories and denote the across-site variability model by ``Gamma.'' All of our data sets consist partially or entirely of protein coding regions, and it is therefore natural to consider partitioning strategies that allow the evolutionary process to vary according to codon position. We use ``Codon'' to denote an across-site variability model that partitions alignment sites as follows: sites from protein coding regions are categorized according to which of the three codon positions they correspond to, and sites from noncoding regions (if any) are assigned a separate category. Thus, there are three or four total categories (depending on whether the data come entirely or partially from protein coding regions), and the Codon model posits a separate substitution rate and set of relative character exchange rates for each category. As a more flexible alternative, we again partition sites as in the aforementioned Codon model and allow each partition to have its own set of relative character exchange rates, but we also allow substitution rate variation within each partition according to an independent \citet{Yang1994} model with five rate categories. We call the resulting model for across-site variability the ``Codon + Gamma'' model. 

For all evolutionary parameters, we use vague prior distributions that correspond to default options in the BEAUti software program for setting up data analyses to be performed by BEAST X \citep{BEAST2018}. In particular, equilibrium frequencies, instantaneous rate matrix parameters, and relative substitution rates for partitions under the Codon partitioning scheme are assigned uniform Dirichlet priors. HKY model transition/transversion ratios are \textit{a priori} log-normal with mean 1.0 and standard deviation 1.25. Finally, the shape/rate parameters for the \citep{Yang1994} discretized gamma model for across-site rate variation have an exponential prior distribution with mean 0.5.

In addition to the aforementioned standard modeling of across-site variation, we employ different infinite mixture models to account for uncertainty in the number of alignment partitions and the assignment of sites to different partitions. The type of infinite mixture model is determined by the Bayesian nonparametric prior we use for the evolutionary parameters, site-to-category assignments, and number of categories. We use a Dirichlet process prior (yielding the ``DP'' model for across-site variability) and an infinite hidden Markov model prior (giving rise to the ``IHMM'' model for across site variability). We also use a hierarchical Dirichlet process prior with alignment sites divided into three or four groups according to the same partitioning scheme employed under the Codon model. We refer to the resulting model for across-site variability as the ``HDP-Codon'' model.  

We do not have strong prior beliefs or information about the number of evolutionary categories, or about the clustering patterns of alignment sites. We therefore assume \textit{a priori} that all infinite mixture model concentration parameters follow diffuse gamma distributions with shape and rate parameters both equal to 0.001. We compose base distributions from which evolutionary parameter values are drawn by specifying independent distributions for equilibrium frequencies, substitution rates and (as applicable) GTR instantaneous rate matrix relative rate parameters and the HKY transition/transversion ratio. We employ Dirichlet distributions for equilibrium frequencies and GTR relative rate parameters, and normal distributions for log-transformed substitution rates and transition/transversion ratios.  

We adopt vague prior distributions for base distribution hyperparameters. In particular, we assume normal distribution means are \textit{a priori} normally distributed with mean 0 and standard deviation 10, and normal distribution precisions are  \textit{a priori} gamma distributed with shape and rate equal to 0.001. We parameterize each Dirichlet prior distribution concentration parameter as a product $\eta \textbf c$, where the scalar $\eta$ is an overall dispersion parameter and the vector $\textbf c=(c_1,\dots,c_d)$ characterizes the relative values of the concentration parameter components. We assign $\textbf c$ a uniform Dirichlet prior distribution, and on $\eta$ we place a gamma prior distribution with shape and rate equal to 0.001. HKY transition/transversion rate base distributions must be handled with extra care, and they are the only base distributions whose hyperparameters we do not jointly infer from the data. HKY model Markov chain transition probabilities can converge to finite values as the transition/transversion rate tends to infinity, so the presence of alignment sites for which there is negligible support for transversions can lead to divergent estimates of transition/transversion rates and base distribution hyperparameters. To allow for large transition/transversion rates to accommodate such alignment sites while ensuring numerical stability, we specify vague base distributions with fixed hyperparameter values. To specify the hyperparameters, we take an empirical Bayes approach and adopt the estimated mean and ten times the estimated standard deviation of transition/transversion rate estimates from HKY + No Variation models.

In all analyses, whether modeling across-site variation via standard approaches or infinite mixture models, we employ a strict molecular clock that assumes the evolutionary rate does not vary among phylogenetic tree branches \citep{Kimura1968}. In the infinite mixture models, we wish to model absolute site-specific substitution rates, so we fix the strict molecular clock rate to 1.0. Under the standard approaches, across-site substitution rate variation is modeled in terms of relative rates, so the strict molecular clock rate is estimated from the data, and we assign it a vague exponential prior with mean 1.0. For the phylogenetic tree, we employ skygrid coalescent-based prior distributions that flexibly model the trajectories of the effective sizes of the populations from which the samples are taken as piece-wise constant functions \citep{Gill2013}. The smoothness of skygrid effective population size trajectories are governed by a precision parameter, to which we assign a diffuse gamma prior with shape and rate equal to 0.001. 

To summarize the clustering pattern inferred under a given infinite mixture model, we use the $k$-means clustering algorithm available in the R \citep{R2021} package MASS \citep{Venables2002} to divide alignment sites into $\tilde{K}$ different categories, where $\tilde{K}$ is the posterior median estimate of the number of evolutionary categories. Each cluster analysis is applied to the site-specific posterior median, $2.5^{\text{th}}$ percentile and $97.5^{\text{th}}$ percentile estimates of all evolutionary model parameters. 

We compare the performance of the various standard models and infinite mixture models by assessing their model fit through estimates of the log marginal likelihood \citep{Newton1994}. Notably, the marginal likelihood implicitly penalizes overfitting and does not systematically prefer more complex models \citep{Jefferys1992}. A greater marginal likelihood corresponds to a better model fit, and two models $M_1$ and $M_2$ can be formally compared by subtracting the log marginal likelihood under model $M_2$ from the log marginal likelihood under model $M_1$ to obtain the log of the Bayes factor in favor for model $M_1$ over $M_2$ \citep{Jeffreys1935, Jeffreys1961}. The evidence in favor of $M_1$ over $M_2$ can be interpreted as follows depending on the value of the log Bayes factor: ``very strong'' if it is greater than 5, ``strong'' if it is between 3 and 5, ``positive'' if it is between 1 and 3, and ``not worth more than a bare mention'' if it is between 0 and 1 \citep{Kass1995}.
To approximate the marginal likelihood, we use an adjusted version \citep{Redelings2005} of the stabilized harmonic mean estimator introduced by \citet{Newton1994}.  

In addition to comparing the model fit, we assess the differences in phylogenetic inferences that result from the various models (see Appendix). We compare summary phylogenetic trees, and we also compare the overall posterior distributions by examining split frequencies \citep{Lakner2008} and two-dimensional representations of phylogenetic treespace \citep{Hillis2005}. Finally, for the best fitting HKY-based models for the different data sets, Figures A16-A20 of the Appendix illustrate the across-site variation in substitution rates and transition/transversion rates.

\subsection{Respiratory Syncytial Virus Subgroup A}

\begin{figure*}[htb]
\centering
  \begin{tabular}{@{}c@{}}
    \includegraphics[width=1.0\textwidth]{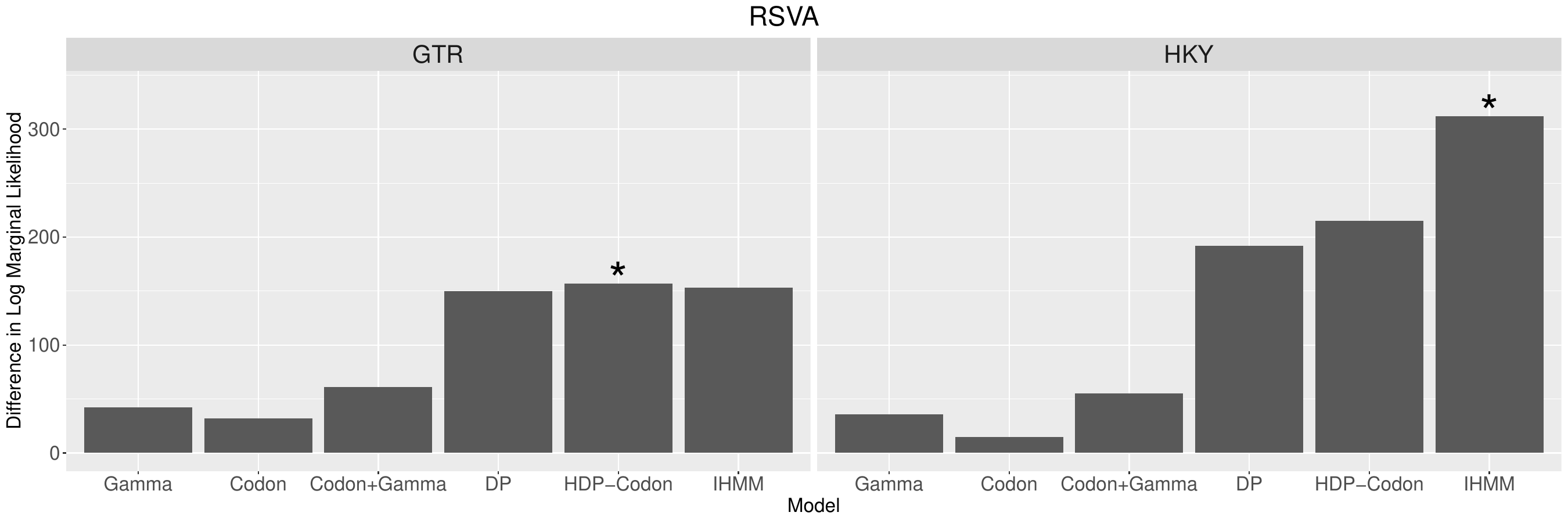} \\
    \includegraphics[width=1.0\textwidth]{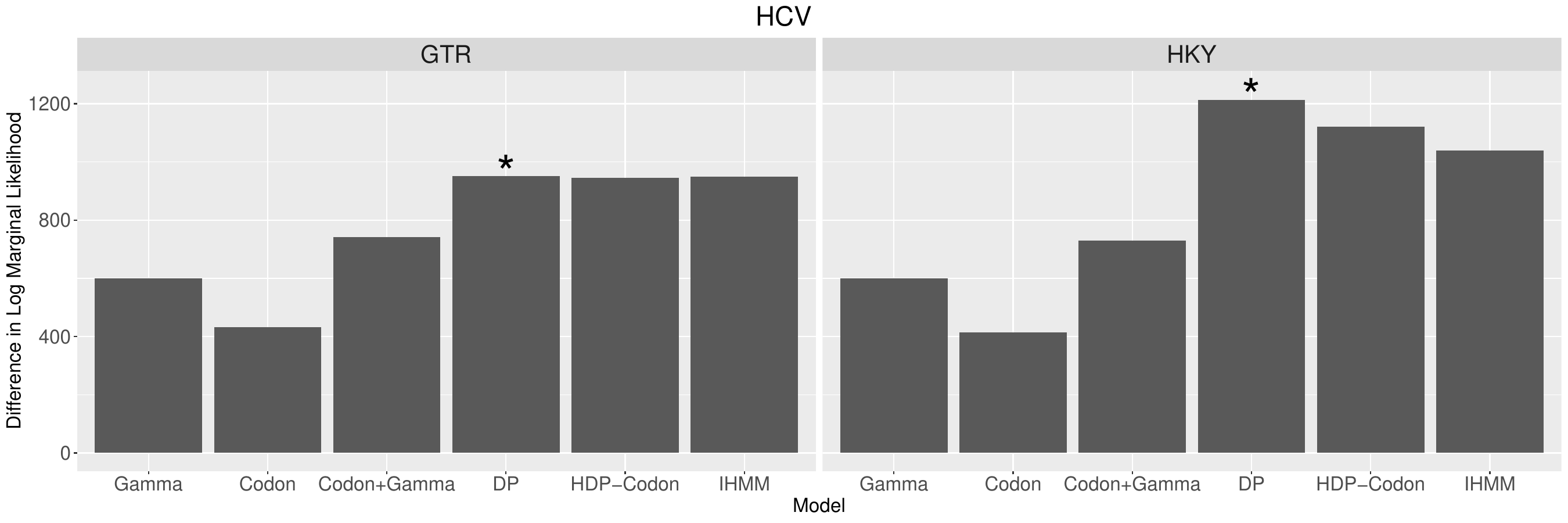} \\
    \includegraphics[width=1.0\textwidth]{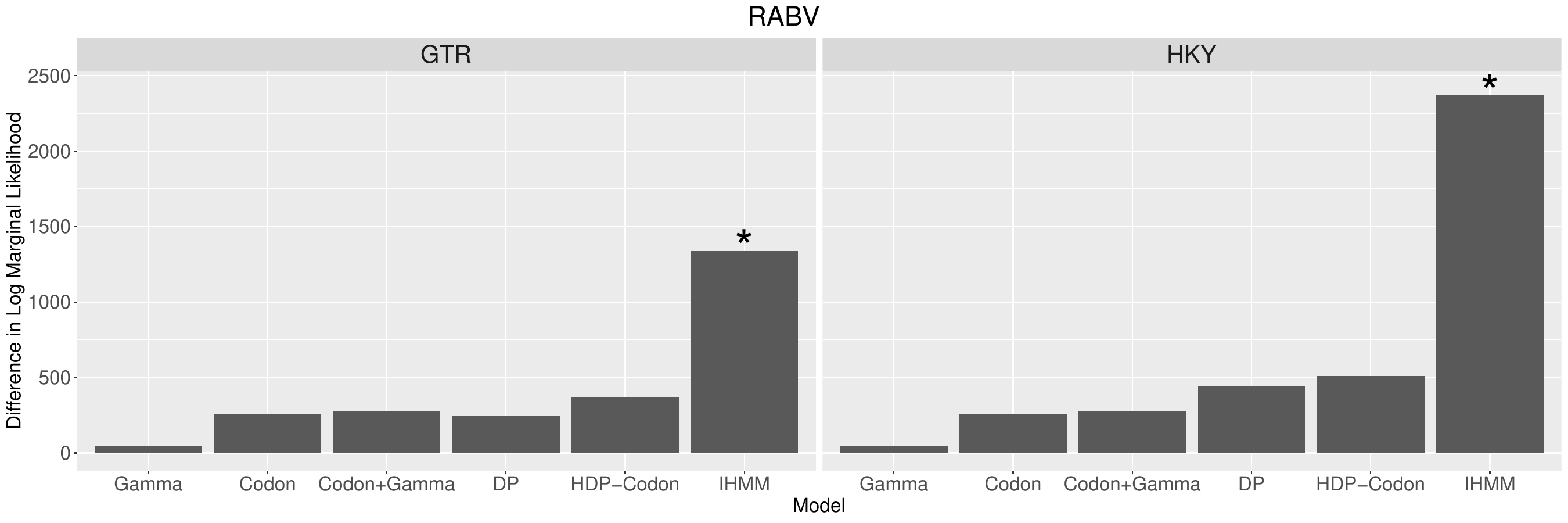} 
  \end{tabular}
  \caption{Performance of different models for across-site variation on respiratory syncytial virus subgroup A (RSVA), hepatitis C subtype 4 (HCV), and rabies virus (RABV) data. For each data set and nucleotide substitution model (GTR or HKY), bars depict improvement in model fit in log marginal likelihood units over a baseline model that assumes no across-site variation in the evolutionary process. The best fitting across-site variation model for each data set and nucleotide substitution model is indicated by a star over the corresponding bar.	
	\label{fig:modelperformance}}
\end{figure*}

We first consider a human respiratory syncytial virus A (RSVA) data set \citep{Zlateva2005} that features 35 sequences of 629 base pairs from the $G$ gene, sampled between 1956 and 2002. The RSVA $G$ gene encodes for the attachment glycoprotein, and for across-site variation models that employ the Codon partitioning scheme, we divide alignment sites into three groups according to the three codon positions. Figure~\ref{fig:modelperformance} shows the improvement in model performance over baseline HKY + No Variation and GTR + No Variation models achieved by using different models for across-site variation. Marginal likelihood estimates for all analyses are reported in Table~\ref{tab:combinedperformance} of the Appendix.

Among the standard approaches, the models without across-site variation clearly perform the worst while the most flexible Codon + Gamma variation scheme yields the best performance. It is interesting to note that, conditional on the substitution model, the Gamma scheme that accounts for uncertainty in substitution rate partitions while using the same relative character exchange rates for all sites performs much better than the Codon scheme, which allows for variation in substitution rates as well as relative character exchange rates but fixes partitions according to codon position. For each of the four standard approaches for modeling across-site variation, using a GTR substitution model leads to a better fit compared to using an HKY substitution model.

\begin{figure*}[htb]
\centering
  \begin{tabular}{@{}c@{}}
  \includegraphics[width=1.0\textwidth]{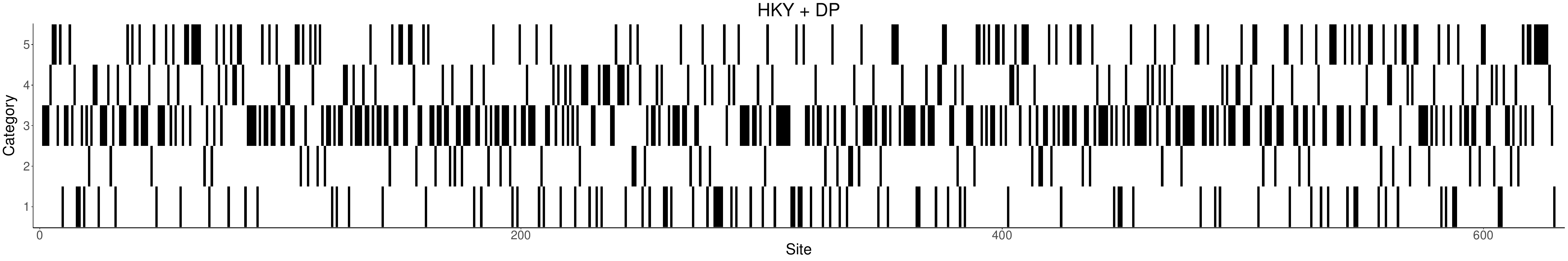} \\
    \includegraphics[width=1.0\textwidth]{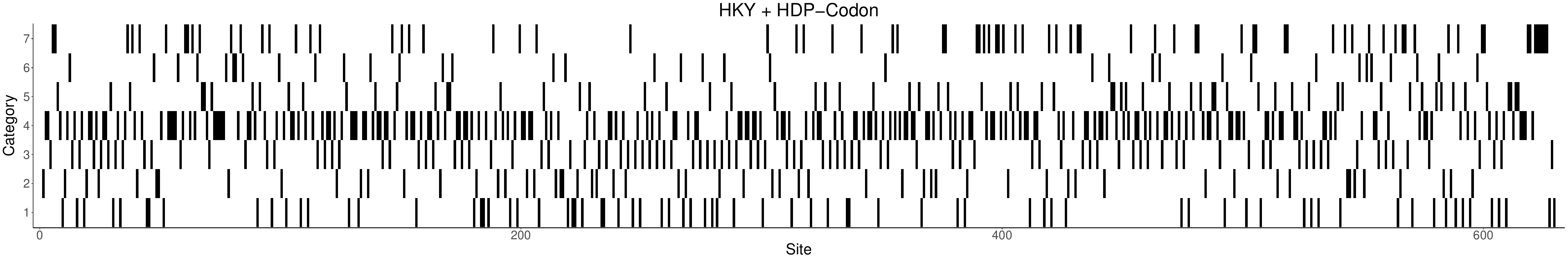} \\
    \includegraphics[width=1.0\textwidth]{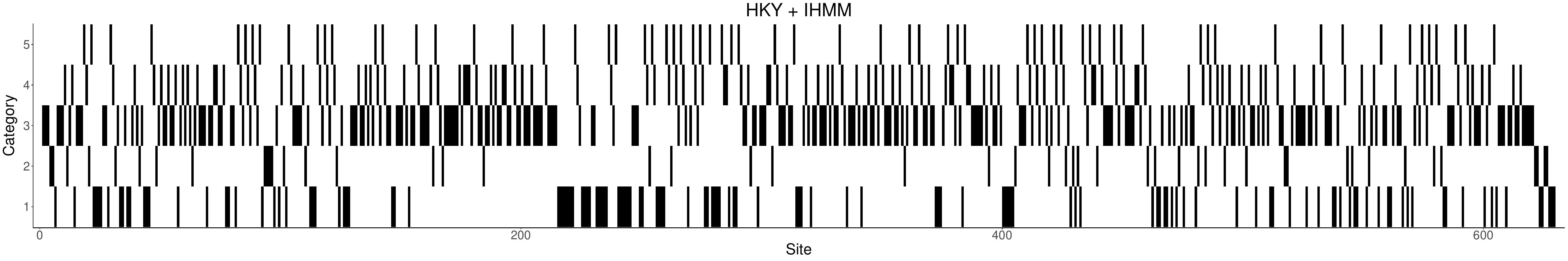} \\
    \includegraphics[width=1.0\textwidth]{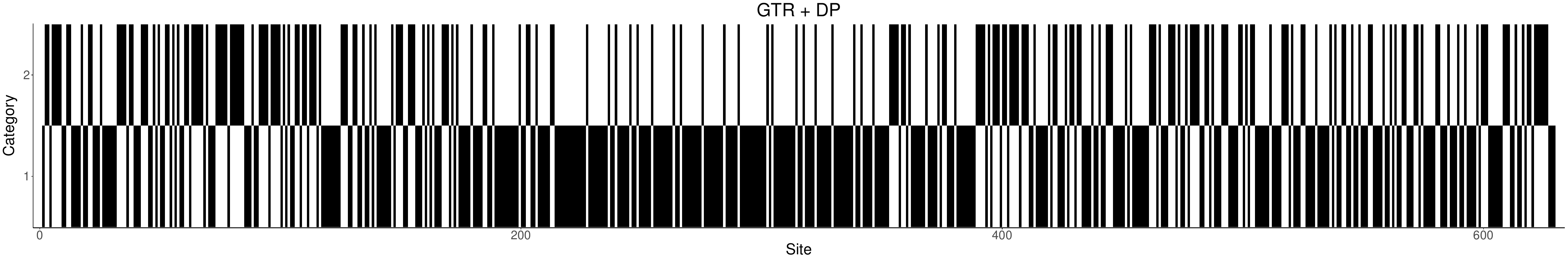} \\
    \includegraphics[width=1.0\textwidth]{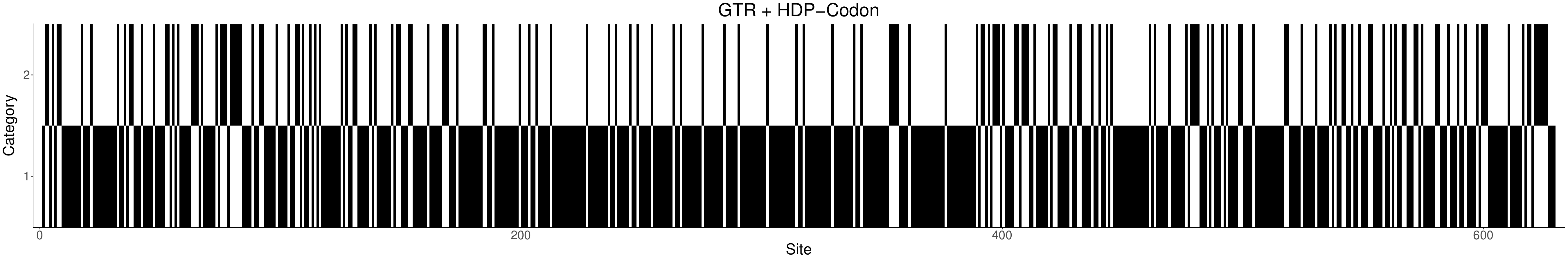} \\
    \includegraphics[width=1.0\textwidth]{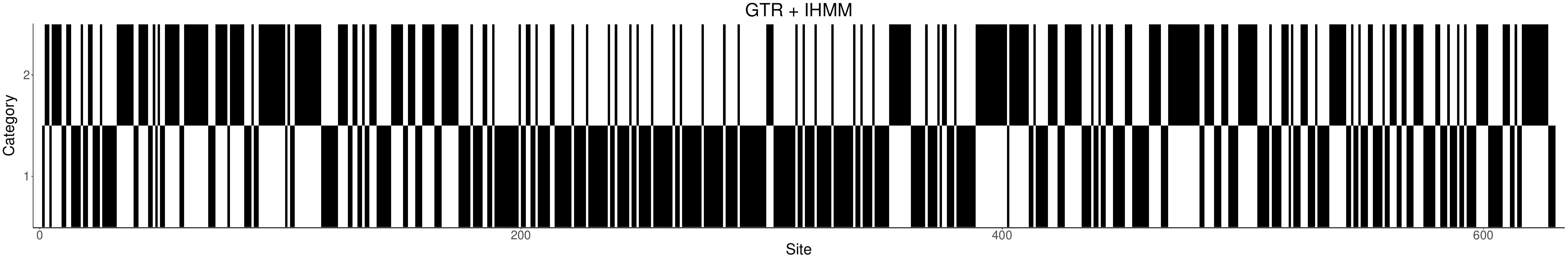} 
  \end{tabular}
  \caption{Summary of alignment site clustering patterns inferred from infinite mixture model analyses of respiratory syncytial virus subgroup A data set. Alignment sites are divided into categories via the $k$-means clustering algorithm applied to site-specific posterior quantile estimates of evolutionary model parameters. The prespecified number of categories for each cluster analysis corresponds to the posterior median estimate of the number of evolutionary categories inferred under the infinite mixture model. 
  In each plot, categories are ordered according to median substitution rate, from lowest to highest.
  The category of each alignment site on a horizontal axis is indicated by a vertical black bar.
	\label{fig:rsvapattern}}
\end{figure*}

All of the infinite mixture models outperform the best of the standard models (GTR + Codon + Gamma) by wide margins. The best model is HKY + IHMM, with a marginal likelihood nearly 100 log units ahead of the second place HKY + HDP, which has a marginal likelihood 23 log units greater than the HKY + DP model. For each infinite mixture model, using the HKY substitution model leads to a better model fit than using the GTR substitution model (in contrast to what we observed under the standard approaches). In fact, the worst fitting model that uses the HKY substitution model has a marginal likelihood 22 log units higher than the best fitting model that uses the GTR substitution model (the GTR + HDP-Codon). The GTR + HDP-Codon has a marginal likelihood 4 log units greater than the GTR + IHMM, which has a marginal likelihood 3 log units greater than the GTR + DP. The posterior estimates of the number of evolutionary categories (Table~\ref{tab:combinednumcategories}) are greater and less precise for models that use the HKY substitution model rather than the GTR substitution model. Thus the infinite mixture models compensate in some sense for a more restrictive substitution model by inferring a larger number of distinct substitution model parameters. The clustering patterns of the alignment sites are summarized in Figure~\ref{fig:rsvapattern}. The pattern variation between models that use the HKY substitution model is greater than the pattern variation between models that use the GTR substitution model. This is consistent with large differences in marginal likelihood between the models that use the former substitution model vs. the relatively small differences in marginal likelihood between models that employ the latter substitution model.

\begin{table*}[htb]
\begin{center}
\begin{tabular}{llccccc}
 & & & \multicolumn{2}{c}{GTR} & \multicolumn{2}{c}{HKY}  \\
\cline{2-7} 
\\\\[-4\medskipamount]
 & & & \multicolumn{1}{c}{Median} & \multicolumn{1}{c}{95$\%$ BCI} & \multicolumn{1}{c}{Median} & \multicolumn{1}{c}{95$\%$ BCI} \\
\cline{2-7}
\\\\[-4\medskipamount]
& DP & & 2 & (2, 2) & 5 & (3, 11) \\[2ex]
RSVA & HDP-Codon & & 2 & (2, 2) & 7 & (4, 11) \\[2ex]
& IHMM & & 2 & (2, 2) & 5 & (5, 7) \\
\\\\[-4\medskipamount]
\cline{2-7}
\cline{2-7}
\\\\[-4\medskipamount]
& DP & & 3 & (3, 3) & 12 & (10, 16) \\[2ex]
HCV & HDP-Codon & & 3 & (3, 3) & 9 & (7, 12) \\[2ex]
& IHMM & & 3 & (3, 3) & 10 & (9, 12) \\
\\\\[-4\medskipamount]
\cline{2-7}
\cline{2-7}
\\\\[-4\medskipamount]
& DP & & 2 & (2, 2) & 2 & (2, 5) \\[2ex]
RABV & HDP-Codon & & 2 & (2, 2) & 4 & (4, 7) \\[2ex]
& IHMM & & 3 & (3, 3) & 5 & (5, 5) \\
\\\\[-4\medskipamount]
\cline{2-7}
\end{tabular}
\end{center}
\caption[Posterior medians and 95$\%$ Bayesian credibility intervals (BCIs) of the number of evolutionary categories inferred from infinite mixture model analyses of respiratory syncytial virus subgroup A (RSVA), hepatitis C subtype 4 (HCV), and rabies virus (RABV) data.]
{Posterior medians and 95$\%$ Bayesian credibility intervals (BCIs) of number of evolutionary categories inferred from infinite mixture model analyses of respiratory syncytial virus subgroup A (RSVA), hepatitis C subtype 4 (HCV), and rabies virus (RABV) data. Rows correspond to different data sets and different infinite mixture models for across-site variation of substitution rates and relative exchange rates of nucleotide bases. Columns correspond to different substitution models.   
\label{tab:combinednumcategories}}
\end{table*}

\subsection{Hepatitis C Subtype 4}

Next, we analyze a hepatitis C subtype 4 (HCV) data set comprising 63 sequences of 412 base pairs from the E1 region \citep{Ray2000}. The sequences all have the same sampling date, which means that the substitution rate cannot be separated from time in the evolutionary model and is thus not identifiable. In order to estimate substitution rate multipliers, we adopt the same resolution as \citet{Wu2013} in their analysis of the data set: we assign the tree root height a restrictive normal prior distribution with mean 1.0 and standard deviation 0.1. The sequences encompass a protein coding region, and we again divide the alignment into three groups according to the three codon positions for across-site variation models that use the Codon partitioning scheme. Improvements in model fit by models for across-site variation over baseline models are shown in Figure~\ref{fig:modelperformance} and complete marginal likelihood estimates are available in Table~\ref{tab:combinedperformance} (Appendix).

The results under the standard approaches mirror those for the RSVA data set. The best model fit is attained under the least restrictive method for modeling across-site variation: the Codon + Gamma scheme.  Also, for each of the four across-site variation models, an underlying GTR substitution model yields a better model fit than an HKY model. 

\begin{figure*}[htb]
\centering
  \begin{tabular}{@{}c@{}}
  \includegraphics[width=1.0\textwidth]{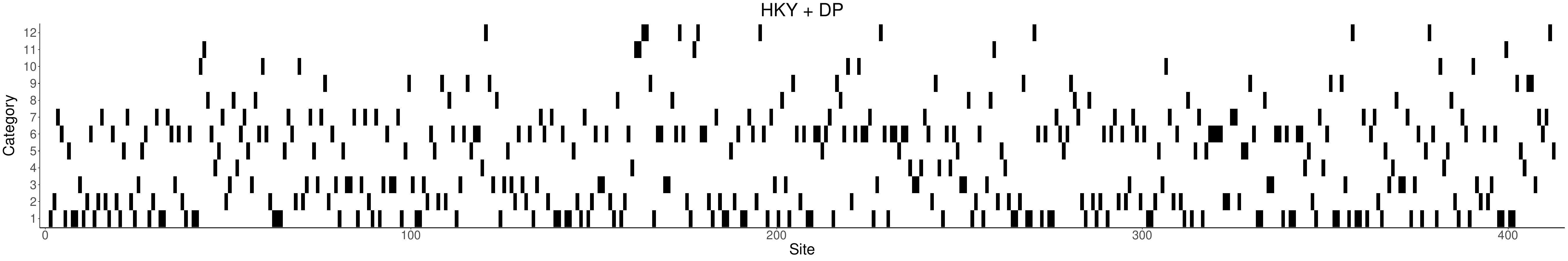} \\
    \includegraphics[width=1.0\textwidth]{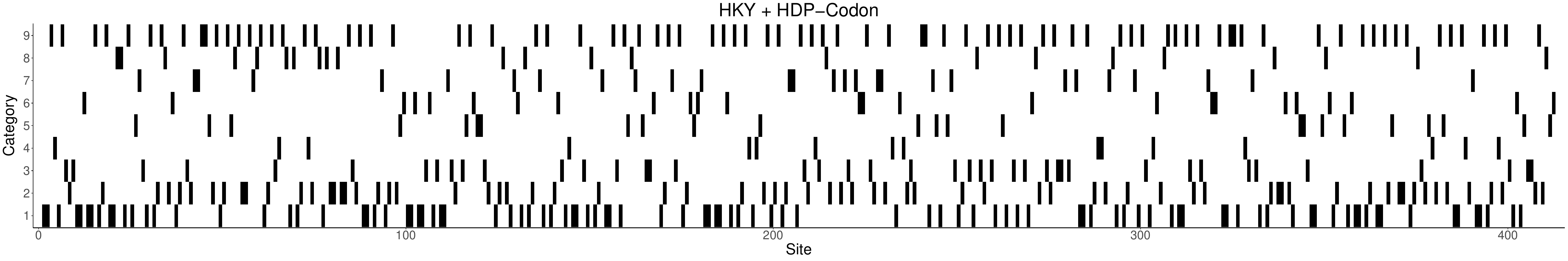} \\
    \includegraphics[width=1.0\textwidth]{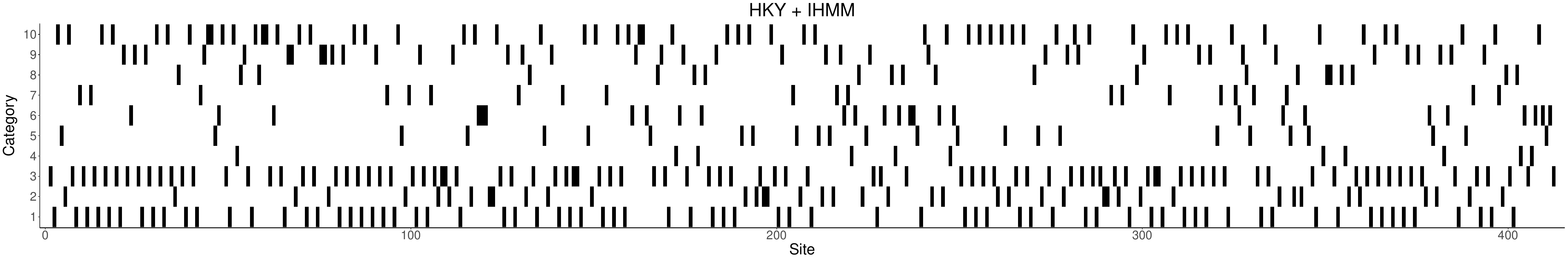} \\
    \includegraphics[width=1.0\textwidth]{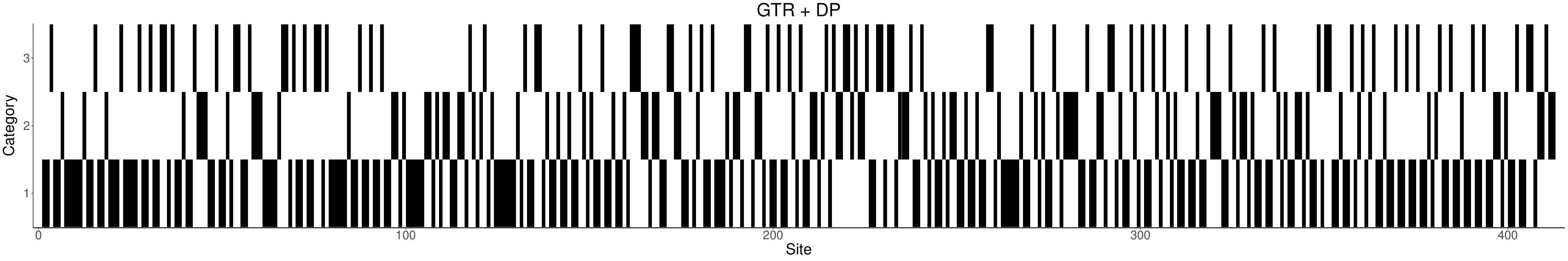} \\
    \includegraphics[width=1.0\textwidth]{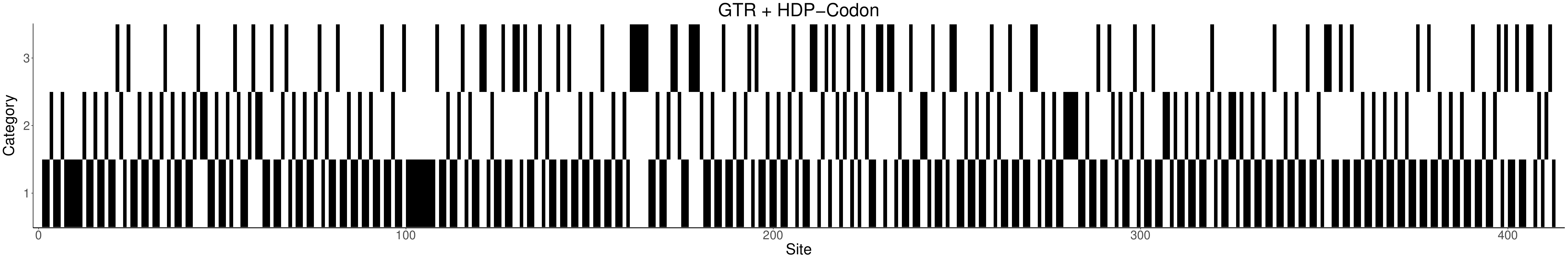} \\
    \includegraphics[width=1.0\textwidth]{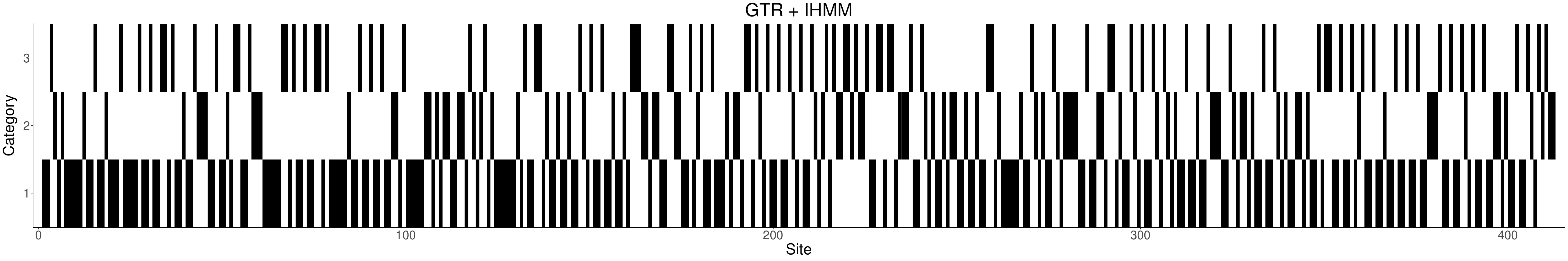} 
  \end{tabular}
  \caption{Summary of alignment site clustering patterns inferred from infinite mixture model analyses of hepatitis C subtype 4 data set. Alignment sites are divided into categories via the $k$-means clustering algorithm applied to site-specific posterior quantile estimates of evolutionary model parameters. The prespecified number of categories for each cluster analysis corresponds to the posterior median estimate of the number of evolutionary categories inferred under the infinite mixture model. 
    In each plot, categories are ordered according to median substitution rate, from lowest to highest.
  The category of each alignment site on a horizontal axis is indicated by a vertical black bar.	
	\label{fig:hcvpattern}}
\end{figure*}

Again, as with the RSVA analyses, every infinite mixture model outperforms the best standard model (the GTR + Codon+ Gamma), and infinite mixture models that employ HKY models uniformly outperform those that use GTR models. In contrast to the RSVA analyses, Dirichlet process mixtures perform best no matter the underlying substitution model. The best fitting model is the HKY + DP, with a marginal likelihood 92 log units greater than the HKY + HDP-Codon, which exceeds the marginal likelihood of the HKY + IHMM by 83 log units. The marginal likelihoods for the GTR model-based mixtures are again relatively close to each other: the GTR + DP's marginal likelihood is 2 log units ahead of the GTR + IHMM's marginal likelihood, which is 4 log units greater than the GTR + HDP-Codon model's marginal likelihood. As in the case of the RSVA analyses, the posterior estimates of the number of evolutionary categories are higher and more variable for infinite mixture models with underlying HKY substitution models (Table~\ref{tab:combinednumcategories}). Figure~\ref{fig:hcvpattern} summarizes the clustering patterns under different mixture models. The clustering patterns vary substantially for models that use the HKY model, reflective of the large differences in their marginal likelihoods. On the other hand, the clustering patterns for models based on the GTR model are very similar, which is in line with their relatively close marginal likelihoods.

\subsection{Rabies Virus}

We analyze a data set comprising 47 rabies virus (RABV) sequences sampled between 1982 and 2004 that was used to study a large-scale outbreak among North American raccoons \citep{Biek2007}. Each sequence includes 1359 bp of the glycoprotein $(G)$ gene, 1365 bp of the nucleoprotein $(N)$ gene, and 87 bp for the noncoding sequence that immediately follows $N$. For across-site variation models that employ the Codon partitioning scheme, we thus divide the sites from the $G$ and $N$ genes into three groups according to codon position, and noncoding sites are assigned a separate fourth group. Table~\ref{tab:combinedperformance} (Appendix) presents the marginal likelihood estimates under different standard and infinite mixture models, and Figure~\ref{fig:modelperformance} depicts the differences in model fit. 

As with the RSVA and HCV analyses, for each standard approach for modeling across-site variation, using the GTR model leads to a better model fit than using the HKY model. Further, the best performance is achieved under Codon + Gamma across-site variation. In contrast to the RSVA and HCV analyses, however, the Codon across-site variation models outperform the Gamma models. 

\begin{figure*}[htb]
\centering
  \begin{tabular}{@{}c@{}}
     \includegraphics[width=1.0\textwidth]{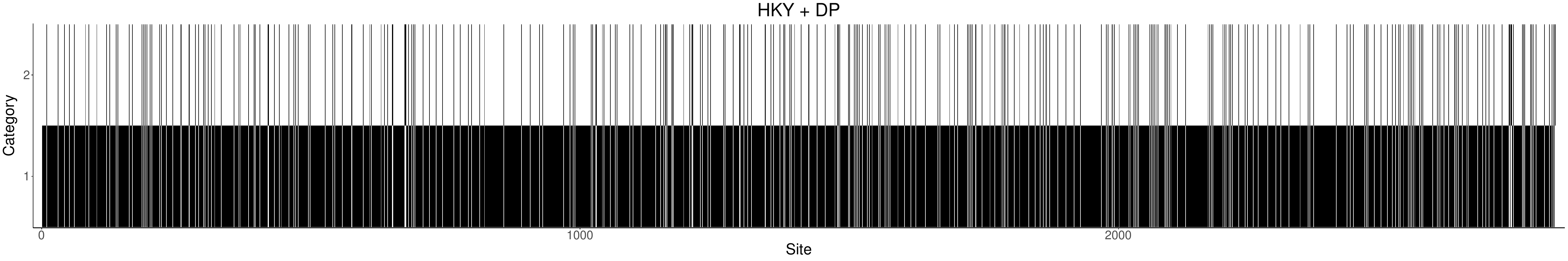} \\
    \includegraphics[width=1.0\textwidth]{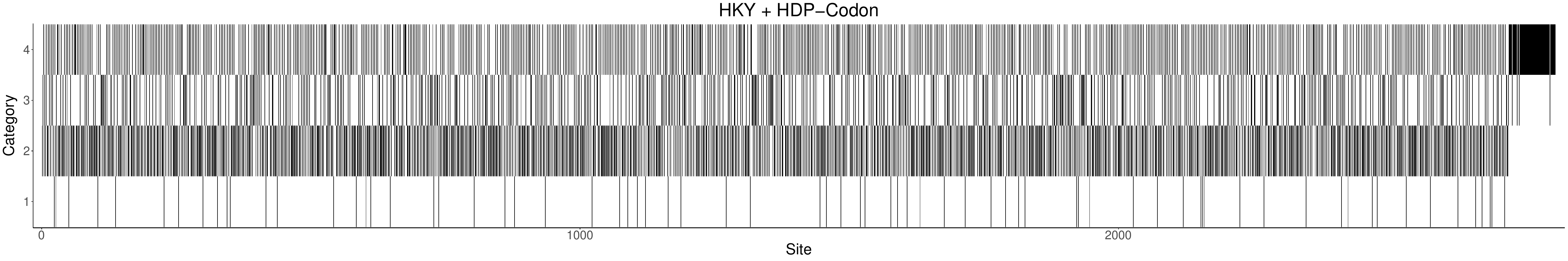} \\
    \includegraphics[width=1.0\textwidth]{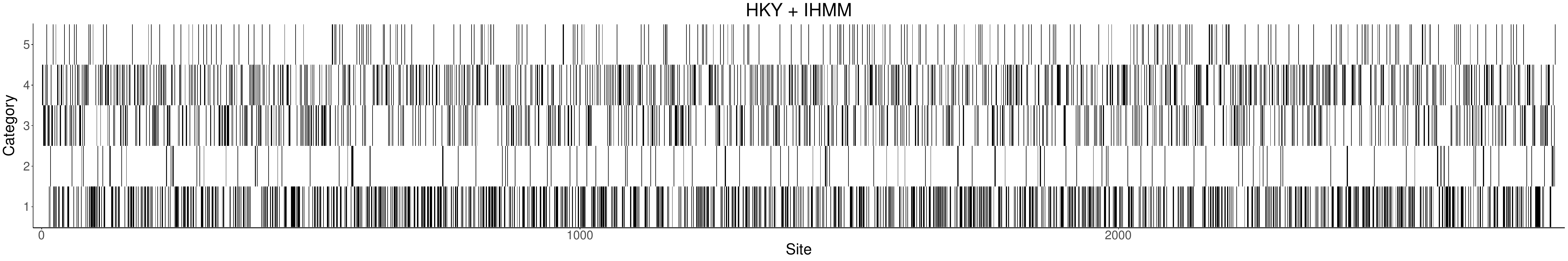} \\
    \includegraphics[width=1.0\textwidth]{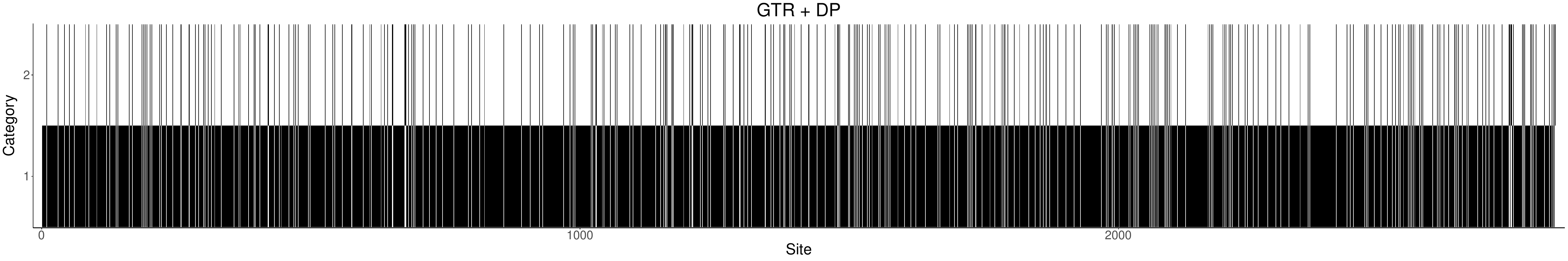} \\
    \includegraphics[width=1.0\textwidth]{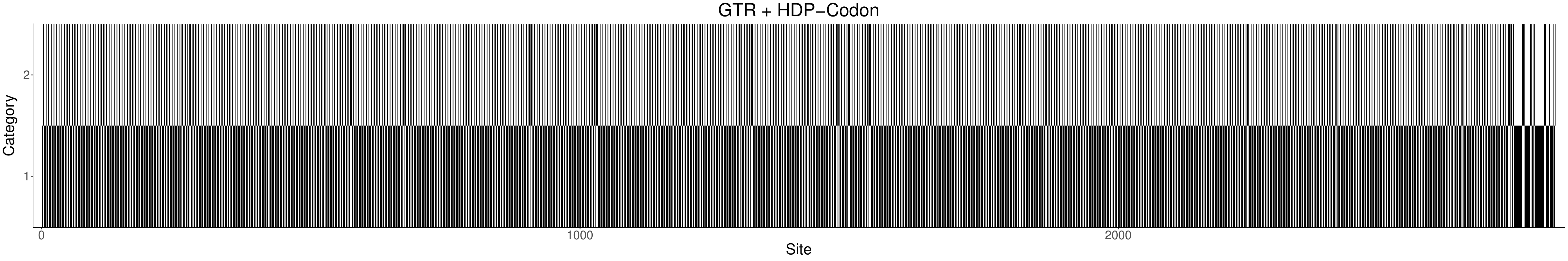} \\
    \includegraphics[width=1.0\textwidth]{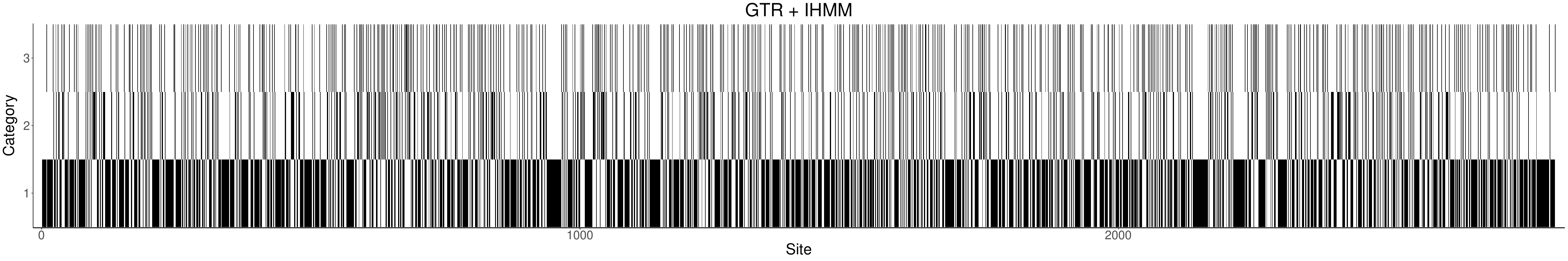} 
  \end{tabular}
  \caption{Summary of alignment site clustering patterns inferred from infinite mixture model analyses of rabies virus data set. Alignment sites are divided into categories via the $k$-means clustering algorithm applied to site-specific posterior quantile estimates of evolutionary model parameters. The pre-specified number of categories for each cluster analysis corresponds to the posterior median estimate of the number of evolutionary categories inferred under the infinite mixture model. 
    In each plot, categories are ordered according to median substitution rate, from lowest to highest.
  The category of each alignment site on a horizontal axis is indicated by a vertical black bar.	
	\label{fig:rabvpattern}}
\end{figure*}

Many more departures from the trends in the RSVA and HCV analyses emerge in infinite mixture model analyses. Models based on the HKY substitution model do not uniformly outperform models based on the GTR model. Indeed, while the HKY + IHMM model achieves the best marginal likelihood by a margin over 1000 log units, the GTR + IHMM model ranks second and has a marginal likelihood 847 log units greater than the third best model. However, conditional on the type of Bayesian nonparametric prior, using an HKY model always yields a better model fit than using a GTR model. While the gaps between the IHMM-based models and the rest are much bigger than any of the other gaps, the difference in marginal likelihood between any two of the infinite mixture models is greater than 50 log units. While five of the six infinite mixture models greatly outperform all of the standard models for across-site variation, the GTR + DP model has a marginal likelihood that is 30 log units less than that of the standard GTR + Codon + Gamma model, and 13 log units less than the standard GTR + Codon model's marginal likelihood. Posterior estimates of the number of evolutionary categories (Table~\ref{tab:combinednumcategories}) are greater and more variable for infinite mixture models that use the HKY substitution model, but the difference is not as great as what we observe in the RSVA and HCV analyses. A summary of the clustering patterns under the different mixture models is depicted in Figure~\ref{fig:rabvpattern}. There is generally substantial variation, but the patterns under the GTR + DP and HKY + DP models are very similar.

\subsection{Posterior Distributions of Phylogenetic Trees}

The substantial variation in model fit among different models for across-site variation in our empirical examples is accompanied by notably different posterior phylogenetic inferences, especially in the cases of the HCV and RABV data sets (see Appendix). Summary trees are widely used to express the essential findings of phylogenetic analyses, and they are often heavily relied upon as bases for downstream evolutionary and epidemiological inferences. Our analyses show that infinite mixture models and standard models for across-site variation can yield maximum clade credibility (MCC) trees with notable differences (Figures A1-A3). The differences in the MCC trees are primarily accounted for by tree nodes with relatively low posterior support, raising the possibility that the uncertainty pertaining to a subset of nodes may make it difficult to arrive at a consensus ``best'' tree, but that the overall posterior distributions are still very similar. However, a more thorough exploration of the posterior via split frequencies (Figures A4-A9) and heatmaps of two dimensional representations of treespace obtained through multidimensional scaling (Figures A10-A15) reveals substantial differences in phylogenetic posterior distributions inferred under different models. Notably, such differences are not just between standard models and infinite mixture models: posterior inferences under different infinite mixture models can also differ substantially. In general, we observe a correspondence between how much the marginal likelihoods for a pair of models differ, and how much the split frequencies and heatmaps that they generate differ.

\section{Discussion}

We introduce a framework for modeling variation of evolutionary processes across multiple sequence alignment sites via a Bayesian infinite mixture model that simultaneously infers assignment of sites to evolutionary categories, the number of evolutionary categories, and the values of evolutionary model parameters that correspond to each category. Importantly, this Bayesian framework naturally accounts for uncertainty in all model parameters, including the aforementioned parameters that specifically account for across-site variation. To build infinite mixture models that can support an arbitrary number of evolutionary categories, our framework relies on Bayesian nonparametric prior distributions. Among such priors, Dirichlet processes have already been employed with great success in evolutionary modeling. Our framework offers novel methods for modeling across-site variation through two different Bayesian nonparametric priors: infinite hidden Markov models and hierarchical Dirichlet processes. 

The Bayesian nonparametric priors proposed here model clustering in different ways and lead to substantial differences in model fit and posterior phylogenetic inferences (see Appendix). We find that the suitability of each prior depends on the data set as well as on the underlying DNA substitution model. Notably, for each of the three types of priors, there is at least one scenario in which it leads to better model fit in terms of marginal likelihood than the other two priors. Infinite hidden Markov models yield the best model fit among analyses of the RSVA data under the HKY substitution model and of the RABV data under both GTR and HKY substitution models. Dirichlet process mixtures achieve the best model fit in analyses of the HCV data under GTR and HKY substitution models. The hierarchical Dirichlet process, with \textit{a priori} groups defined by codon position, leads to the best fit in analyses of the RSVA data under the GTR substitution model. These results underscore the importance of expanding beyond Dirichlet processes, which have thus far formed the basis of infinite mixture models for across-site evolutionary variation, and taking all three Bayesian nonparametric priors into consideration in order to best model across-site variation for different data sets with different evolutionary models. While no specific type of infinite mixture model emerges as a clear favorite, our study does provide support for regular adoption of infinite mixture models over standard models for across-site variation. Indeed, the best infinite mixture models outperform the best standard models in terms of model fit in all scenarios and, with the exception of the Dirichlet process mixture paired with a GTR substitution model in the case of RABV data, infinite mixture models always outperform standard models.

Our exploration in terms of data set sizes and complexity remains limited, but future application of our methodology on a variety of data sets will contribute to our understanding of the relative strengths and weaknesses of different types of infinite mixtures, and it can also provide opportunities to apply the methodology in novel ways. For example, all data sets that we analyze consist mostly or entirely of protein coding regions, suggesting hierarchical Dirichlet processes with \textit{a priori} groups determined by codon positions. Hierarchical Dirichlet processes can be specified differently for sequence data that offer alternative ``natural'' groupings of sites. In the case of genome data, for instance, sites can be grouped according to gene. It may also be fruitful to apply independent mixture models based on different kinds of nonparametric priors to different parts of an alignment. 

While moving from standard models toward infinite mixture models is a big step toward ``letting the data decide,'' our framework still relies on more \textit{a priori} modeling assumptions than ideal. For instance, we must pre-specify a particular substitution model. For every combination of data set and Bayesian nonparametric prior we examine, using an underlying HKY substitution model results in a substantially better model fit than using a less restrictive GTR model. This makes for an interesting contrast with standard modeling: for every combination of data set and standard model for across-site variation, using an underlying GTR substitution model leads to a better model fit. The sensitivity of the results underlines the importance of the substitution model. The trends we observe in our examples regarding the HKY and GTR models may not hold for other data sets, and it is also possible that other underlying substitution models may be preferable. \citet{Wu2013} present an appealing alternative to conditioning on any given substitution model: they account for substitution model uncertainty in Dirichlet process mixtures through a stochastic procedure that selects among a range of substitution models. We plan on implementing such an approach that enables substitution model selection for mixtures based on infinite hidden Markov models and hierarchical Dirichlet processes. In addition, when adhering to a single underlying substitution model, we could consider making its parametrization more data driven by using random-effect substitution models, for which efficient implementations exist in our Bayesian inference framework \citep{Magee:2024aa}.

There is much potential for further development of infinite mixture models for across-site variation. There is a wide range of Bayesian nonparametric prior distributions that model clustering in different ways than the three types of priors that we employ. These include Pitman-Yor processes \citep{Pitman1997}, dependent Dirichlet processes \citep{MacEachern1999, MacEachern2000}, spatial Dirichlet processes \citep{Gelfand2005}, nested Dirichlet processes \citep{Rodriguez2008}, sticky hierarchical Dirichlet process hidden Markov models \citep{Fox2011}, and nested hierarchical Dirichlet processes \citep{Paisley2014}. In certain scenarios, models based on such priors may very well outperform any of the models that we consider. In another direction, our framework infers partitions under the assumption that overall substitution rates and relative molecular character exchange rates are constant within each partition. \citet{Wu2013} demonstrate improved inference through decoupling the clustering of substitution rates and relative exchange rates by modeling them with independent Dirichlet processes. We anticipate that similar gains can be achieved through analogous decoupled clustering in models based on other Bayesian nonparametric priors.

While our focus has been on modeling evolutionary variation across multiple sequence alignment sites, evolutionary processes often also exhibit variation among phylogenetic tree branches. Numerous molecular clock models that posit branch-specific substitution rate variation have been developed and enjoy widespread use \citep{Thorne1998, Huelsenbeck2000b, Drummond2006, Lepage2006, Drummond2010, Bletsa2019, Didelot2021}, including a Dirichlet process mixture model \citep{Heath2012}. Other advances have united branch- and site-specific variation of substitution rates \citep{Tuffley1998, Galtier2001, Huelsenbeck2002, Zhou2010b} and all substitution model parameters \citep{Guindon2004, Gascuel2007, Whelan2008, Baele2021}. Simultaneous modeling of branch- and site-specific variation of evolutionary processes via infinite mixtures represents a promising next step. For instance, hierarchical Dirichlet processes offer a natural framework to allow across-site variation dynamics to differ among phylogenetic tree branches while still capturing essential aspects of an overall shared structure.

An important future direction that will allow for more insight into how to best model evolutionary variation is development of better methods for marginal likelihood estimation that are computationally feasible for infinite mixture models. Several studies \citep{Xie2011, Baele2012, Baele2012b, Fourment2020} have shown that estimation procedures such as path sampling \citep{Gelman1998,Lartillot2006,Baele2012}, stepping-stone sampling \citep{Xie2011}, and generalized stepping-stone sampling \citep{Fan2011,Baele2016} outperform the stabilized harmonic mean estimator that we have used. Developing and evaluating similarly sophisticated methods for infinite mixture models in phylogenetics is, however, a difficult task \citep{Fourment2020, Hairault2022}. While there has been limited development and evaluation of marginal likelihood estimation procedures specifically for Dirichlet process mixtures \citep{Basu2003, Lartillot2004, Hairault2022}, to the best of our knowledge there is no such existing work for hierarchical Dirichlet processes and infinite hidden Markov models. The AICM \citep{Raftery2007}, a simulation based analogue of of the classic Akaike information criterion (AIC) \citep{Akaike1973}, can also be used for model comparison, and \citet{Baele2012} have proposed it as a computationally efficient alternative when the best marginal likelihood estimation procedures are not feasible. However, the asymptotic theory that serves as the basis for the AIC does not hold for mixture models \citep{Watanabe2010, Gelman2014}, and recent work suggests that the AIC is not suitable for assessing mixture models in phylogenetics \citep{Susko2020, Crotty2022, Liu2023}.

The development of Bayesian nonparametrics has largely been motivated by scenarios with a growing amount of data that necessitate increasing model complexity to adequately capture structure and patterns as data accrue \citep{Ghahramani2013}. Such scenarios have taken center stage in genomic epidemiology, where advances in sequencing capabilities have enabled genomic surveillance in pathogen outbreaks and epidemics in close to real-time \citep{Quick2016, Douglas2021}. Phylogenetic inference has proven to be an integral tool in genomic epidemiology, providing evolutionary and epidemiological insights that cannot be obtained through other methods \citep{Attwood2022}, and researchers have developed frameworks for ``online'' phylogenetic inference that can efficiently deliver updated real-time inferences as new data become available \citep{Fourment2018, Gill2020}. Such online inference frameworks will undoubtedly benefit from the integration of infinite mixture models that not only outperform standard approaches on fixed data sets, but can dynamically adjust to maintain high performance as data sets grow.

In order for infinite mixture models for across-site variation to have maximal impact in ``real time'' inference settings and in general, it is essential that they are accompanied by efficient algorithms for posterior simulation. To this end, our implementation exploits a cost-effective ``data squashing'' MCMC sampling scheme for mixture model parameters \citep{Guha2010} and takes advantage of the opportunity to evaluate the data likelihood for different combinations of alignment site and evolutionary category in parallel via an interface with BEAGLE \citep{Ayres2019}. However, efficient posterior sampling remains challenging. For example, we have observed inconsistent performance in MCMC convergence and mixing across replicates of the same analysis. The data squashing strategy is powerful in its generality, subsuming Gibbs sampling as a special case while providing the opportunity to update parameter values for a large number of sites at once. We plan to investigate the data squashing tuning parameters in more detail and devise strategies to optimize them for different data sets and mixture models. We will also explore the possibility of adapting other promising strategies, such as ensemble approaches that enable multiple Markov chains that are exploring different regions of the posterior distribution to interact \citep{Lindsey2022}.

\section*{Supplementary Material}

BEAST X XML input files that feature all data analyzed in this study are available at \url{https://github.com/mandevgill/infinitemixturemodels}.

\section*{Acknowledgments}

The research leading to these results has received funding from the European Research Council under the European Union's Horizon 2020 research and innovation programme (grant agreement no. 725422-ReservoirDOCS) and from the European Union's Horizon 2020 project MOOD (grant agreement no. 874850).
The Artic Network receives funding from the Wellcome Trust through project 206298/Z/17/Z. PL acknowledges support by the Research Foundation - Flanders (‘Fonds voor Wetenschappelijk Onderzoek - Vlaanderen’, G0D5117N, G0B9317N and G051322N). 
MAS acknowledges support from US National Institutes of Health grants U19 AI135995, R01 AI153044 and R01 AI162611. MSG acknowledges support from the Centers for Disease Control and Prevention, Department of Health and Human Services, under contract NU50CK000626.

\clearpage
\bibliographystyle{mbe}
\bibliography{list_of_references}

\begin{thebibliography}{105}
\providecommand{\natexlab}[1]{#1}

\bibitem[{Akaike(1973)}]{Akaike1973}
Akaike H. 1973.
\newblock Information theory and an extension of the maximum likelihood
  principle.
\newblock In: Petrov BN, Csaki F, editors, Proceedings of the Second
  International Symposium on Information Theory. pp. 267--281.

\bibitem[{Aldous(1985)}]{Aldous1985}
Aldous D. 1985.
\newblock Ecole d'Ete de Probabilities de Saint-Flour XIII-1983,
  Springer-Verlag, chapter Exchangeability and related topics, pp. 1--198.

\bibitem[{Antoniak(1974)}]{Antoniak1974}
Antoniak C. 1974.
\newblock Mixtures of {D}irichlet processes with applications to {B}ayesian
  nonparametric problems.
\newblock \emph{The Annals of Statistics}. 2:1152--1174.

\bibitem[{Attwood et~al.(2022)Attwood, Hill, Aanensen, Connor, and
  Pybus}]{Attwood2022}
Attwood SW, Hill SC, Aanensen DM, Connor TR, Pybus OG. 2022.
\newblock Phylogenetic and phylodynamic approaches to understanding annd
  combating the early {SARS-CoV-2} pandemic.
\newblock \emph{Nature Reviews Genetics}. 23:547--562.

\bibitem[{Ayres et~al.(2019)Ayres, Cummings, Baele, Darling, Lewis, Swofford,
  Huelsenbeck, Lemey, Rambaut, and Suchard}]{Ayres2019}
Ayres DL, Cummings MP, Baele G, Darling AE, Lewis PO, Swofford DL, Huelsenbeck
  JP, Lemey P, Rambaut A, Suchard MA. 2019.
\newblock {BEAGLE} 3: Improved performance, scaling, and usability for a
  high-performance computing library for statistical phylogenetics.
\newblock \emph{Systematic Biology}. 68:1052--1061.

\bibitem[{Baele et~al.(2021)Baele, Gill, Bastide, Lemey, and
  Suchard}]{Baele2021}
Baele G, Gill MS, Bastide P, Lemey P, Suchard MA. 2021.
\newblock Markov-modulated continuous-time markov chains to identify site- and
  branch-specific evolutionary variation in {BEAST}.
\newblock \emph{Systematic Biology}. 70:181--189.

\bibitem[{Baele et~al.(2012{\natexlab{a}})Baele, Lemey, Bedford, Rambaut,
  Suchard, and Alekseyenko}]{Baele2012}
Baele G, Lemey P, Bedford T, Rambaut A, Suchard MA, Alekseyenko AV.
  2012{\natexlab{a}}.
\newblock Improving the accuracy of demographic and molecular clock model
  comparison while accommodating phylogenetic uncertainty.
\newblock \emph{Molecular Biology and Evolution}. 29:2157--67.

\bibitem[{Baele et~al.(2016)Baele, Lemey, and Suchard}]{Baele2016}
Baele G, Lemey P, Suchard MA. 2016.
\newblock Genealogical working distributions for {B}ayesian model testing with
  phylogenetic uncertainty.
\newblock \emph{Systematic Biology}. 65:250--264.

\bibitem[{Baele et~al.(2012{\natexlab{b}})Baele, Li, Drummond, Suchard, and
  Lemey}]{Baele2012b}
Baele G, Li WLS, Drummond AJ, Suchard MA, Lemey P. 2012{\natexlab{b}}.
\newblock Accurate model selection of relaxed molecular clocks in {B}ayesian
  phylogenetics.
\newblock \emph{Molecular Biology and Evolution}. 20:239--243.

\bibitem[{Basu and Chib(2003)}]{Basu2003}
Basu S, Chib S. 2003.
\newblock Marginal likelihood and {B}ayes factors for {D}irichlet process
  mixture models.
\newblock \emph{Journal of the American Statistical Association}. 98:224--235.

\bibitem[{Baum and Petrie(1966)}]{Baum1966}
Baum LE, Petrie T. 1966.
\newblock Statistical inference for probabilistic functions of finite state
  {M}arkov chains.
\newblock \emph{Annals of Mathematical Statistics}. 37:1554--1563.

\bibitem[{Beal et~al.(2002)Beal, Ghahramani, and Rasmussen}]{Beal2002}
Beal MJ, Ghahramani Z, Rasmussen C. 2002.
\newblock The infinite hidden {M}arkov model.
\newblock In: Dietterich TG, Becker S, Ghahramani Z, editors, Advances in
  Neural Information Processing Systems. MIT Press, volume~14, pp. 577--584.

\bibitem[{Biek et~al.(2007)Biek, Henderson, Waller, Rupprecht, and
  Real}]{Biek2007}
Biek R, Henderson J, Waller L, Rupprecht C, Real L. 2007.
\newblock A high-resolution genetic signature of demographic and spatial
  expansion in epizootic rabies virus.
\newblock \emph{Proceedings of the National Academy of Sciences}.
  104:7993--7998.

\bibitem[{Blackwell and MacQueen(1973)}]{Blackwell1973}
Blackwell D, MacQueen J. 1973.
\newblock Ferguson distributions via polya urn schemes.
\newblock \emph{The Annals of Statistics}. 1:353--355.

\bibitem[{Bletsa et~al.(2019)Bletsa, Suchard, Ji, Gryseels, Vrancken, Baele,
  Worobey, and Lemey}]{Bletsa2019}
Bletsa M, Suchard MA, Ji X, Gryseels S, Vrancken B, Baele G, Worobey M, Lemey
  P. 2019.
\newblock Divergence dating using mixed effects clock modelling: an application
  to {HIV-1}.
\newblock \emph{Virus Evolution}. 5:vez036.

\bibitem[{Bruno(1996)}]{Bruno1996}
Bruno WJ. 1996.
\newblock Modeling residue usage in aligned protein sequences via maximum
  likelihood.
\newblock \emph{Molecular Biology and Evolution}. 13:1368--1374.

\bibitem[{Crotty and Holland(2022)}]{Crotty2022}
Crotty SM, Holland BR. 2022.
\newblock Comparing partitioned models to mixture models: do information
  criteria apply?
\newblock \emph{Systematic Biology}. 71:1541--1548.

\bibitem[{Didelot et~al.(2021)Didelot, Siveroni, and Volz}]{Didelot2021}
Didelot X, Siveroni I, Volz EM. 2021.
\newblock Additive uncorrelated relaxed clock models for the dating of genomic
  epidemiology phylogenies.
\newblock \emph{Molecular Biology and Evolution}. 38:307--317.

\bibitem[{Douglas et~al.(2021)Douglas, Geoghegan, Hadfield, Bouckaert, Storey,
  Ren, de~Ligt, French, and Welch}]{Douglas2021}
Douglas J, Geoghegan JL, Hadfield J, Bouckaert R, Storey M, Ren X, de~Ligt J,
  French N, Welch D. 2021.
\newblock Real-time genomics for tracking severe acute respiratory syndrome
  coronavirus 2 border incursions after virus elimination, {N}ew {Z}ealand.
\newblock \emph{Emerging Infectious Diseases}. 27:2361--2368.

\bibitem[{Drummond et~al.(2006)Drummond, Ho, Phillips, and
  Rambaut}]{Drummond2006}
Drummond A, Ho S, Phillips M, Rambaut A. 2006.
\newblock Relaxed phylogenetics and dating with confidence.
\newblock \emph{PLoS Biology}. 4:e88.

\bibitem[{Drummond and Suchard(2010)}]{Drummond2010}
Drummond AJ, Suchard MA. 2010.
\newblock Bayesian random local clocks, or one rate to rule them all.
\newblock \emph{BMC Biology}. 8.

\bibitem[{Fan et~al.(2011)Fan, Wu, Chen, Kuo, and Lewis}]{Fan2011}
Fan Y, Wu R, Chen MH, Kuo L, Lewis PO. 2011.
\newblock Choosing among partition models in {B}ayesian phylogenetics.
\newblock \emph{Molecular Biology and Evolution}. 28:523--532.

\bibitem[{Felsenstein(1981)}]{Felsenstein81}
Felsenstein J. 1981.
\newblock Evolutionary trees from {DNA} sequences: a maximum likelihood
  approach.
\newblock \emph{Journal of Molecular Evolution}. 13:93--104.

\bibitem[{Felsenstein(2004)}]{Felsenstein2004}
Felsenstein J. 2004.
\newblock Inferring Phylogenies.
\newblock Sunderland, MA: Sinauer Associates, Inc.

\bibitem[{Felsenstein and Churchill(1996)}]{Felsenstein1996}
Felsenstein J, Churchill G. 1996.
\newblock A hidden {M}arkov model approach to variation among sites in rate of
  evolution.
\newblock \emph{Molecular Biology and Evolution}. 13:93--104.

\bibitem[{Ferguson(1973)}]{Ferguson1973}
Ferguson T. 1973.
\newblock A {B}ayesian analysis of some nonparametric problems.
\newblock \emph{The Annals of Statistics}. 1:209--230.

\bibitem[{Fourment et~al.(2018)Fourment, Claywell, Dinh, McCoy, Matsen~IV, and
  Darling}]{Fourment2018}
Fourment M, Claywell BC, Dinh V, McCoy C, Matsen~IV FA, Darling AE. 2018.
\newblock Effective online {B}ayesian phylogenetics via sequential {M}onte
  {C}arlo with guided proposals.
\newblock \emph{Systematic Biology}. 67:490--502.

\bibitem[{Fourment et~al.(2020)Fourment, Magee, Whidden, Bilge, Matsen~IV, and
  Minin}]{Fourment2020}
Fourment M, Magee AF, Whidden C, Bilge A, Matsen~IV FA, Minin VN. 2020.
\newblock 19 dubious ways to compute the marginal likelihood of a phylogenetic
  tree topology.
\newblock \emph{Systematic Biology}. 69:209--220.

\bibitem[{Fox et~al.(2011)Fox, Sudderth, Jordan, and Willsky}]{Fox2011}
Fox EB, Sudderth EB, Jordan MI, Willsky AS. 2011.
\newblock A sticky {HDP-HMM} with application to speaker diarization.
\newblock \emph{The Annals of Applied Statistics}. 5:1020--1056.

\bibitem[{Galtier(2001)}]{Galtier2001}
Galtier N. 2001.
\newblock Maximum-likelihood phylogenetic analysis under a covarion-like model.
\newblock \emph{Molecular Biology and Evolution}. 18:866--873.

\bibitem[{Gascuel and Guindon(2007)}]{Gascuel2007}
Gascuel O, Guindon S. 2007.
\newblock Reconstructing evolution: new mathematical and computational
  advances, Oxford University Press, chapter Modelling the variability of
  evolutionary processes.

\bibitem[{Gelfand et~al.(2005)Gelfand, Kottas, and MacEachern}]{Gelfand2005}
Gelfand AE, Kottas A, MacEachern SN. 2005.
\newblock Bayesian nonparametric spatial modeling with {D}irichlet process
  mixing.
\newblock \emph{Journal of the American Statistical Association}.
  100:1021--1035.

\bibitem[{Gelman et~al.(2014)Gelman, Hwang, and Vehtari}]{Gelman2014}
Gelman A, Hwang J, Vehtari A. 2014.
\newblock Understanding predictive information criteria for {B}ayesian models.
\newblock \emph{Statistics and Computing}. 24:997--1016.

\bibitem[{Gelman and Meng(1998)}]{Gelman1998}
Gelman A, Meng XL. 1998.
\newblock Simulating normalizing constants: from importance sampling to bridge
  sampling to path sampling.
\newblock \emph{Statistical Science}. 13:163--185.

\bibitem[{Ghahramani(2013)}]{Ghahramani2013}
Ghahramani Z. 2013.
\newblock Bayesian non-parametrics and the probabilistic approach to modelling.
\newblock \emph{Philosophical Transactions of the Royal Society A}.
  371:20110553.

\bibitem[{Gill et~al.(2013)Gill, Lemey, Faria, Rambaut, Shapiro, and
  Suchard}]{Gill2013}
Gill MS, Lemey P, Faria NR, Rambaut A, Shapiro B, Suchard MA. 2013.
\newblock Improving {B}ayesian population dynamics inference: a
  coalescent-based model for multiple loci.
\newblock \emph{Molecular Biology and Evolution}. 30:713--724.

\bibitem[{Gill et~al.(2020)Gill, Lemey, Suchard, Rambaut, and Baele}]{Gill2020}
Gill MS, Lemey P, Suchard MA, Rambaut A, Baele G. 2020.
\newblock Online {B}ayesian phylodynamic inference in {BEAST} with application
  to epidemic reconstruction.
\newblock \emph{Molecular Biology and Evolution}. 37:1832--1842.

\bibitem[{Golding(1983)}]{Golding1983}
Golding GB. 1983.
\newblock Estimates of {DNA} and protein sequence divergence: an examination of
  some assumptions.
\newblock \emph{Molecular Biology and Evolution}. 1:125--142.

\bibitem[{Guha(2010)}]{Guha2010}
Guha S. 2010.
\newblock Posterior simulation in countable mixture models for large datasets.
\newblock \emph{Journal of the American Statistical Association}. 105:775--786.

\bibitem[{Guindon et~al.(2004)Guindon, Rodrigo, Dyer, and
  Huelsenbeck}]{Guindon2004}
Guindon S, Rodrigo AG, Dyer KA, Huelsenbeck JP. 2004.
\newblock Modeling the site-specific variation of selection patterns along
  lineages.
\newblock \emph{Proceedings of the National Academy of Sciences}.
  101:12957--12962.

\bibitem[{Hairault et~al.(2022)Hairault, Robert, and Rousseau}]{Hairault2022}
Hairault A, Robert CP, Rousseau J. 2022.
\newblock Evidence estimation in finite and infinite mixture models and
  applications.
\newblock ArXiv:2205.05416.

\bibitem[{Hasegawa et~al.(1985)Hasegawa, Kishino, and Yano}]{Hasegawa1985}
Hasegawa M, Kishino H, Yano T. 1985.
\newblock Dating the human-ape splitting by a molecular clock of mitochondrial
  {DNA}.
\newblock \emph{Journal of Molecular Evolution}. 22:160--174.

\bibitem[{Hastings(1970)}]{Hastings1970}
Hastings W. 1970.
\newblock {M}onte {C}arlo sampling methods using {M}arkov chains and their
  applications.
\newblock \emph{Biometrika}. 57:97--109.

\bibitem[{Heath et~al.(2012)Heath, Holder, and Huelsenbeck}]{Heath2012}
Heath TA, Holder MT, Huelsenbeck JP. 2012.
\newblock A {D}irichlet process prior for estimating lineage-specific
  substitution rates.
\newblock \emph{Molecular Biology and Evolution}. 29:939--955.

\bibitem[{Hillis et~al.(2005)Hillis, Heath, and St.~John}]{Hillis2005}
Hillis DM, Heath TA, St~John K. 2005.
\newblock Analysis and visualization of tree space.
\newblock \emph{Systematic Biology}. 54:471--482.

\bibitem[{Huelsenbeck(2002)}]{Huelsenbeck2002}
Huelsenbeck JP. 2002.
\newblock Testing a covariotide model of {DNA} substitution.
\newblock \emph{Molecular Biology and Evolution}. 19:698--707.

\bibitem[{Huelsenbeck et~al.(2000)Huelsenbeck, Larget, and
  Swofford}]{Huelsenbeck2000b}
Huelsenbeck JP, Larget B, Swofford DL. 2000.
\newblock A compound {P}oisson orocess for relaxing the molecular clock.
\newblock \emph{Genetics}. 154:1879--1892.

\bibitem[{Huelsenbeck and Nielsen(1999)}]{Huelsenbeck1999}
Huelsenbeck JP, Nielsen R. 1999.
\newblock Variation in the pattern of nucleotide substitution across sites.
\newblock \emph{Journal of Molecular Evolution}. 48:86--93.

\bibitem[{Huelsenbeck and Suchard(2007)}]{Huelsenbeck2007b}
Huelsenbeck JP, Suchard MA. 2007.
\newblock A nonparametric method for accommodating and testing across-site rate
  variation.
\newblock \emph{Systematic Biology}. 56:975--987.

\bibitem[{Jefferys and Berger(1992)}]{Jefferys1992}
Jefferys W, Berger J. 1992.
\newblock Ockham's razor and {B}ayesian analysis.
\newblock \emph{American Statistician}. 80:64--72.

\bibitem[{Jeffreys(1935)}]{Jeffreys1935}
Jeffreys H. 1935.
\newblock Some tests of significance, treated by the theory of probability.
\newblock \emph{Mathematical Proceedings of the Cambridge Philosophical
  Society}. 31:203--222.

\bibitem[{Jeffreys(1961)}]{Jeffreys1961}
Jeffreys H. 1961.
\newblock Theory of Probability.
\newblock Oxford University Press.

\bibitem[{Jukes and Cantor(1969)}]{Jukes1969}
Jukes T, Cantor C. 1969.
\newblock Evolution of protein molecules.
\newblock In: Munro H, editor, Mammalian Protein Metabolism. Academic Press,
  pp. 21--132.

\bibitem[{Kass and Raftery(1995)}]{Kass1995}
Kass R, Raftery A. 1995.
\newblock {B}ayes factors.
\newblock \emph{Journal of the American Statistical Association}. 90:773--795.

\bibitem[{Kimura(1968)}]{Kimura1968}
Kimura M. 1968.
\newblock Evolutionary rate at the molecular level.
\newblock \emph{Nature}. 217:624--626.

\bibitem[{Kimura(1980)}]{Kimura1980}
Kimura M. 1980.
\newblock A simple method for estimating evolutionary rates of base
  substitutions through comparative studies of nucleotide sequences.
\newblock \emph{Journal of Molecular Evolution}. 16:111--120.

\bibitem[{Lakner et~al.(2008)Lakner, van~der Mark, Huelsenbeck, and
  Ronquist}]{Lakner2008}
Lakner C, van~der Mark P, Huelsenbeck B J P annd~Larget, Ronquist F. 2008.
\newblock Efficiency of {M}arkov chain {M}onte {C}arlo tree proposals in
  {B}ayesian phylogenetics.
\newblock \emph{Systematic Biology}. 57:86--103.

\bibitem[{Lanave et~al.(1984)Lanave, Preparata, Saccone, and
  Serio}]{Lanave1984}
Lanave C, Preparata G, Saccone C, Serio G. 1984.
\newblock A new method for calculating evolutionary substitution rates.
\newblock \emph{Jounal of Molecular Evolution}. 20:86--93.

\bibitem[{Lartillot and Philippe(2004)}]{Lartillot2004}
Lartillot N, Philippe H. 2004.
\newblock A {B}ayesian mixture model for across-site heterogeneities in the
  amino-acid replacement process.
\newblock \emph{Molecular Biology and Evolution}. 21:2004.

\bibitem[{Lartillot and Philippe(2006)}]{Lartillot2006}
Lartillot N, Philippe H. 2006.
\newblock Computing {B}ayes factors using thermodynamic integration.
\newblock \emph{Systematic Biology}. 55:195--207.

\bibitem[{Lepage et~al.(2006)Lepage, Lawi, Tupper, and Bryant}]{Lepage2006}
Lepage T, Lawi S, Tupper P, Bryant D. 2006.
\newblock Continuous and tractable models for the variation of evolutionary
  rates.
\newblock \emph{Mathematical Biosciences}. 199:216--233.

\bibitem[{Lindsey et~al.(2022)Lindsey, Weare, and Zhang}]{Lindsey2022}
Lindsey M, Weare J, Zhang A. 2022.
\newblock Ensemble {M}arkov chain {M}onte {C}arlo with teleporting walkers.
\newblock \emph{SIAM/ASA Journal on Uncertainty Quantification}. 10:860--885.

\bibitem[{Liu et~al.(2023)Liu, Charleston, Richards, and Holland}]{Liu2023}
Liu Q, Charleston MA, Richards SA, Holland BR. 2023.
\newblock Performance of {A}kaike information criterion and {B}ayesian
  information criterion in selecting partition models and mixture models.
\newblock \emph{Systematic Biology}. 72:92--105.

\bibitem[{MacEachern(1999)}]{MacEachern1999}
MacEachern SN. 1999.
\newblock Dependent nonparametric processes.
\newblock In: ASA {P}roceedings of the {S}ection on {B}ayesian {S}tatistical
  {S}cience. pp. 50--55.

\bibitem[{MacEachern(2000)}]{MacEachern2000}
MacEachern SN. 2000.
\newblock Dependent {D}irichlet processes.
\newblock Technical report, Ohio State University, Deparment of Statistics.

\bibitem[{Magee et~al.(2024)Magee, Holbrook, Pekar, Caviedes-Solis, Matsen~Iv,
  Baele, Wertheim, Ji, Lemey, and Suchard}]{Magee:2024aa}
Magee AF, Holbrook AJ, Pekar JE, Caviedes-Solis IW, Matsen~Iv FA, Baele G,
  Wertheim JO, Ji X, Lemey P, Suchard MA. 2024.
\newblock Random-effects substitution models for phylogenetics via scalable
  gradient approximations.
\newblock \emph{Syst Biol}. 73:562--578.

\bibitem[{Metropolis et~al.(1953)Metropolis, Rosenbluth, Rosenbluth, Teller,
  and Teller}]{Metropolis1953}
Metropolis N, Rosenbluth A, Rosenbluth M, Teller A, Teller E. 1953.
\newblock Equation of state calculation by fast computing machines.
\newblock \emph{Journal of Chemical Physics}. 21:1087--1092.

\bibitem[{Neal(2000)}]{Neal2000}
Neal RM. 2000.
\newblock {M}arkov chain sampling methods for {D}irichlet process mixture
  models.
\newblock \emph{Journal of Computational and Graphical Statistics}. 9:249--265.

\bibitem[{Nei et~al.(1976)Nei, Chakraborty, and Fuerst}]{Nei1976}
Nei M, Chakraborty R, Fuerst PA. 1976.
\newblock Infinite allele model with varying mutation rate.
\newblock \emph{Proceedings of the National Academy of Sciences}.
  73:4164--4168.

\bibitem[{Newton and Raftery(1994)}]{Newton1994}
Newton M, Raftery A. 1994.
\newblock Approximate {B}ayesian inference with the weighted likelihood
  bootstrap.
\newblock \emph{Journal of the Royal Statistical Society, Series B}. 56:3--48.

\bibitem[{Nielsen(1997)}]{Nielsen1997}
Nielsen R. 1997.
\newblock Site-by-site estimation of the rate of evolution and the correlation
  of rates in mitochondrial {DNA}.
\newblock \emph{Systematic Biology}. 46:346--353.

\bibitem[{Pagel and Meade(2004)}]{PagelMeade2004}
Pagel M, Meade A. 2004.
\newblock A phylogenetic mixture model for detecting pattern-heterogeneity in
  gene sequence or character-state data.
\newblock \emph{Systematic Biology}. 53:571--581.

\bibitem[{Paisley et~al.(2014)Paisley, Wang, Blei, and Jordan}]{Paisley2014}
Paisley J, Wang C, Blei DM, Jordan MI. 2014.
\newblock Nested hierarchical {D}irichlet processes.
\newblock \emph{IEEE transactions on pattern analysis and machine
  intelligence}. 37:256--270.

\bibitem[{Pitman and Yor(1997)}]{Pitman1997}
Pitman J, Yor M. 1997.
\newblock The two-parameter {P}oisson-{D}irichlet distribution derived from a
  stable subordinator.
\newblock \emph{The Annals of Probability}. 25:855--900.

\bibitem[{Quick et~al.(2016)Quick, Loman, Durrafour et~al.}]{Quick2016}
Quick J, Loman N, Durrafour S, et~al. (101 co-authors). 2016.
\newblock Real-time, portable genome sequencing for {E}bola surveillance.
\newblock \emph{Nature}. 530:228--232.

\bibitem[{{R Core Team}(2021)}]{R2021}
{R Core Team}. 2021.
\newblock R: A Language and Environment for Statistical Computing.
\newblock R Foundation for Statistical Computing, Vienna, Austria.

\bibitem[{Raftery et~al.(2007)Raftery, Newton, Satagopan, and
  Krivitsky}]{Raftery2007}
Raftery A, Newton M, Satagopan J, Krivitsky P. 2007.
\newblock Bayesian Statistics, Oxford University Press, chapter Estimating the
  integrated likelihood via posterior simulation using the harmonic mean
  identity, pp. 1--45.

\bibitem[{Ray et~al.(2000)Ray, Arthur, Carella, Bukh, and Thomas}]{Ray2000}
Ray SC, Arthur RR, Carella A, Bukh J, Thomas DL. 2000.
\newblock Genetic epidemiology of hepatitis {C} virus throughout {E}gypt.
\newblock \emph{Journal of Infectious Diseases}. 182:698--707.

\bibitem[{Redelings and Suchard(2005)}]{Redelings2005}
Redelings BD, Suchard MA. 2005.
\newblock Joint {B}ayesian estimation of alignment and phylogeny.
\newblock \emph{Systematic Biology}. 54:401--418.

\bibitem[{Rodriguez et~al.(2008)Rodriguez, Dunson, and Gelfand}]{Rodriguez2008}
Rodriguez A, Dunson DB, Gelfand AE. 2008.
\newblock The nested {D}irichlet process.
\newblock \emph{Journal of the American Statistical Association}.
  103:1131--11144.

\bibitem[{Sethuraman(1994)}]{Sethuraman1994}
Sethuraman J. 1994.
\newblock A constructive definition of {D}irichlet priors.
\newblock \emph{Statistica Sinica}. 4:639--650.

\bibitem[{Suchard et~al.(2018)Suchard, Lemey, Baele, Ayres, Drummond, and
  Rambaut}]{BEAST2018}
Suchard MA, Lemey P, Baele G, Ayres DL, Drummond AJ, Rambaut A. 2018.
\newblock Bayesian phylogenetic and phylodynamic data integration using {BEAST}
  1.10.
\newblock \emph{Virus Evolution}. 4:vey016.

\bibitem[{Susko and Roger(2020)}]{Susko2020}
Susko E, Roger AJ. 2020.
\newblock On the use of information criteria for model selection in
  phylogenetics.
\newblock \emph{Molecular Biology and Evolution}. 37:549--562.

\bibitem[{Swofford et~al.(1996)Swofford, Olsen, Waddell, and
  Hillis}]{Swofford1996}
Swofford DL, Olsen GJ, Waddell PJ, Hillis DM. 1996.
\newblock Molecular Systematics, Sinauer Associates, Inc., chapter Phylogenetic
  Inference, pp. 407--514.
\newblock 2nd edition.

\bibitem[{Tamura(1992)}]{Tamura1992}
Tamura K. 1992.
\newblock Estimation of the number of nucleotide substitutions when there are
  stronng transition-transversion and {G + C}-content biases.
\newblock \emph{Molecular Biology and Evolution}. 9:678--687.

\bibitem[{Tamura and Nei(1993)}]{Tamura1993}
Tamura K, Nei M. 1993.
\newblock Estimation of the number of nucelotide substitutions in the control
  region of mitochondrial {DNA} in humans and chimpanzees.
\newblock \emph{Molecular Biology and Evolution}. 10:512--526.

\bibitem[{Tavare(1986)}]{Tavare1986}
Tavare S. 1986.
\newblock Some probabilistic and statistical problems on the analysis of dna
  sequences.
\newblock \emph{Lectures on Mathematics in the Life Sciences}. 17:57--86.

\bibitem[{Teh et~al.(2006)Teh, Jordan, Beal, and Blei}]{Teh2006}
Teh YW, Jordan MI, Beal MJ, Blei DM. 2006.
\newblock Hierarchical {D}irichlet processes.
\newblock \emph{Journal of the American Statistical Association}.
  101:1566--1581.

\bibitem[{Thorne et~al.(1998)Thorne, Kishino, and Painter}]{Thorne1998}
Thorne J, Kishino H, Painter I. 1998.
\newblock Estimating the rate of evolution of the rate of molecular evolution.
\newblock \emph{Molecular Biology and Evolution}. 15:1647--1657.

\bibitem[{Tuffley and Steel(1998)}]{Tuffley1998}
Tuffley C, Steel M. 1998.
\newblock Modeling the covarion hypothesis of nucleotide substitution.
\newblock \emph{Mathematical Biosciences}. 147:63--91.

\bibitem[{Venables and Ripley(2002)}]{Venables2002}
Venables WN, Ripley BD. 2002.
\newblock Modern Applied Statistics with {S}.
\newblock New York: Springer, fourth edition.

\bibitem[{Venditti et~al.(2008)Venditti, Meade, and Pagel}]{Venditti2008}
Venditti C, Meade A, Pagel M. 2008.
\newblock Phylogenetic mixture models can reduce node-density artifacts.
\newblock \emph{Systematic Biology}. 57:286--293.

\bibitem[{Waddell and Steel(1997)}]{Waddell1997}
Waddell PJ, Steel MA. 1997.
\newblock General time-reversible distances with unequal rates across sites:
  mixing and inverse {G}aussian distributions with invariant sites.
\newblock \emph{Molecular Phylogenetics and Evolution}. 8:398--414.

\bibitem[{Warren et~al.(2017)Warren, Geneva, and Lanfear}]{Warren2017}
Warren DL, Geneva AJ, Lanfear R. 2017.
\newblock {RWTY: (R We There Yet):} an {R} package for examining convergence of
  {B}ayesian phylogenetic analyses.
\newblock \emph{Molecular Biology and Evolution}. 34:1016--1020.

\bibitem[{Watanabe(2010)}]{Watanabe2010}
Watanabe S. 2010.
\newblock Asymptotic equivalence of {B}ayess cross validation and widely
  applicable information criterion in singular learning theory.
\newblock \emph{Journal of Machine Learning Research}. 11:3571--3594.

\bibitem[{Whelan(2008)}]{Whelan2008}
Whelan S. 2008.
\newblock Spatial and temporal heterogeneity in nuceleotide sequence evolution.
\newblock \emph{Molecular Biology and Evolution}. 25:1683--1694.

\bibitem[{Wu et~al.(2013)Wu, Suchard, and Drummond}]{Wu2013}
Wu CH, Suchard MA, Drummond AJ. 2013.
\newblock {B}ayesian selection of nucleotide substitution model and their site
  assignments.
\newblock \emph{Molecular Biology and Evolution}. 30:669--688.

\bibitem[{Xie et~al.(2011)Xie, Lewis, and Kuo}]{Xie2011}
Xie W, Lewis PO, Kuo MH L annd~Chen. 2011.
\newblock Improving marginal likelihood estimation for {B}ayesian phylogenetic
  model selection.
\newblock \emph{Systematic Biology}. 60:150--160.

\bibitem[{Yang and Le~Cam(2000)}]{YangLeCam2000}
Yang GL, Le~Cam L. 2000.
\newblock Asymptotics in Statistics: Some Basic Concepts.
\newblock Berlin: Springer.

\bibitem[{Yang(1993)}]{Yang1993}
Yang Z. 1993.
\newblock Maximum-likelihood estimation of phylogeny from {DNA} sequences when
  substitution rates differ over sites.
\newblock \emph{Molecular Biology and Evolution}. 10:1396--1401.

\bibitem[{Yang(1994)}]{Yang1994}
Yang Z. 1994.
\newblock Maximum likelihood phylogenetic estimation from {DNA} sequences with
  variable rates over sites: approximate methods.
\newblock \emph{Journal of Molecular Evolution}. 39:306--314.

\bibitem[{Yang(1995)}]{Yang1995}
Yang Z. 1995.
\newblock A space-time process model for the evolution of {DNA} sequences.
\newblock \emph{Genetics}. 139:993--1005.

\bibitem[{Yang(2006)}]{Yang2006}
Yang Z. 2006.
\newblock Computational Molecular Evolution.
\newblock Oxford University Press.

\bibitem[{Zhou et~al.(2010)Zhou, Brinkmannn, Rodrigue, Lartillot, and
  Philippe}]{Zhou2010b}
Zhou Y, Brinkmannn H, Rodrigue N, Lartillot N, Philippe H. 2010.
\newblock A {D}irichlet process covarion mixture model and its assessments
  using posterior predictive discrepancy tests.
\newblock \emph{Molecular Biology and Evolution}. 27:371--384.

\bibitem[{Zlateva et~al.(2005)Zlateva, Lemey, Mo{\"e}s, Vandamme, and
  Van~Ranst}]{Zlateva2005}
Zlateva K, Lemey P, Mo{\"e}s E, Vandamme AM, Van~Ranst M. 2005.
\newblock Genetic variability and molecular evolution of the human respiratory
  syncytial virus subgroup {B} atttachment {G} protein.
\newblock \emph{J. Virol.} 79:9157--9167.

\end{thebibliography}
%\bibliography{../list_of_references}

\clearpage

\appendix

\section{Appendix}

\renewcommand{\thefigure}{A\arabic{figure}}
\setcounter{figure}{0}
\setcounter{table}{0}
\renewcommand{\thetable}{A\arabic{table}}

\subsection{Performance of Across-Site Variation Models in Empirical Examples}

\begin{table*}[htb]
\begin{center}
\begin{tabular}{lccccccc}
 & & \multicolumn{2}{c}{RSVA} & \multicolumn{2}{c}{HCV} & \multicolumn{2}{c}{RABV} \\
\cline{1-8} 
\\\\[-4\medskipamount]
 & & \multicolumn{1}{c}{GTR} & \multicolumn{1}{c}{HKY}  & \multicolumn{1}{c}{GTR} & \multicolumn{1}{c}{HKY} & \multicolumn{1}{c}{GTR} & \multicolumn{1}{c}{HKY}\\
\hline 
\\\\[-4\medskipamount]
No Variation & & -3134 & -3147 & -6768 & -6769 & -6828 & -6846 \\[2ex]
Gamma & & -3092 & -3111 & -6167 & -6169 & -6785 & -6802 \\[2ex]
Codon & & -3102 & -3132 & -6335 & -6355 & -6569 & -6591 \\[2ex]
Codon $+$ Gamma & & -3073 & -3092 & -6026 & -6039 & -6552 & -6572 \\[2ex]
DP & & -2984 & -2955 & \textbf{-5817} & \textbf{-5555} & -6582 & -6400 \\[2ex]
HDP-Codon & & \textbf{-2977} & -2932 & -5823 & -5647 & -6460 & -6336 \\[2ex]
IHMM & & -2981 & \textbf{-2835} & -5819 & -5730 & \textbf{-5489} & \textbf{-4477} \\
\\\\[-4\medskipamount]
\hline
\end{tabular}
\end{center}
\caption[Performance of different models for across-site variation on respiratory syncytial virus subgroup A (RSVA), hepatitis C subtype 4 (HCV), and rabies virus (RABV) data.]
{Log marginal likelihoods for analyses of respiratory syncytial virus subgroup A (RSVA), hepatitis C subtype 4 (HCV), and rabies virus (RABV) data under different evolutionary models. Rows correspond to different approaches for modeling across-site variation of substitution rates and relative exchange rates of nucleotide bases. Columns correspond to different data sets and different nucleotide substitution models. A greater log marginal likelihood indicates a better model fit (the log marginal likelihood associated with best fitting model in each column is bolded). \label{tab:combinedperformance}}
\end{table*}

\subsection{Comparison of Summary Trees under Best Infinite Mixture Model vs. Standard Model}

We illustrate the impact of choice of across-site variation model on trees that summarize the posterior sample of phylogenetic trees from a Bayesian analysis. For each viral data set, we compare the maximum clade credibility (MCC) trees under the best fitting infinite mixture model and the best fitting standard model that uses the same underlying nucleotide substitution model as the infinite mixture model. To arrive at an MCC tree, each clade is assigned a score that reflects how frequently it appears in the posterior sample of phylogenetic trees. Each tree in the sample is assigned a score based on the scores of its clades, and the tree with the highest score is the MCC tree. We depict the internal nodes of MCC trees as circles of varying size, with larger circles corresponding to greater posterior support. Nodes are annotated with shaded bars that represent the 95$\%$ highest posterior density estimates of node heights.

\begin{figure*}[htb]
\centering
\begin{subfigure}{0.49\textwidth}
  \centering
     \includegraphics[width=1.0\linewidth]{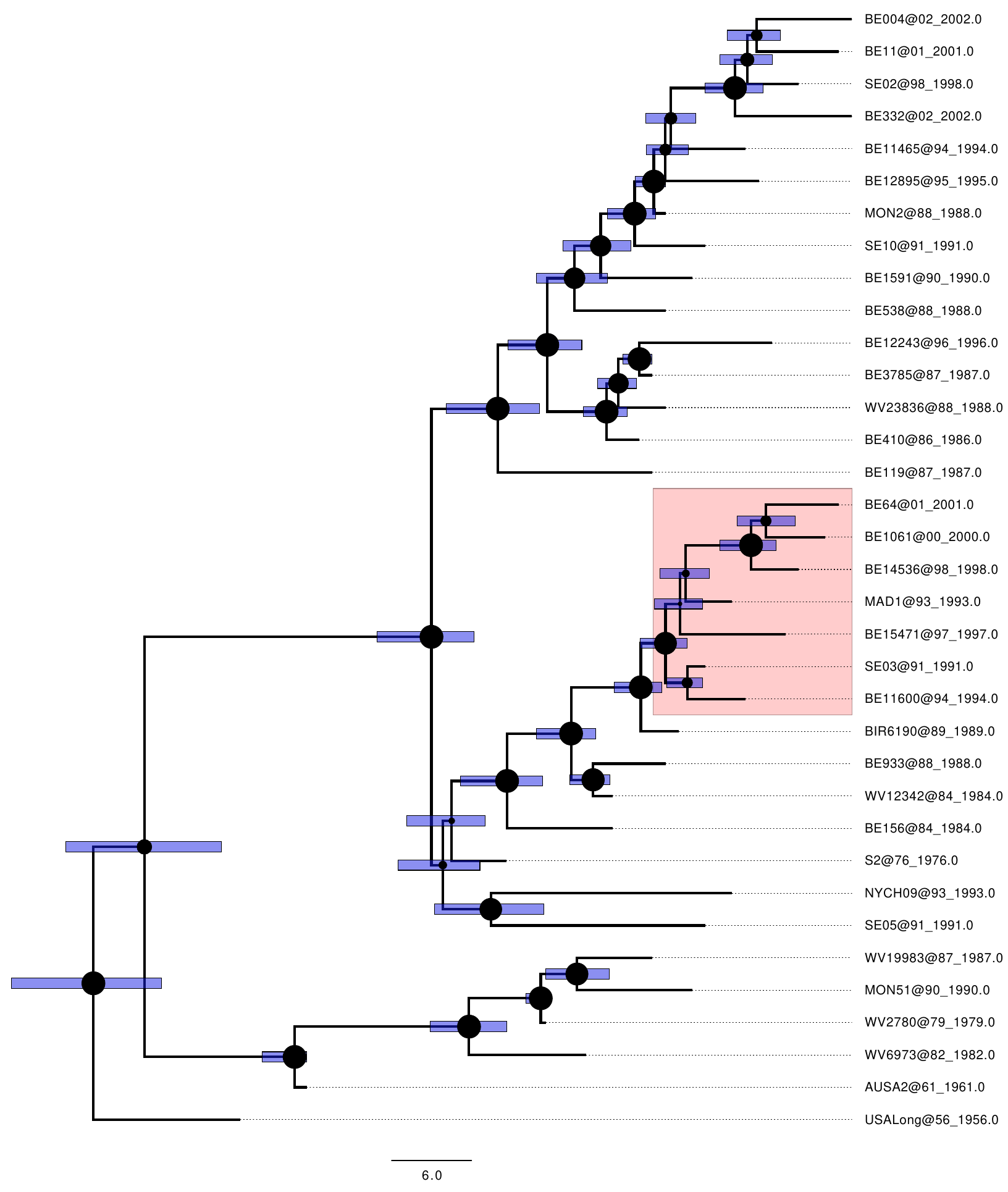}
        \label{fig:mccrsva1}
  \end{subfigure}
\begin{subfigure}{0.49\textwidth}
  \centering
   \includegraphics[width=1.0\linewidth]{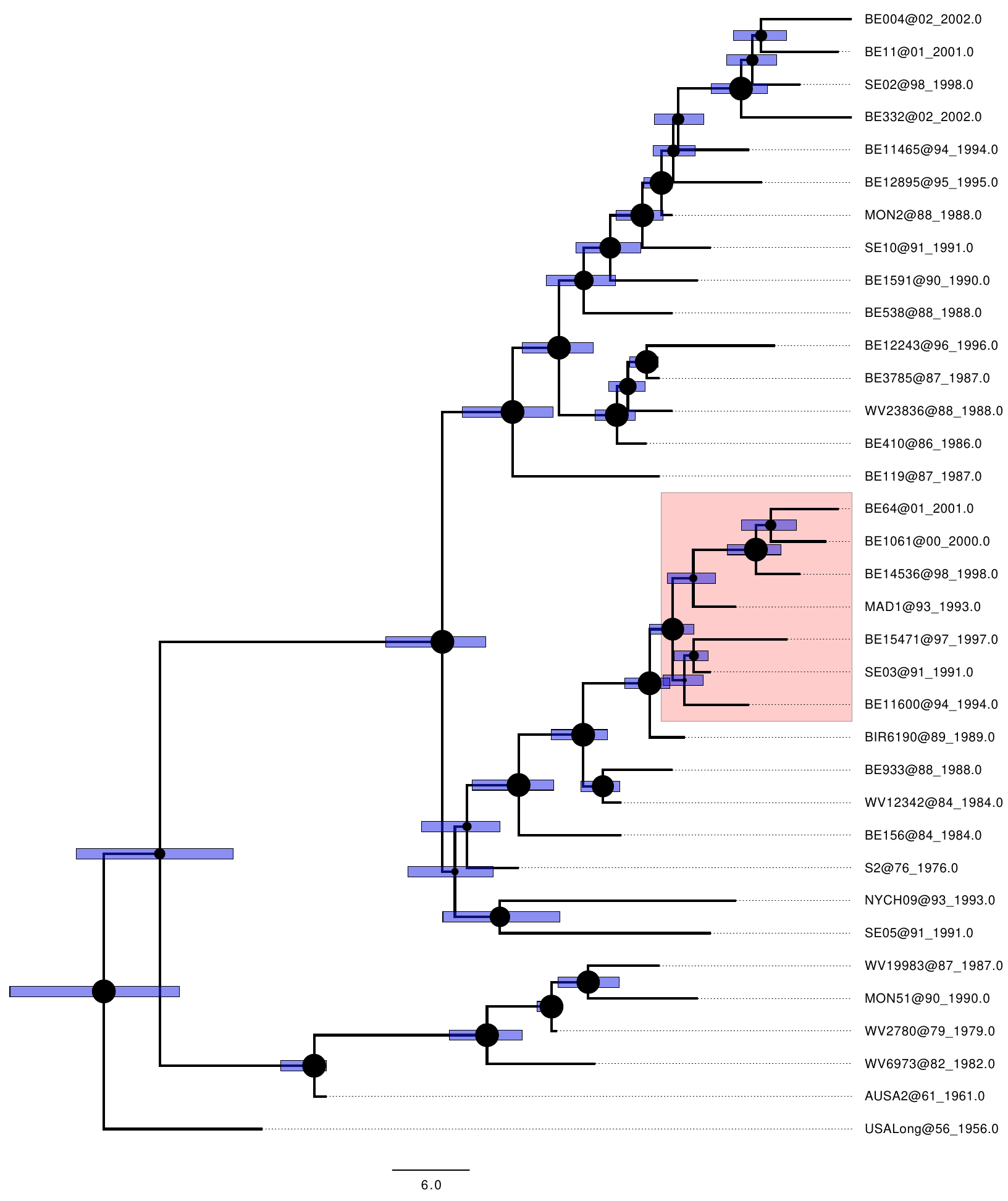}
        \label{fig:mccrsva2}
\end{subfigure}	 
  \caption{Comparison of Maximum clade credibility (MCC) trees for analyses of respiratory syncytial virus subgroup A data. The MCC tree on the left is based on the HKY + Codon + Gamma model and the MCC tree on the right is based on the HKY + IHMM model. The topological differences between the two trees can be seen in the clades shaded in red. Internal tree nodes are depicted as circles of varying size corresponding to the level of posterior support. Shaded bars represent 95$\%$ highest posterior density estimates of node heights.}
     \label{fig:mccrsva}
\end{figure*}

\begin{figure*}[htb]
\centering
\begin{subfigure}{0.5\textwidth}
  \centering
    \includegraphics[width=1.0\linewidth]{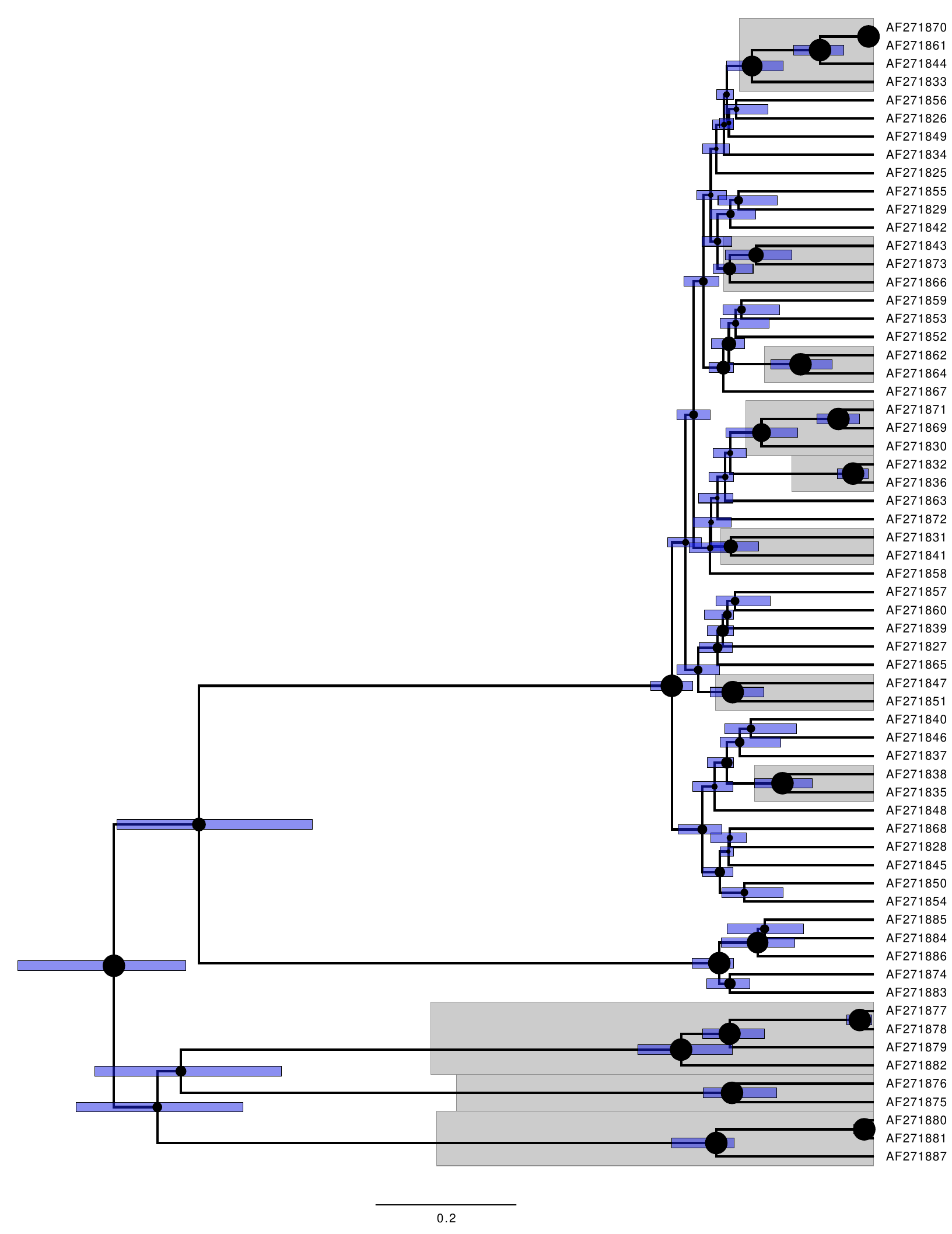}
  \end{subfigure}%
\begin{subfigure}{0.5\textwidth}
  \centering
    \includegraphics[width=1.0\linewidth]{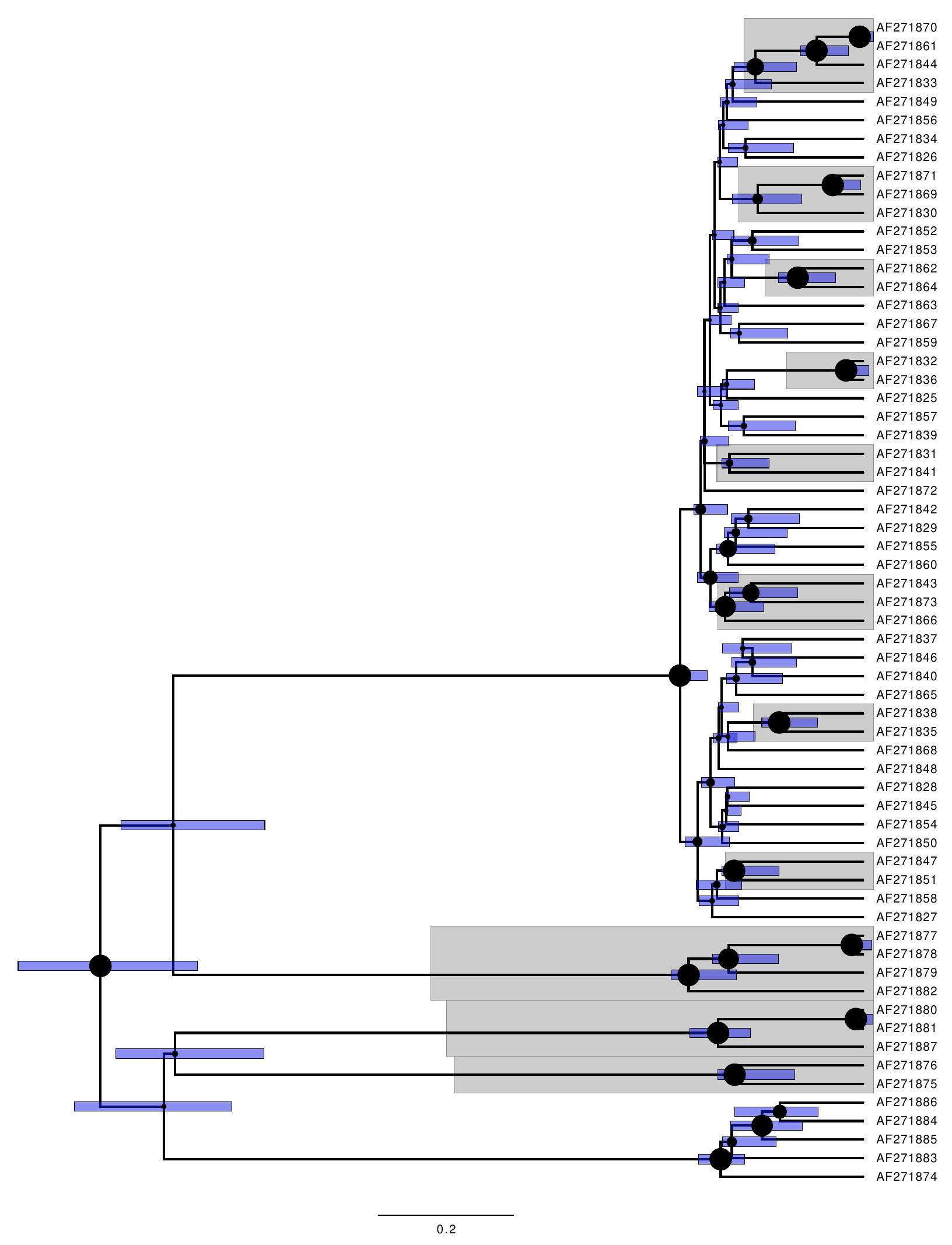}
\end{subfigure}	 
  \caption{Comparison of Maximum clade credibility (MCC) trees for analyses of hepatitis C subtype 4 data. The MCC tree on the left is based on the HKY + Codon + Gamma model and the MCC tree on the right is based on the HKY + DP model. To ease the comparison, clades that feature the same evolutionary relationships in both trees are shaded in gray. Internal tree nodes are depicted as circles of varying size corresponding to the level of posterior support. Shaded bars represent 95$\%$ highest posterior density estimates of node heights.}
     \label{fig:mcchcv}
\end{figure*}

\begin{figure*}[htb]
\centering
\begin{subfigure}{0.5\textwidth}
  \centering
    \includegraphics[width=1.0\linewidth]{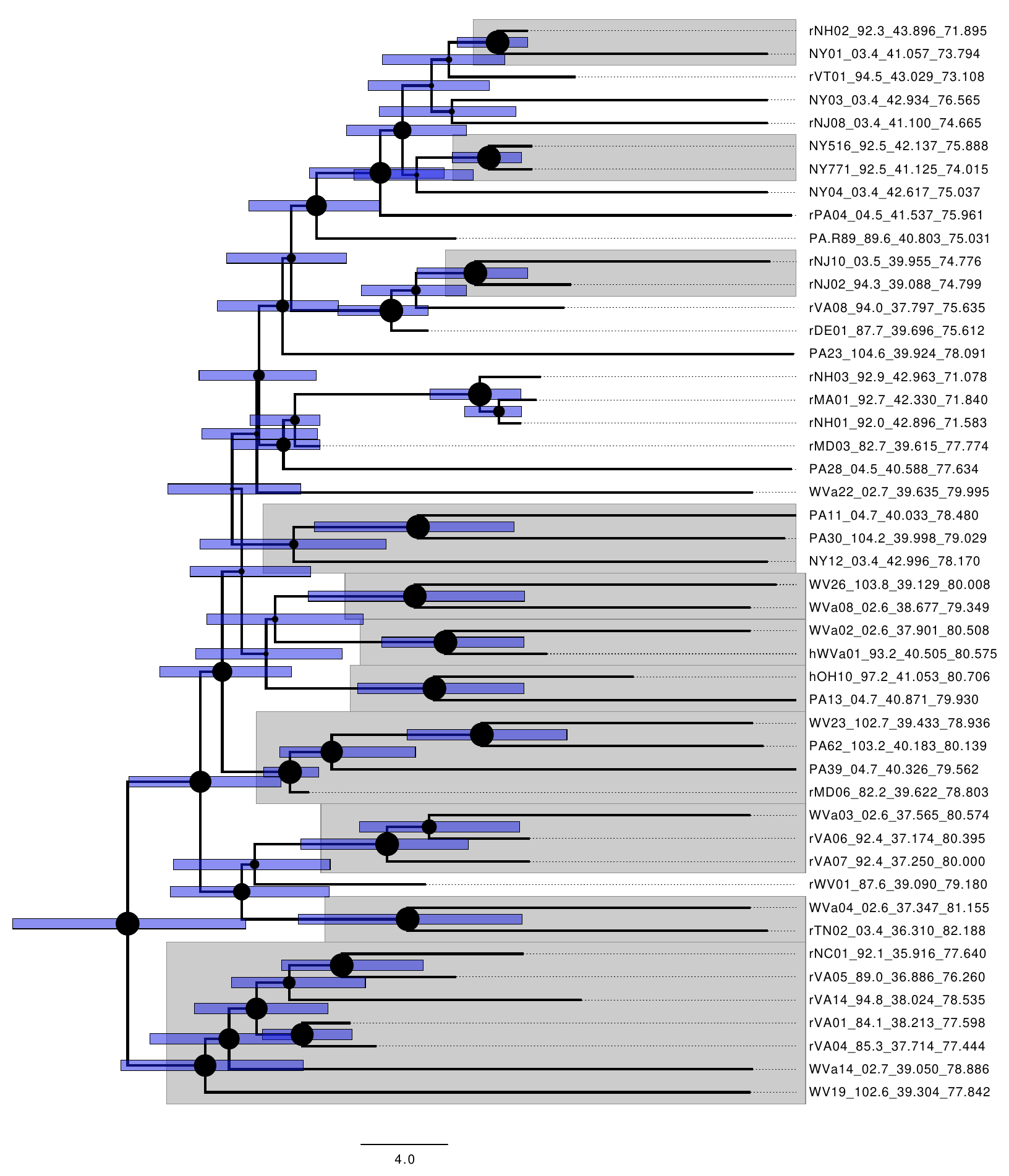}
  \end{subfigure}%
\begin{subfigure}{0.5\textwidth}
  \centering
   \includegraphics[width=1.0\linewidth]{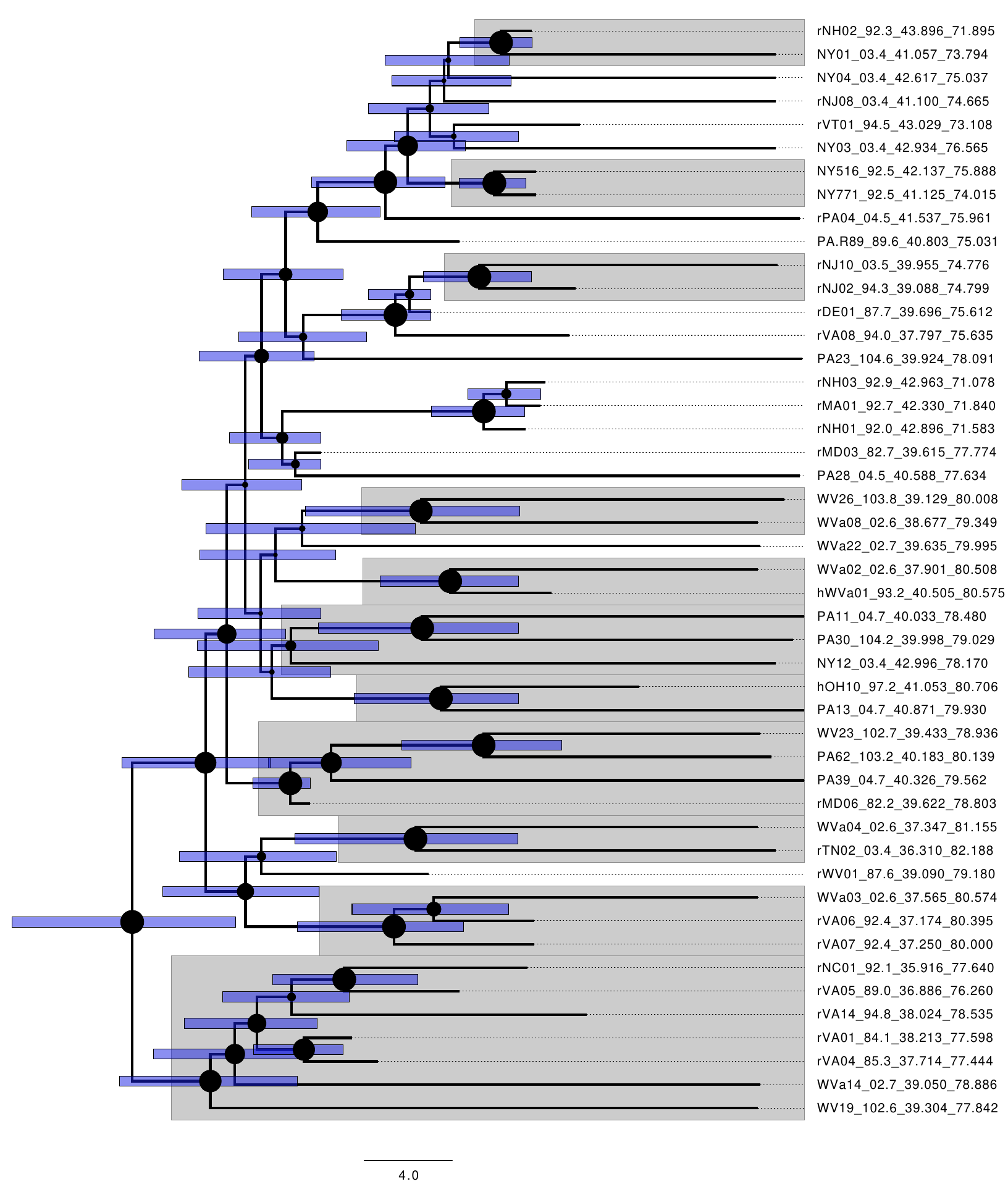}
\end{subfigure}	 
  \caption{Comparison of Maximum clade credibility (MCC) trees for analyses of rabies virus data. The MCC tree on the left is based on the HKY + Codon + Gamma model and the MCC tree on the right is based on the HKY + IHMM model. To ease the comparison, clades that feature the same evolutionary relationships in both trees are shaded in gray. Internal tree nodes are depicted as circles of varying size corresponding to the level of posterior support. Shaded bars represent 95$\%$ highest posterior density estimates of node heights.}
     \label{fig:mccrabv}
\end{figure*}

Figure~\ref{fig:mccrsva} shows the MCC trees resulting from analyses of respiratory syncytial virus subgroup A (RSVA) data \citep{Zlateva2005} under the HKY  + Codon + Gamma and HKY + IHMM models. The evolutionary relationships in the two trees match closely, and the clade highlighted in red exhibits the topological differences between the two trees. Figure~\ref{fig:mcchcv} presents MCC trees inferred from hepatitis C subtype 4 (HCV) data \citep{Ray2000} under the HKY + Codon + Gamma and HKY + DP models. These trees depict many different evolutionary relationships, and to make it easier to see how the trees disagree, we have shaded clades with the \textit{same} evolutionary relationships in both trees in gray. MCC trees from analyes of rabies virus (RABV) data \citep{Biek2007} under the HKY + Codon + Gamma and HKY + IHMM models are shown in Figure~\ref{fig:mccrabv}. As in Figure~\ref{fig:mcchcv}, the MCC trees exhibit many differences, and clades that have the same evolutionary relationships in both trees are shaded in gray.

\clearpage

\subsection{Comparison of Phylogenetic Tree Posterior Distributions in Analyses of Viral Data Sets}

We compare the posterior distributions of phylogenetic trees in analyses of viral data sets under different models for across-site evolutionary variation. First, we examine the frequencies of splits in posterior samples. We consider analyses of the RSVA, HCV, and RABV data sets. For each combination of data set and nucleotide substitution model (HKY or GTR), we make pairwise comparisons of split frequencies under four different models for across site variation: the best fitting standard model (Codon + Gamma) and the DP, HDP + Codon, and IHMM infinite mixture models. We use the RWTY (R We There Yet) software package \citep{Warren2017}, to create plots of split frequencies for different pairs of models and also compute their correlation and the average standard deviation of split frequencies (ASDSF) \citep{Lakner2008}. Results are shown in Figures A4-A9. While the posterior split frequencies are similar under different models for across-site variation, there are some notable differences, especially for the HCV and RABV analyses. In general, we observe a correspondence between the degree of difference between the posterior split frequencies and between the marginal likelihoods under a given pair of models.

To visualize the extent to which different models sample from different regions of phylogenetic tree space, we use RWTY to map phylogenetic trees into two dimensions via multidimensional scaling \citep{Hillis2005} and create heatmaps of posterior samples under different models. Each of Figures A10-A15 depicts heatmaps for one of the three viral data sets under four different models: the DP, HDP + Codon, IHMM, and best fitting standard model (Codon + Gamma) with the same underlying nucleotide substitution model (HKY or GTR). In each figure, the treespace is generated using all four sets of posterior samples so that the heatmaps are comparable. Heatmaps from different figures are not directly comparable. We observe a general correspondence between the degree of difference between the heatmaps and between the marginal likelihoods under a given pair of models.

\begin{figure*}[htb]
\centering
	 \includegraphics[width=1.0\textwidth, trim={0.5cm 0.5cm 0.5cm 0.5cm},clip]{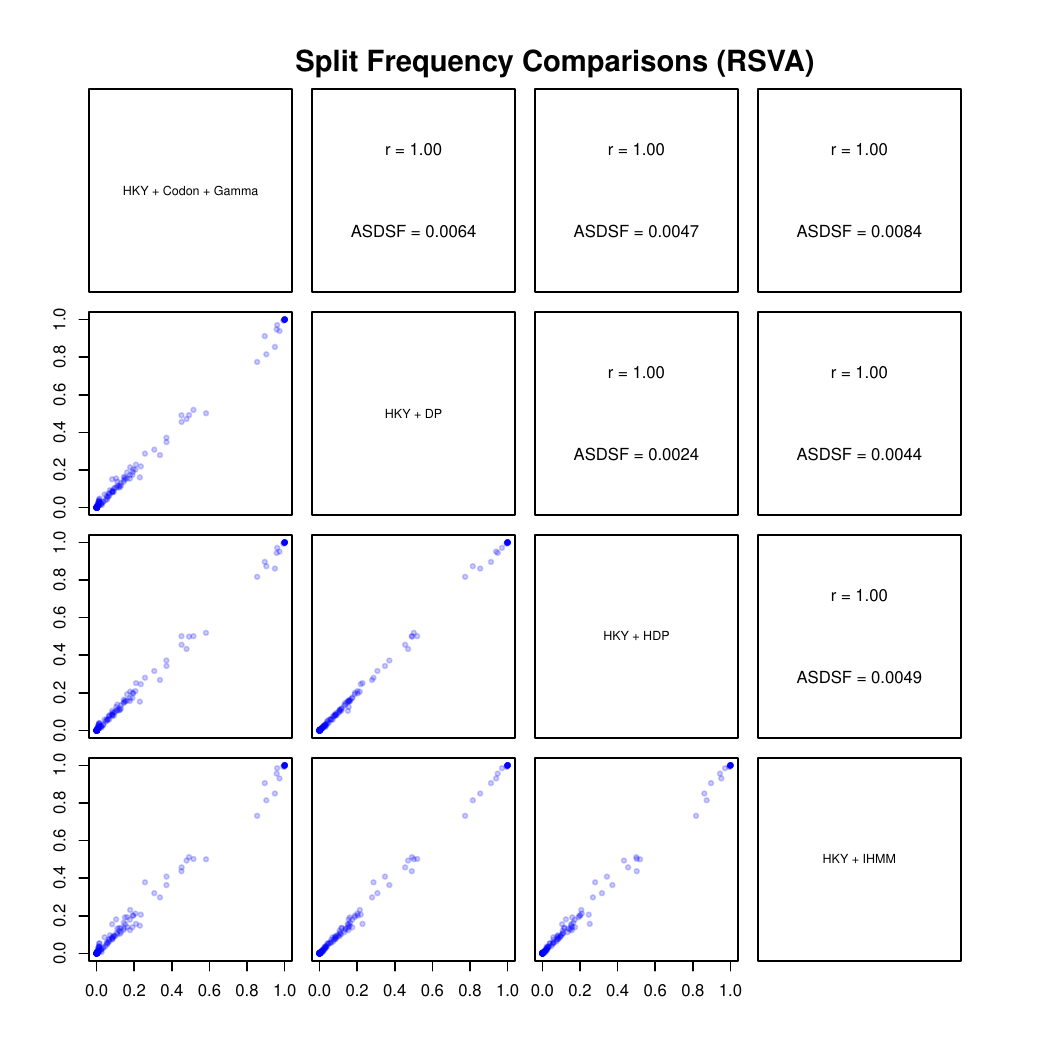} 
  \caption{Comparison of split frequencies in posterior samples from analyses of respiratory syncytial virus subgroup A data using different models for across-site evolutionary variation. All models use the HKY nucleotide substitution model. For each pair of models, the frequencies of clades in the posterior samples are plotted against each other in the plots below the diagonal of the figure, and the correlation and the average standard deviation of split frequencies (ASDSF) of the pair are shown above the diagonal. The diagonal entries correspond to different models for across-site evolutionary variation and indicate the specific pairwise comparisons that are made below and above the diagonal.}
     \label{fig:splitfreqrsvahky}
\end{figure*}

\begin{figure*}[htb]
\centering
	 \includegraphics[width=1.0\textwidth, trim={0.5cm 0.5cm 0.5cm 0.5cm},clip]{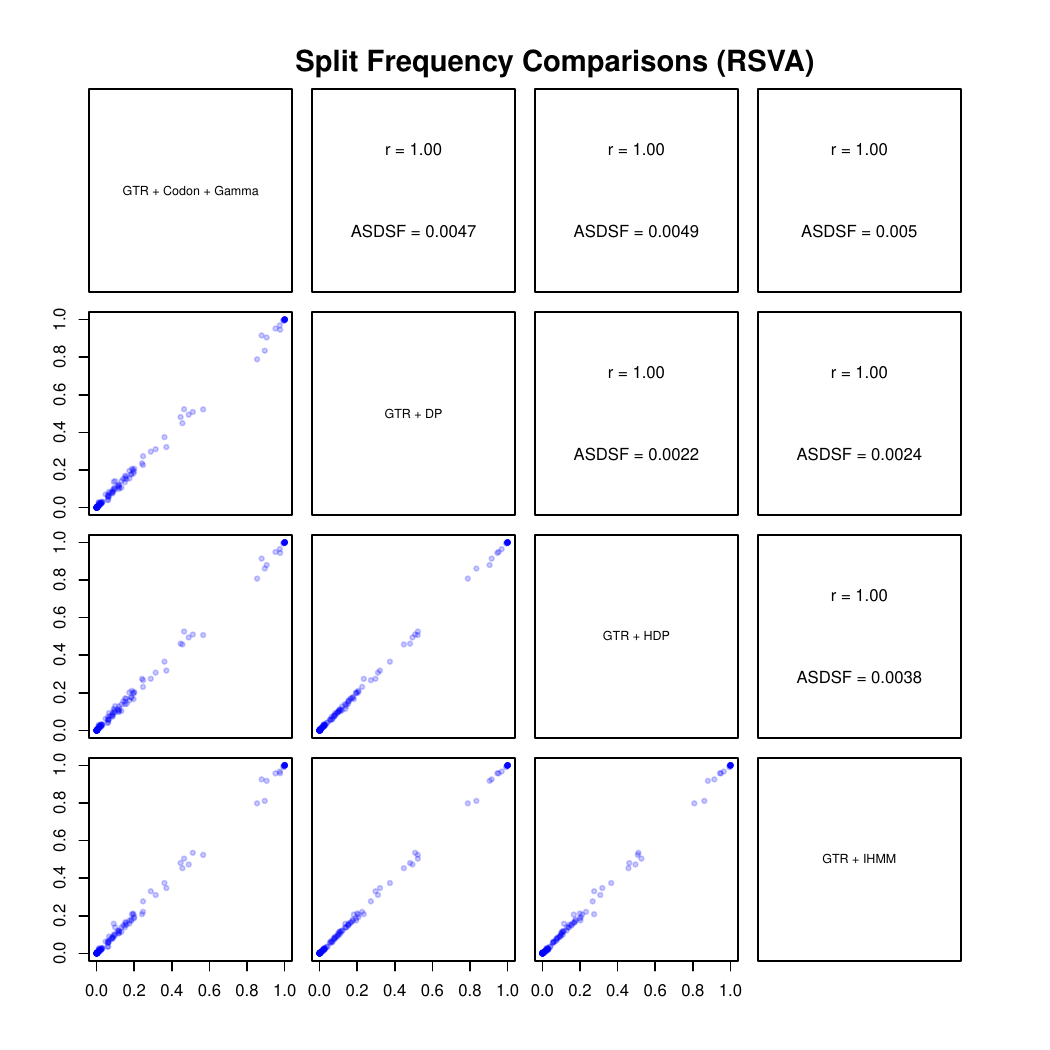} 
  \caption{Comparison of split frequencies in posterior samples from analyses of respiratory syncytial virus subgroup A data using different models for across-site evolutionary variation. All models use the GTR nucleotide substitution model. For each pair of models, the frequencies of clades in the posterior samples are plotted against each other in the plots below the diagonal of the figure, and the correlation and the average standard deviation of split frequencies (ASDSF) of the pair are shown above the diagonal. The diagonal entries correspond to different models for across-site evolutionary variation and indicate the specific pairwise comparisons that are made below and above the diagonal.}
     \label{fig:splitfreqrsvagtr}
\end{figure*}

\begin{figure*}[htb]
\centering
	 \includegraphics[width=1.0\textwidth, trim={0.5cm 0.5cm 0.5cm 0.5cm},clip]{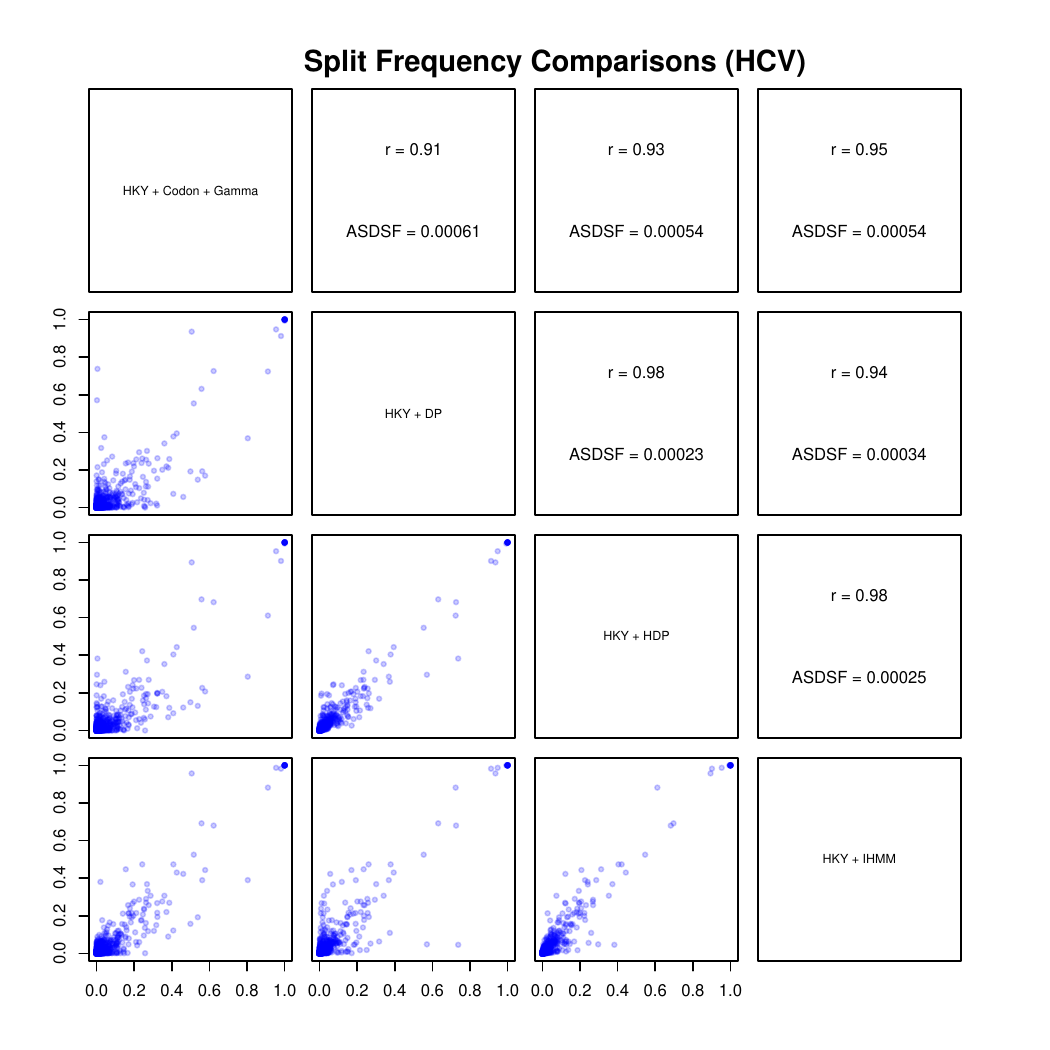} 
  \caption{Comparison of split frequencies in posterior samples from analyses of hepatitis C subtype 4 data using different models for across-site evolutionary variation. All models use the HKY nucleotide substitution model. For each pair of models, the frequencies of clades in the posterior samples are plotted against each other in the plots below the diagonal of the figure, and the correlation and the average standard deviation of split frequencies (ASDSF) of the pair are shown above the diagonal. The diagonal entries correspond to different models for across-site evolutionary variation and indicate the specific pairwise comparisons that are made below and above the diagonal.}
     \label{fig:splitfreqhcvhky}
\end{figure*}

\begin{figure*}[htb]
\centering
	 \includegraphics[width=1.0\textwidth, trim={0.5cm 0.5cm 0.5cm 0.5cm},clip]{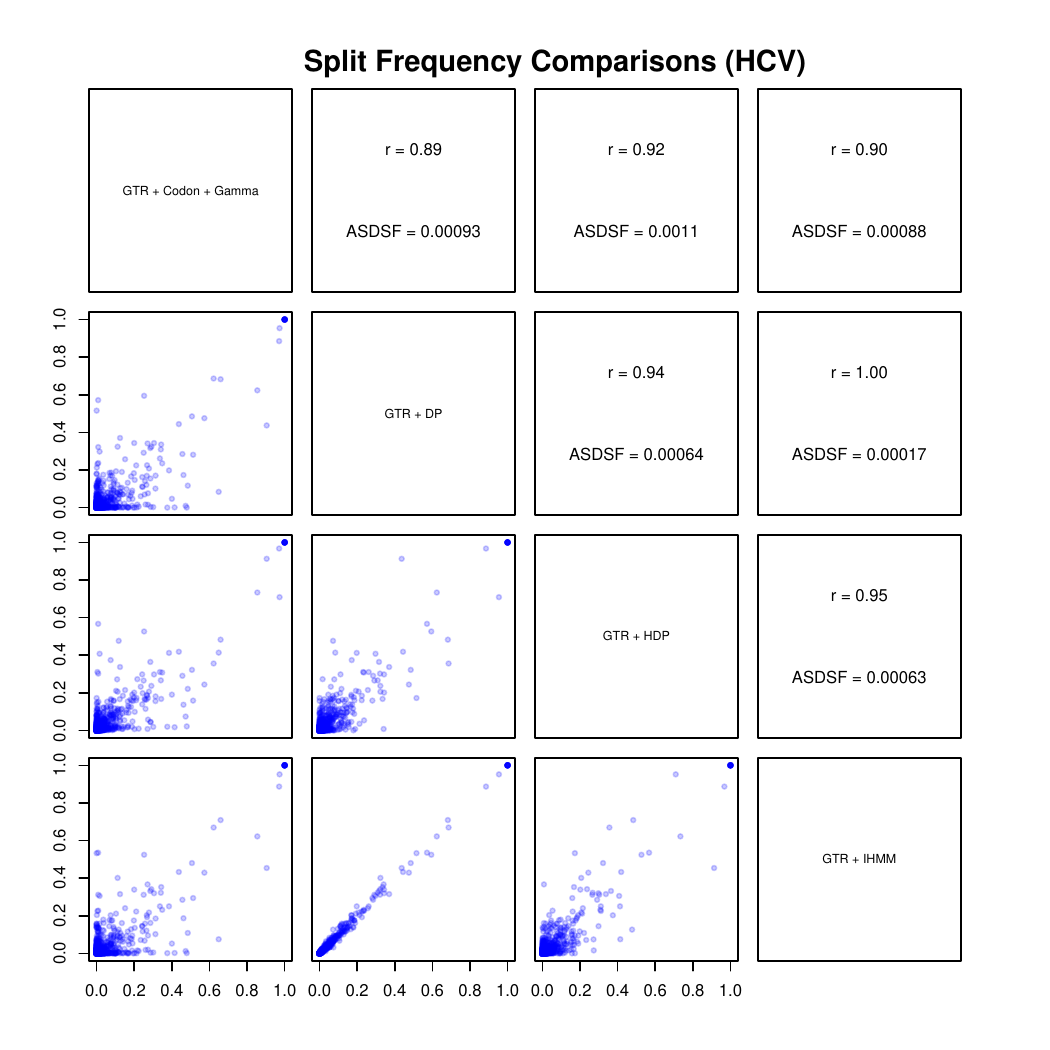} 
  \caption{Comparison of split frequencies in posterior samples from analyses of hepatitis C subtype 4 data using different models for across-site evolutionary variation. All models use the GTR nucleotide substitution model. For each pair of models, the frequencies of clades in the posterior samples are plotted against each other in the plots below the diagonal of the figure, and the correlation and the average standard deviation of split frequencies (ASDSF) of the pair are shown above the diagonal. The diagonal entries correspond to different models for across-site evolutionary variation and indicate the specific pairwise comparisons that are made below and above the diagonal.}
     \label{fig:splitfreqhcvgtr}
\end{figure*}

\begin{figure*}[htb]
\centering
	 \includegraphics[width=1.0\textwidth, trim={0.5cm 0.5cm 0.5cm 0.5cm},clip]{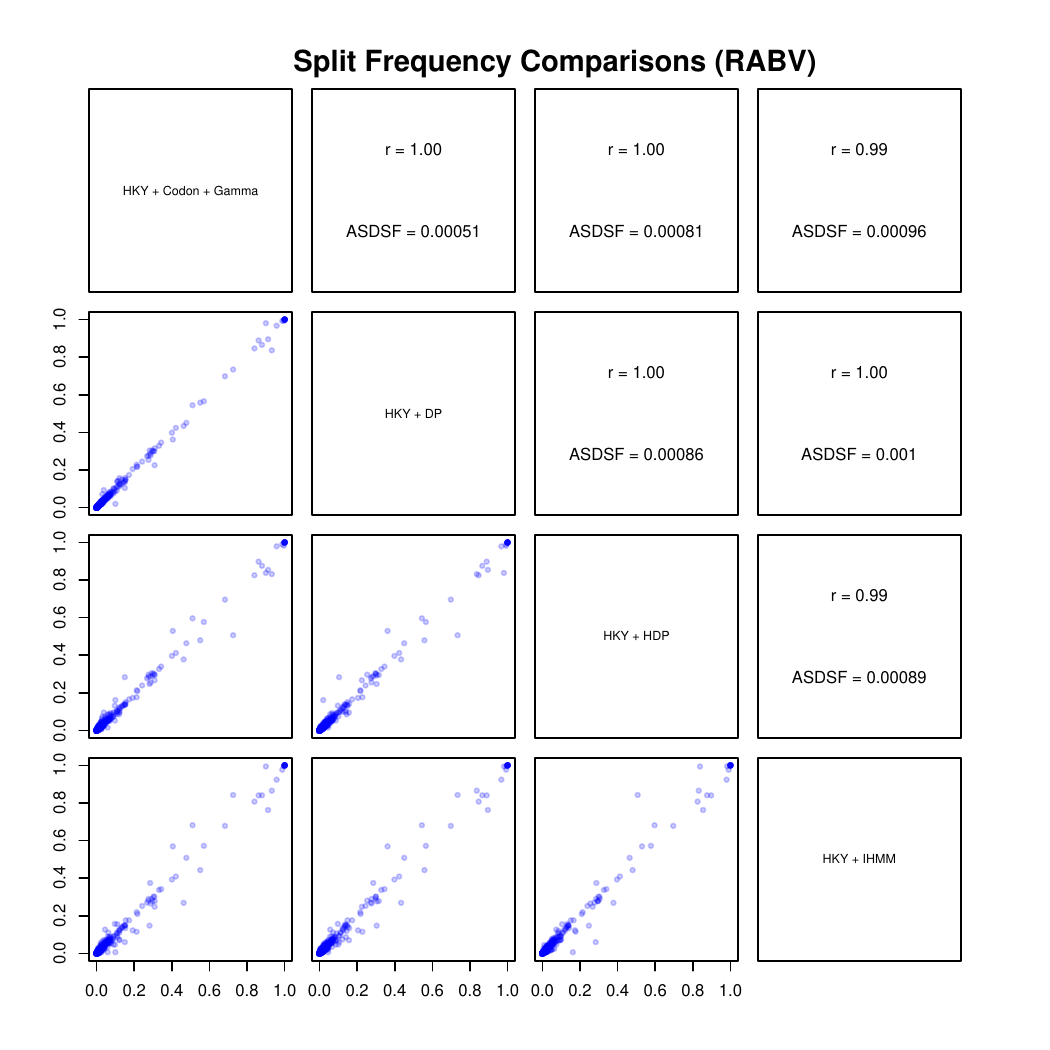} 
  \caption{Comparison of split frequencies in posterior samples from analyses of rabies virus data using different models for across-site evolutionary variation. All models use the HKY nucleotide substitution model. For each pair of models, the frequencies of clades in the posterior samples are plotted against each other in the plots below the diagonal of the figure, and the correlation and the average standard deviation of split frequencies (ASDSF) of the pair are shown above the diagonal. The diagonal entries correspond to different models for across-site evolutionary variation and indicate the specific pairwise comparisons that are made below and above the diagonal.}
     \label{fig:splitfreqrabvhky}
\end{figure*}

\begin{figure*}[htb]
\centering
	 \includegraphics[width=1.0\textwidth, trim={0.5cm 0.5cm 0.5cm 0.5cm},clip]{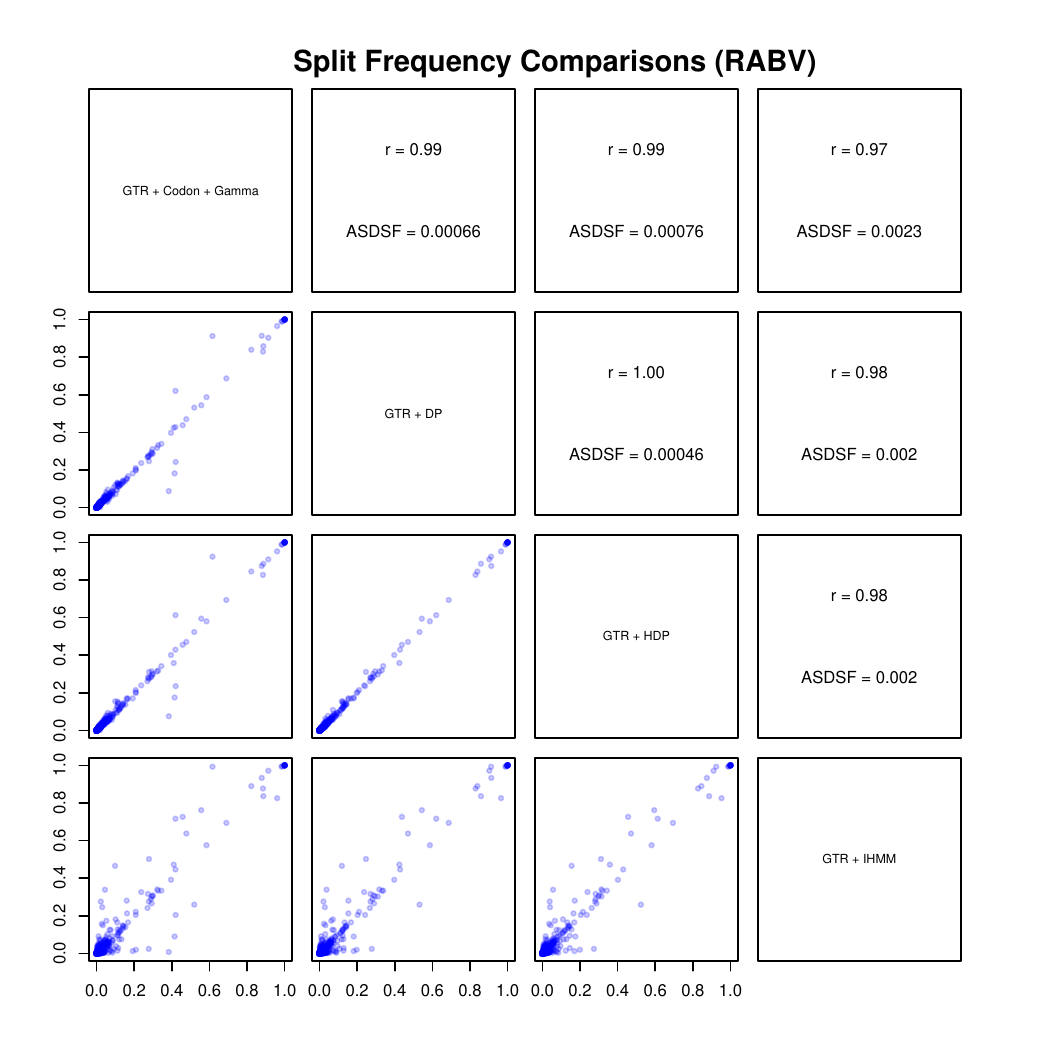} 
  \caption{Comparison of split frequencies in posterior samples from analyses of rabies virus data using different models for across-site evolutionary variation. All models use the GTR nucleotide substitution model. For each pair of models, the frequencies of clades in the posterior samples are plotted against each other in the plots below the diagonal of the figure, and the correlation and the average standard deviation of split frequencies (ASDSF) of the pair are shown above the diagonal. The diagonal entries correspond to different models for across-site evolutionary variation and indicate the specific pairwise comparisons that are made below and above the diagonal.}
     \label{fig:splitfreqrabvgtr}
\end{figure*}

\begin{figure*}[htb]
\centering
	 \includegraphics[width=1.0\textwidth, trim={0.3cm 0.5cm 0.3cm 0.7cm},clip]{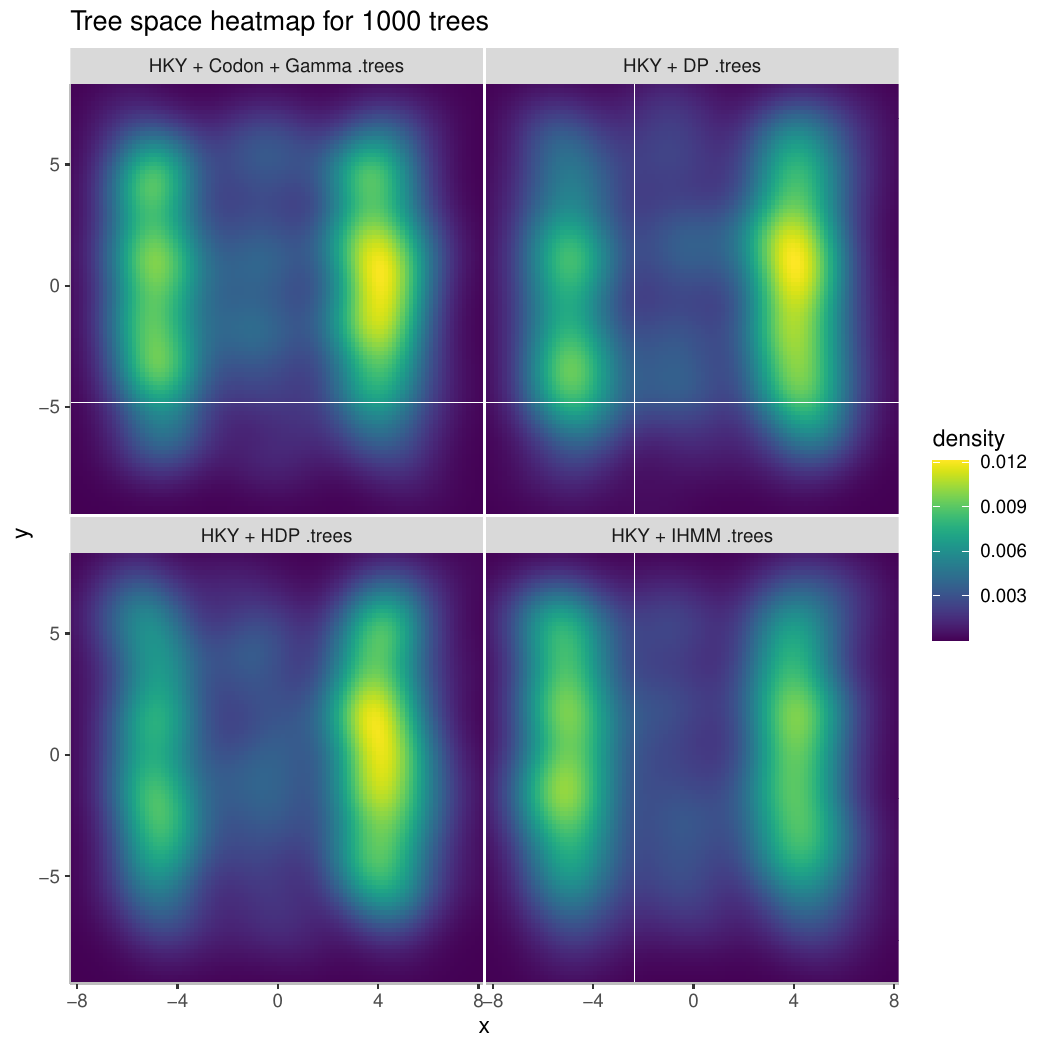} 
  \caption{Comparison of phylogenetic treespace sampled under different models for across-site evolutionary variation in analyses of respiratory syncytial virus subgroup A data. All models use the HKY nucleotide substitution model. The posterior samples for each model and the best fitting standard models (Codon + Gamma) are depicted as a heatmap of a two dimensional representation of treespace based on multidimensional scaling.}
     \label{fig:heatmaprsvahky}
\end{figure*}

\begin{figure*}[htb]
\centering
	 \includegraphics[width=1.0\textwidth, trim={0.3cm 0.5cm 0.3cm 0.7cm},clip]{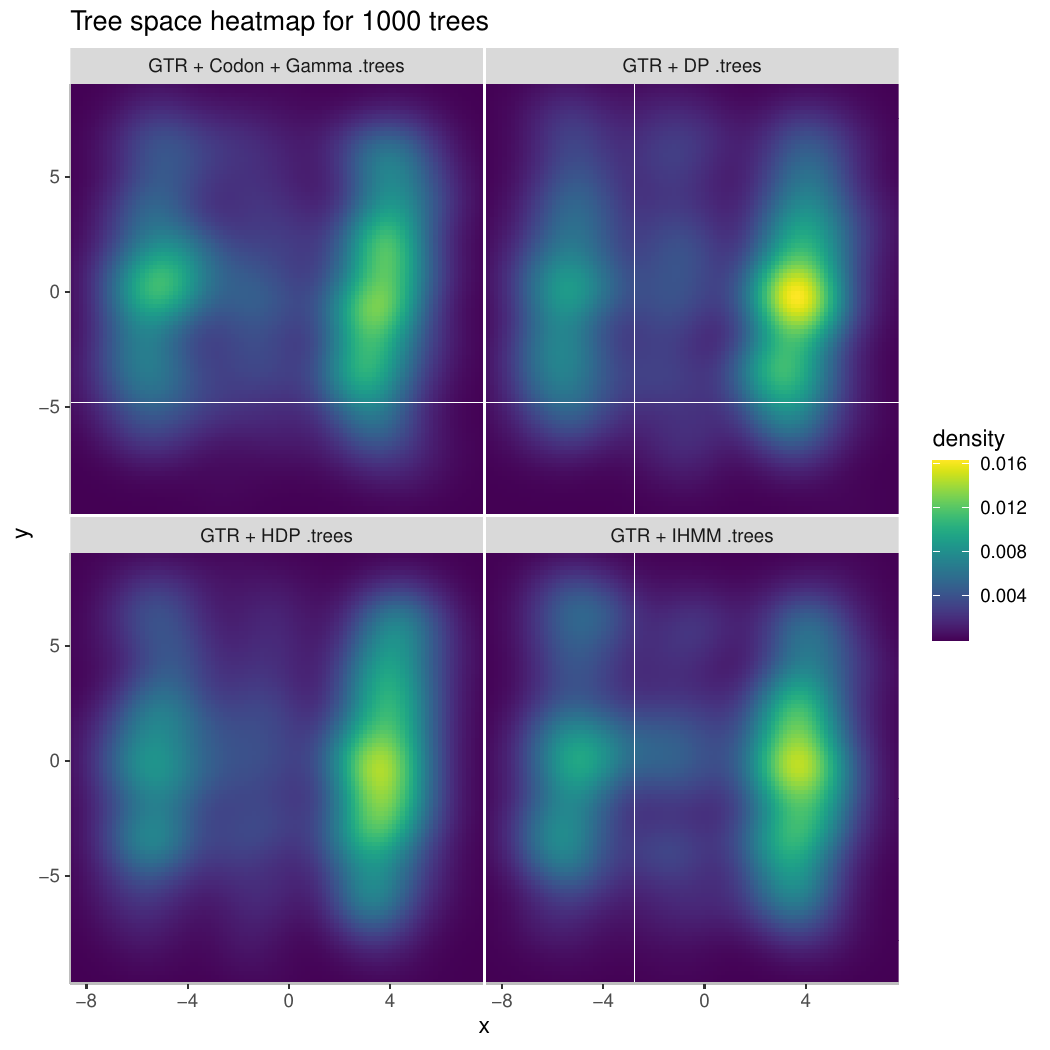} 
  \caption{Comparison of phylogenetic treespace sampled under different models for across-site evolutionary variation in analyses of respiratory syncytial virus subgroup A data. All models use the GTR nucleotide substitution model. The posterior sample for each model is depicted as a heatmap of a two dimensional representation of treespace based on multidimensional scaling.}
     \label{fig:heatmaprsvagtr}
\end{figure*}

\begin{figure*}[htb]
\centering
	 \includegraphics[width=1.0\textwidth, trim={0.3cm 0.5cm 0.3cm 0.7cm},clip]{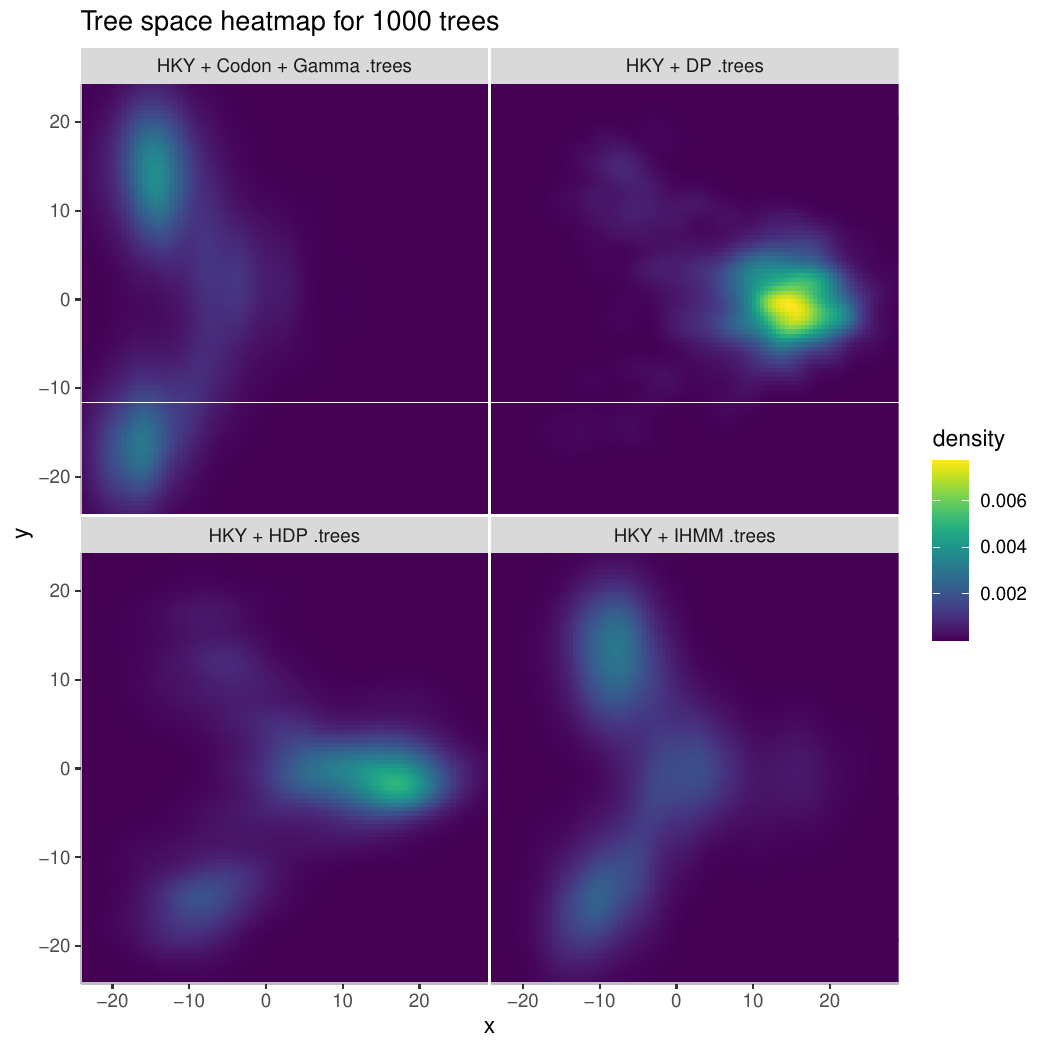} 
  \caption{Comparison of phylogenetic treespace sampled under different models for across-site evolutionary variation in analyses of hepatitis C subtype 4 data. All models use the HKY nucleotide substitution model. The posterior sample for each model is depicted as a heatmap of a two dimensional representation of treespace based on multidimensional scaling.}
     \label{fig:heatmaphcvhky}
\end{figure*}

\begin{figure*}[htb]
\centering
	 \includegraphics[width=1.0\textwidth, trim={0.3cm 0.5cm 0.3cm 0.7cm},clip]{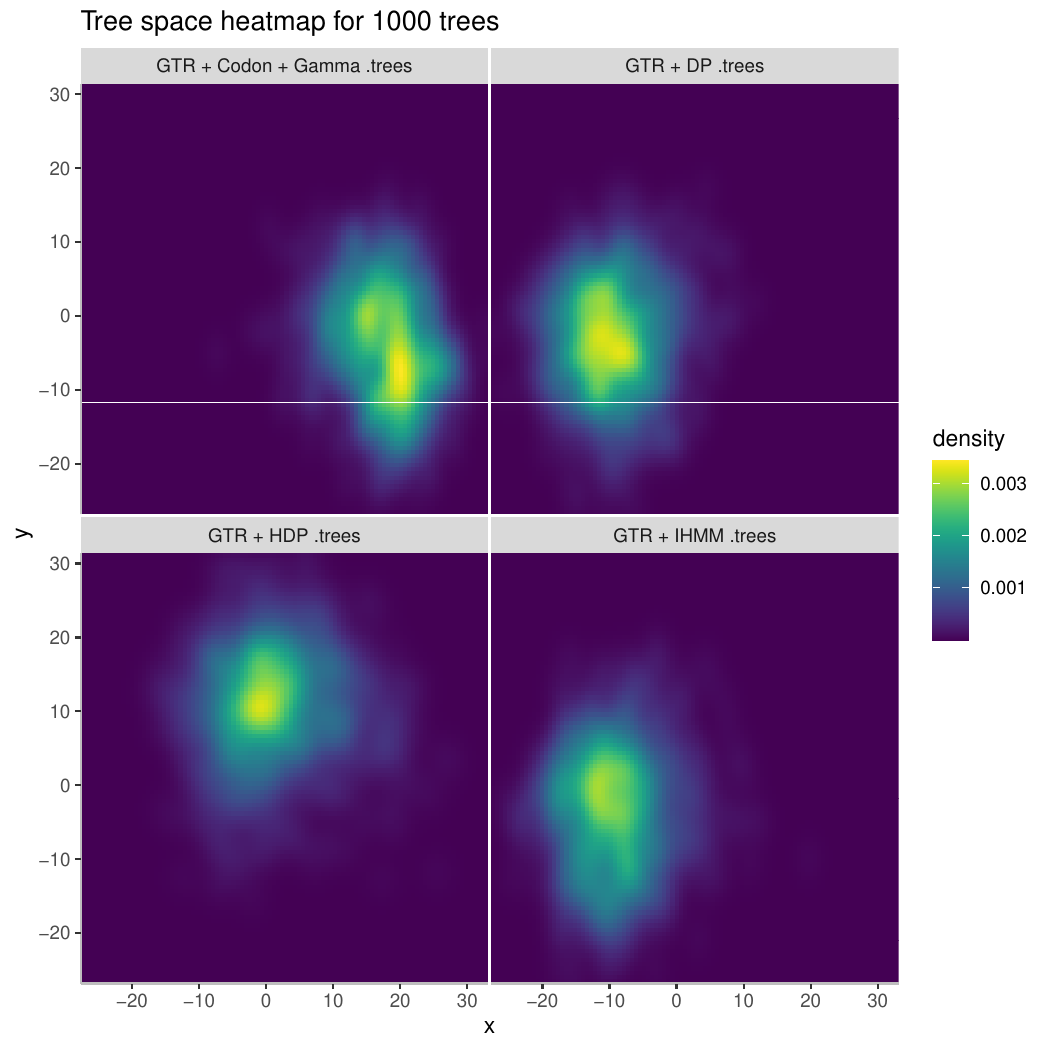} 
  \caption{Comparison of phylogenetic treespace sampled under different models for across-site evolutionary variation in analyses of hepatitis C subtype 4 data. All models use the GTR nucleotide substitution model. The posterior sample for each model is depicted as a heatmap of a two dimensional representation of treespace based on multidimensional scaling.}
     \label{fig:heatmaphcvgtr}
\end{figure*}

\begin{figure*}[htb]
\centering
	 \includegraphics[width=1.0\textwidth, trim={0.3cm 0.5cm 0.3cm 0.7cm},clip]{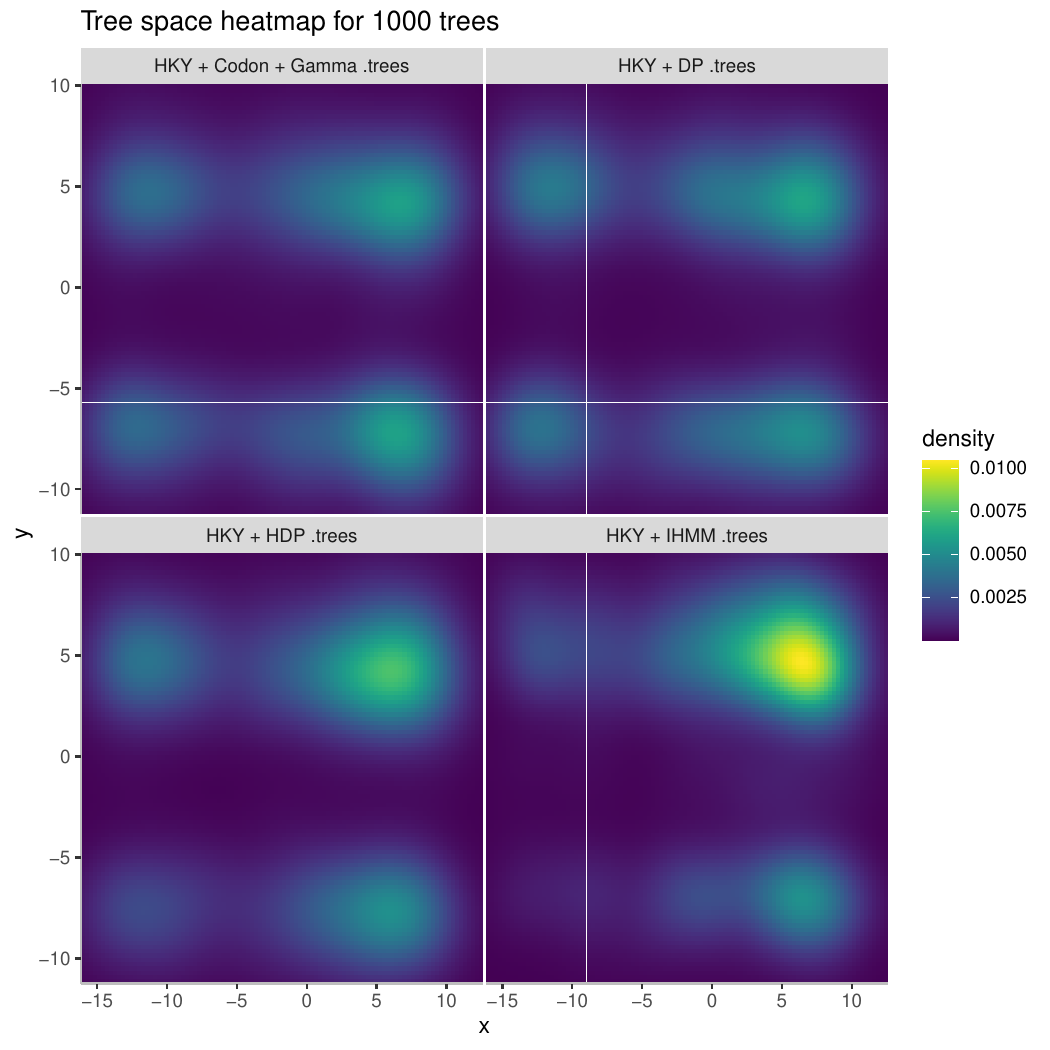} 
  \caption{Comparison of phylogenetic treespace sampled under different models for across-site evolutionary variation in analyses of rabies virus data. All models use the HKY nucleotide substitution model. The posterior sample for each model is depicted as a heatmap of a two dimensional representation of treespace based on multidimensional scaling.}
     \label{fig:heatmaprabvhky}
\end{figure*}

\begin{figure*}[htb]
\centering
	 \includegraphics[width=1.0\textwidth, trim={0.3cm 0.5cm 0.3cm 0.7cm},clip]{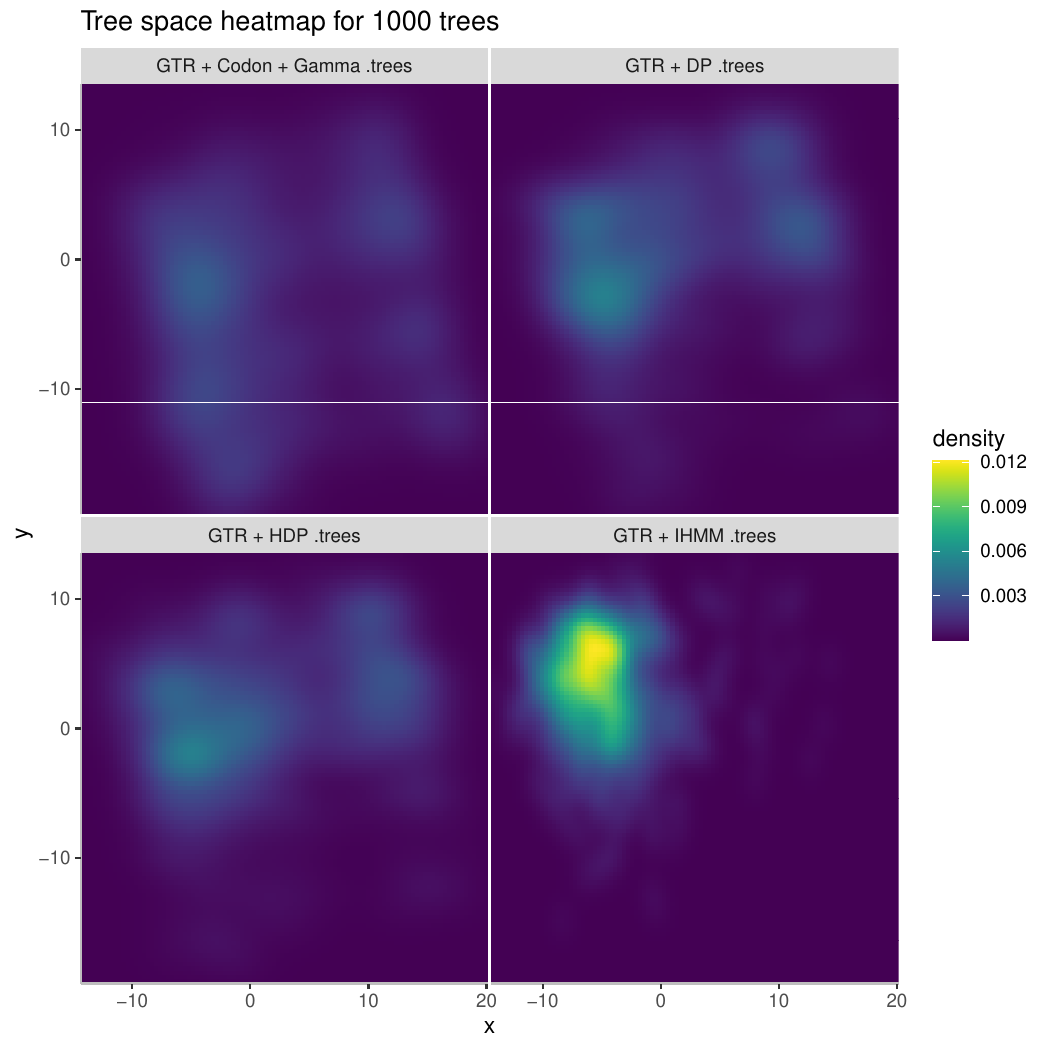} 
  \caption{Comparison of phylogenetic treespace sampled under different models for across-site evolutionary variation in analyses of rabies virus data. All models use the GTR nucleotide substitution model. The posterior sample for each model is depicted as a heatmap of a two dimensional representation of treespace based on multidimensional scaling.}
     \label{fig:heatmaprabvgtr}
\end{figure*}

\clearpage

\subsection{Across-Site Variation in Substitution Rates and Transition/Transversion Rates in Analyses Based on HKY Models}

We illustrate the variation in substitution rates and transition/transversion rates across alignment sites in the best fitting HKY-based models for the respiratory syncytial virus subgroup A (Figure A16), hepatitis C subtype 4 (Figure A17), and rabies virus data sets (Figures A18-A20). For each alignment site of each data set, we depict the posterior means of the rates as black dots and the corresponding 95$\%$ Bayesian credibility intervals as gray bars.

\begin{figure*}[htb]
\centering
	\begin{tabular}{@{}c@{}}
  \includegraphics[width=1.0\textwidth]{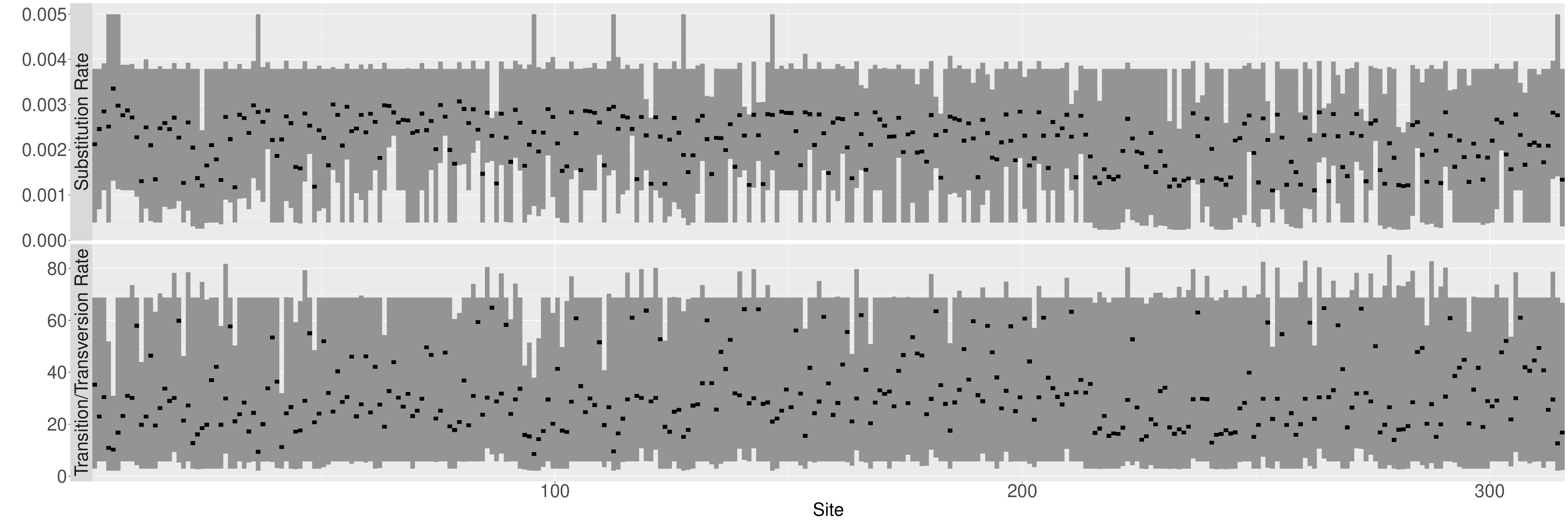} \\
    \includegraphics[width=1.0\textwidth]{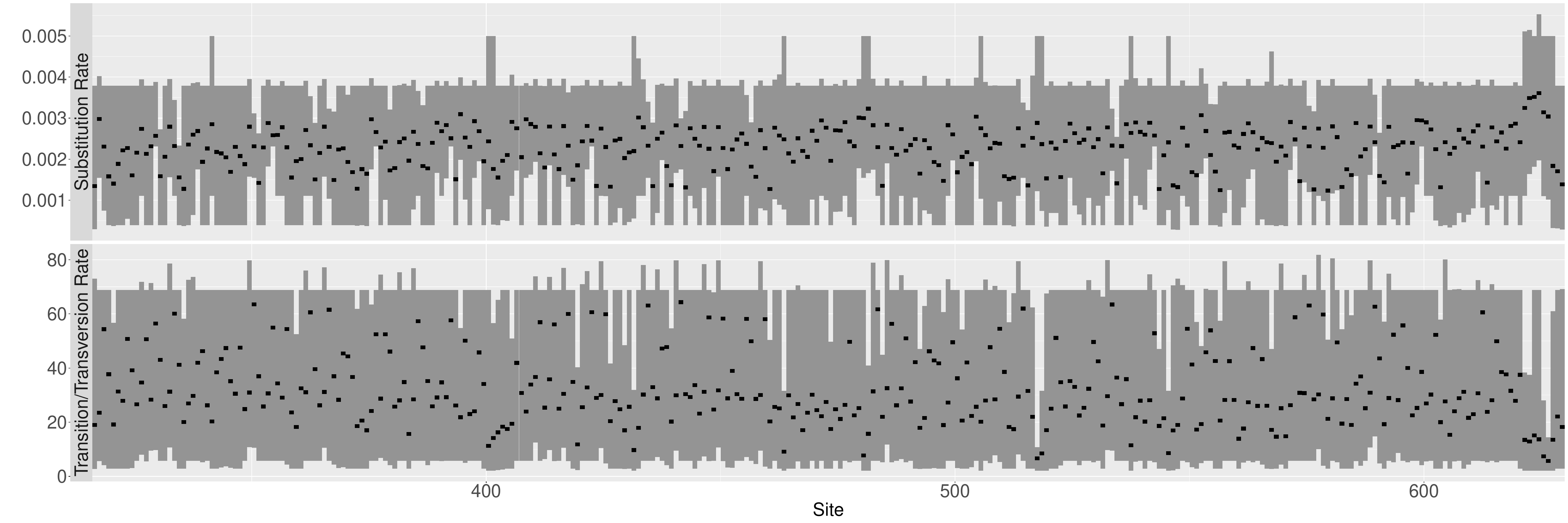}     
 	 \end{tabular}

  \caption{Site-specific variation in substitution rate and transition/transversion rate in IHMM+HKY analysis of respiratory syncytial virus subgroup A data. The black dots represent posterior mean estimates and gray bars represent 95$\%$ Bayesian credibility intervals.}
     \label{fig:siteratesrsva_ihmm}
\end{figure*}

\begin{figure*}[htb]
\centering
	\begin{tabular}{@{}c@{}}
  \includegraphics[width=1.0\textwidth]{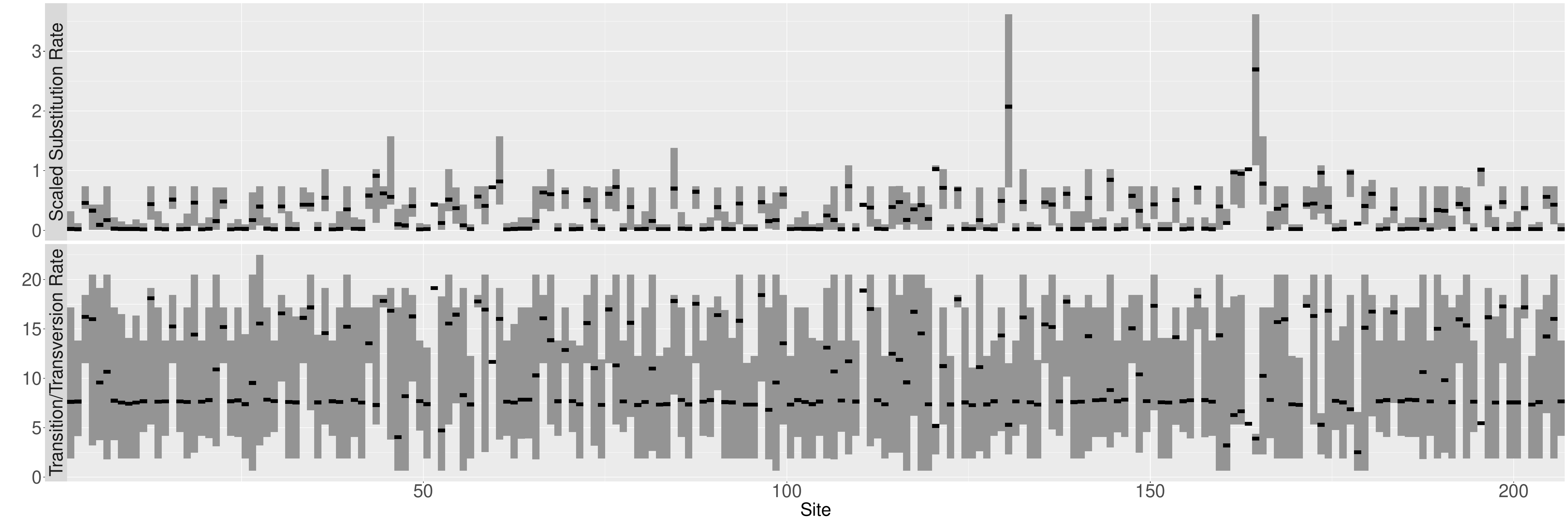} \\
    \includegraphics[width=1.0\textwidth]{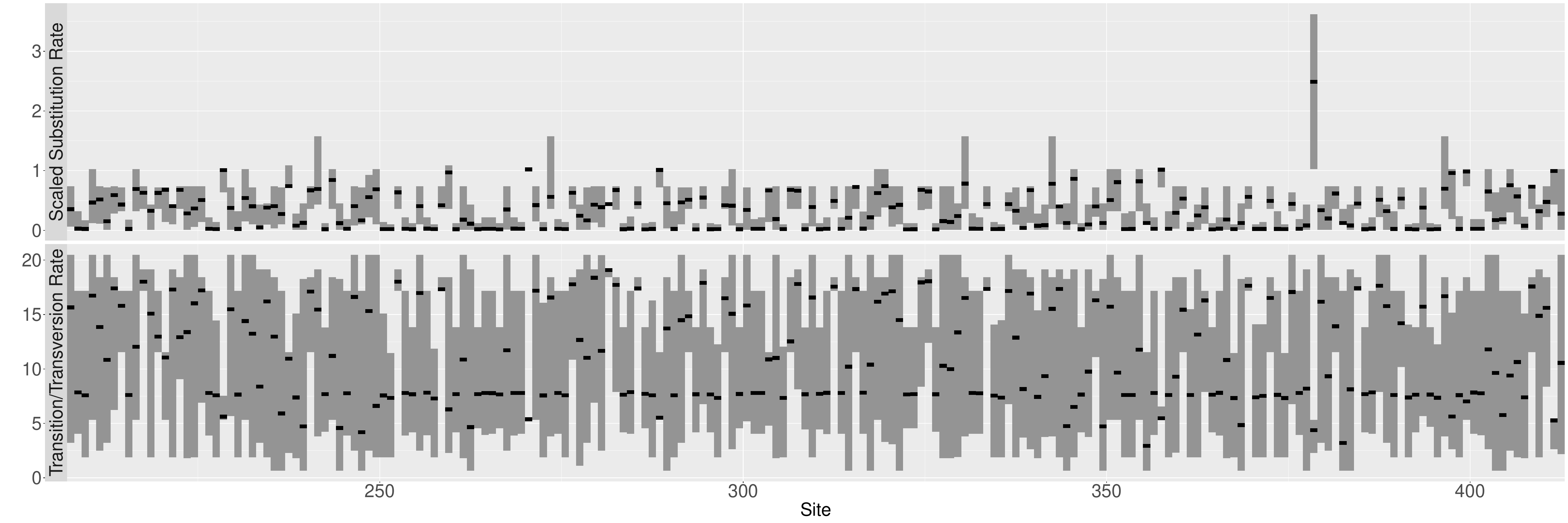}     
 	 \end{tabular}
  \caption{Site-specific variation in scaled substitution rate and transition/transversion rate in DP+HKY analysis of hepatitis C subtype 4 data. The black dots represent posterior mean estimates and gray bars represent 95$\%$ Bayesian credibility intervals.}
     \label{fig:siterateshcv_dp}
\end{figure*}

\begin{figure*}[htb]
\centering
	\begin{tabular}{@{}c@{}}
  \includegraphics[width=1.0\textwidth]{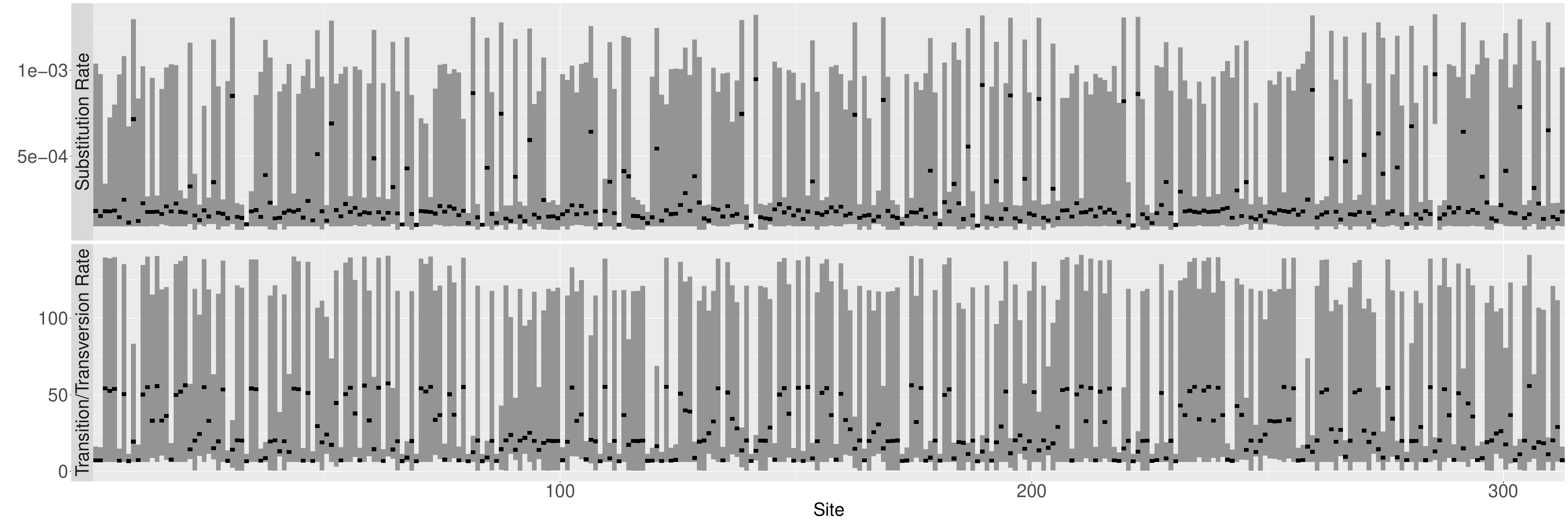} \\
    \includegraphics[width=1.0\textwidth]{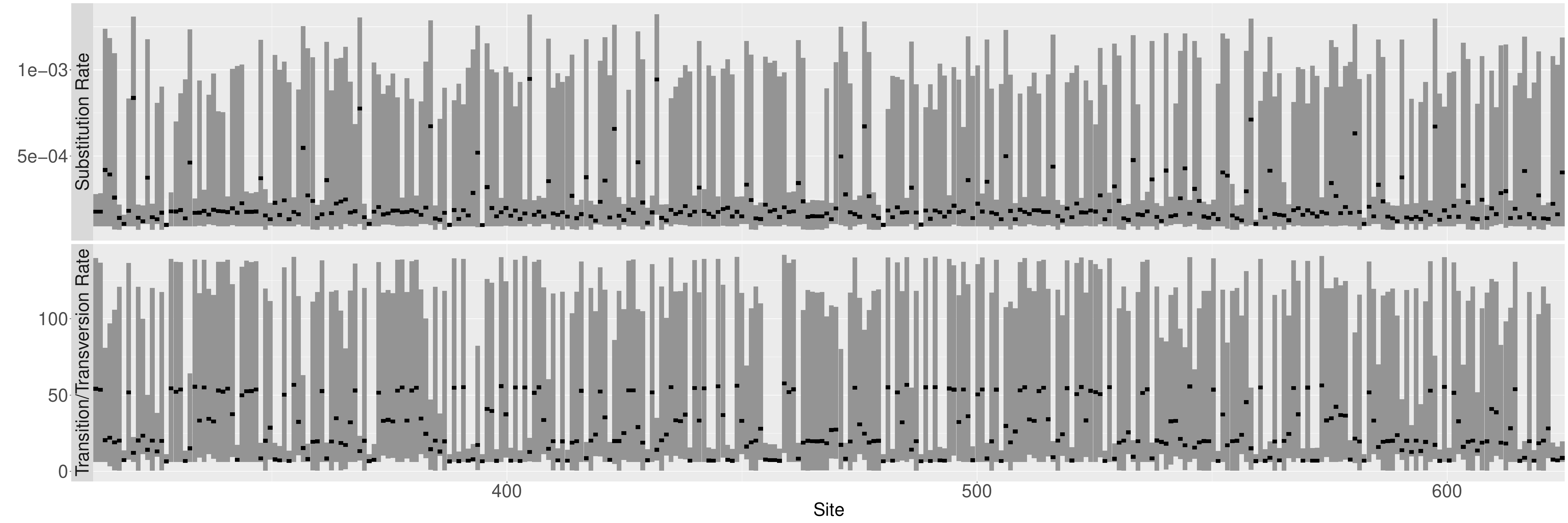} \\     
        \includegraphics[width=1.0\textwidth]{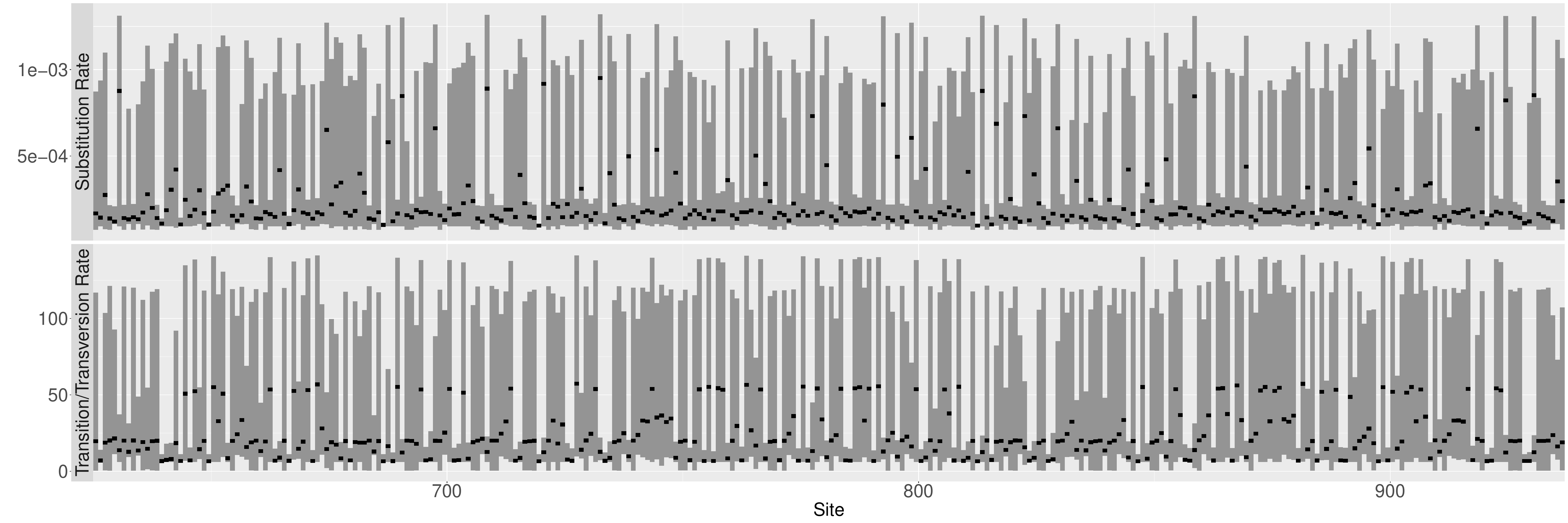} \\    
    \includegraphics[width=1.0\textwidth]{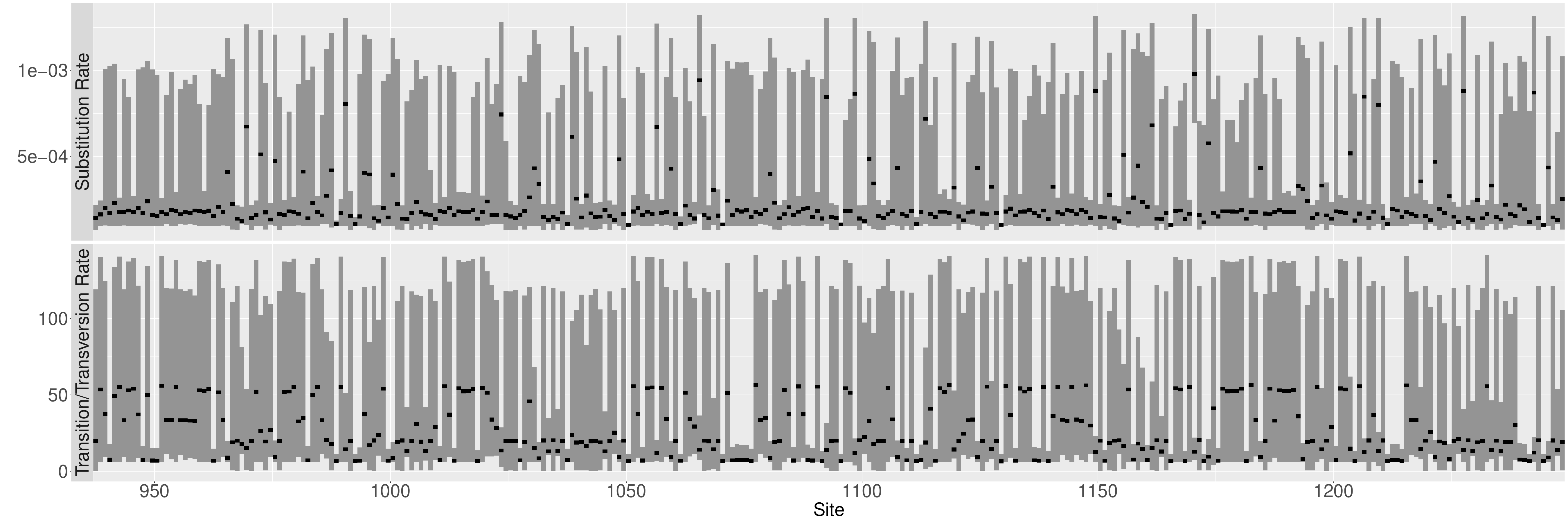}   
 	 \end{tabular}
  \caption{Site-specific variation in substitution rate and transition/transversion rate for sites 1-1248 in IHMM+HKY analysis of rabies virus data. The black dots represent posterior mean estimates and gray bars represent 95$\%$ Bayesian credibility intervals.}
     \label{fig:siteratesrabv_ihmm1}
\end{figure*}

\begin{figure*}[htb]
\centering
	\begin{tabular}{@{}c@{}}
  \includegraphics[width=1.0\textwidth]{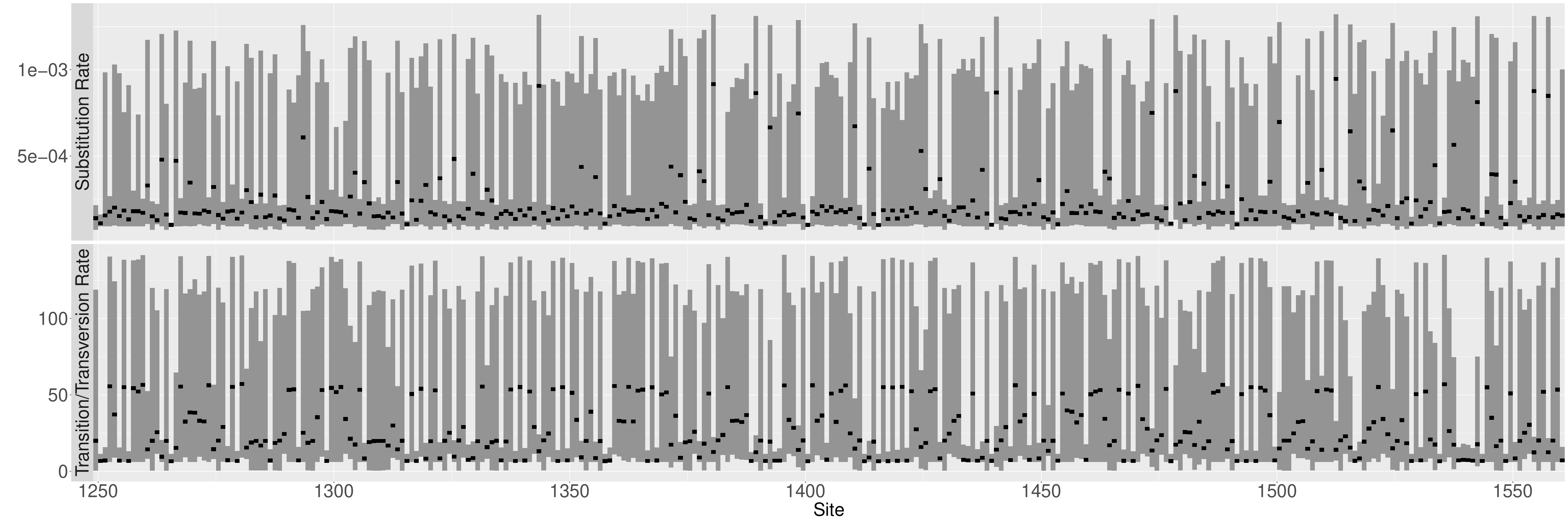} \\
    \includegraphics[width=1.0\textwidth]{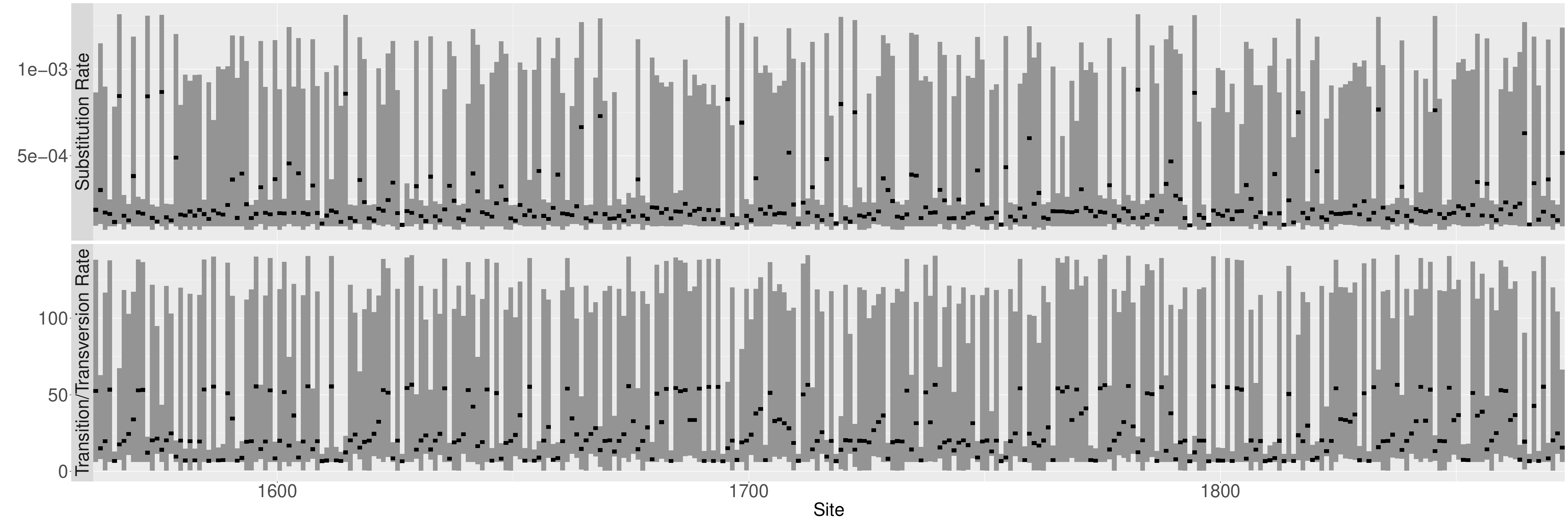} \\     
        \includegraphics[width=1.0\textwidth]{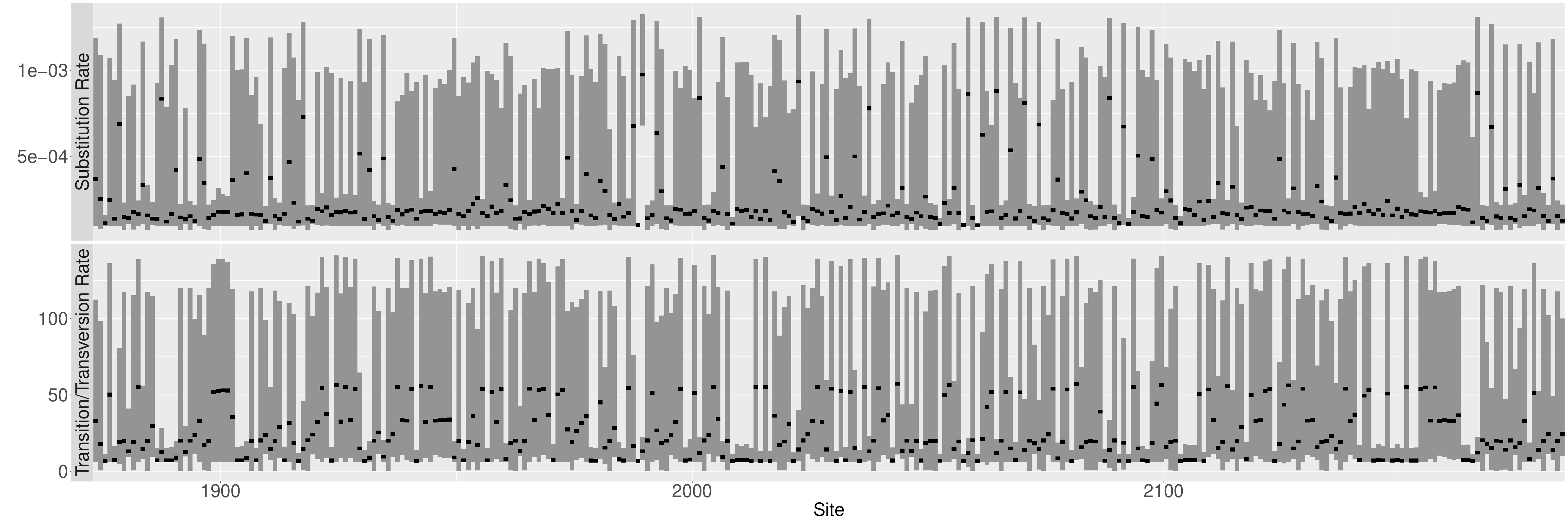} \\    
    \includegraphics[width=1.0\textwidth]{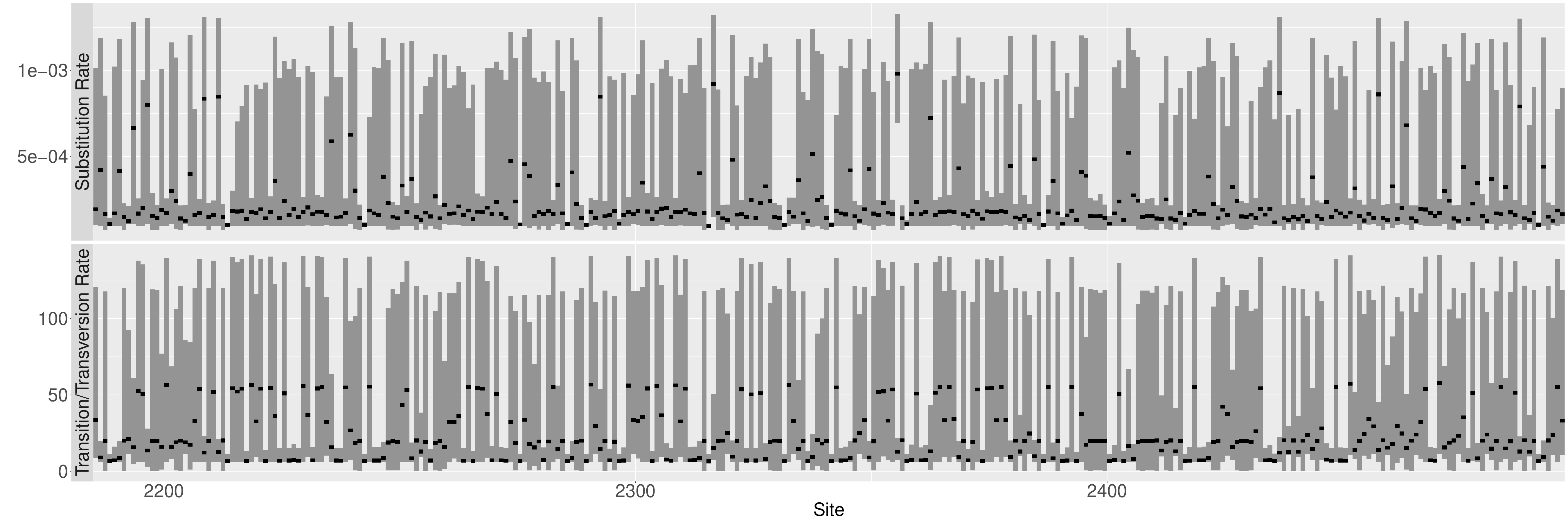}   
 	 \end{tabular}
  \caption{Site-specific variation in substitution rate and transition/transversion rate for sites 1249-2496 in IHMM+HKY analysis of rabies virus data. The black dots represent posterior mean estimates and gray bars represent 95$\%$ Bayesian credibility intervals.}
     \label{fig:siteratesrabv_ihmm2}
\end{figure*}

\begin{figure*}[htb]
\centering
	\begin{tabular}{@{}c@{}}
  \includegraphics[width=1.0\textwidth]{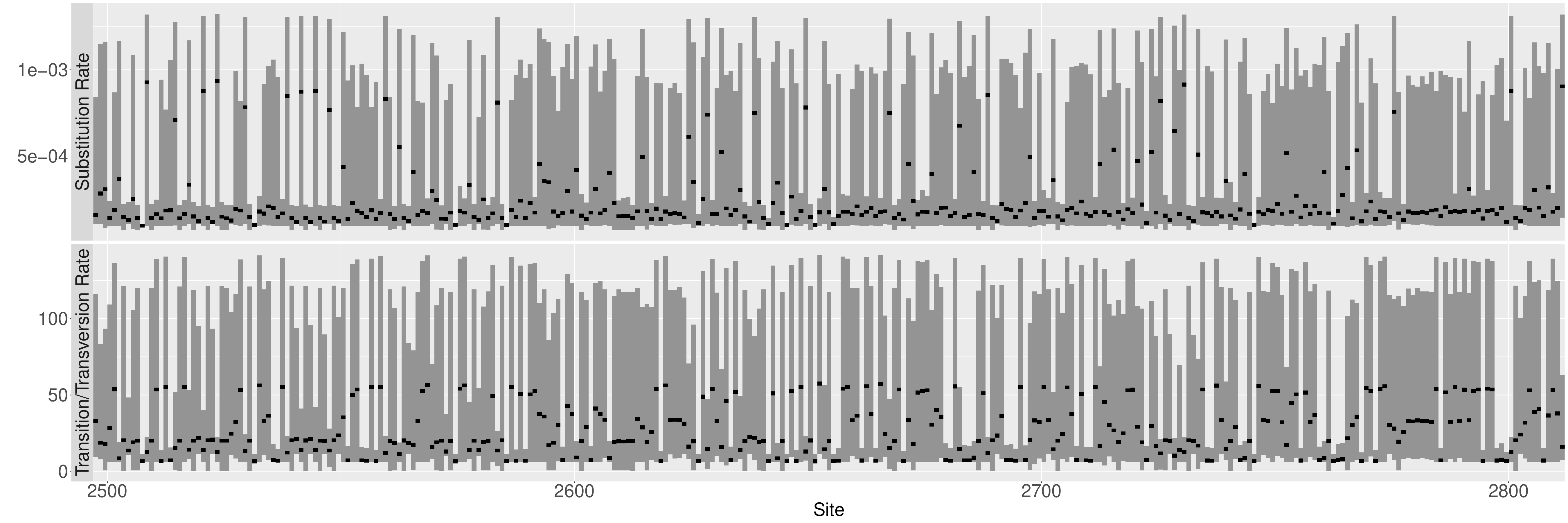} 
 	 \end{tabular}
  \caption{Site-specific variation in substitution rate and transition/transversion rate for sites 2497-2811 in IHMM+HKY analysis of rabies virus data. The black dots represent posterior mean estimates and gray bars represent 95$\%$ Bayesian credibility intervals.}
     \label{fig:siteratesrabv_ihmm3}
\end{figure*}

\end{document}